\documentclass{aa}

\usepackage{amsmath}
\usepackage{graphicx}
\usepackage{caption}
\usepackage{subcaption}
\usepackage[varg]{txfonts}
\usepackage[printonlyused]{acronym}
\usepackage{xr-hyper}
\usepackage{hyperref}
\hypersetup{
    colorlinks=true,
    linkcolor=blue,
    citecolor=blue,
    }
\usepackage{ulem}
\usepackage{nicematrix}

\acrodef{tde}[TDE]{Tidal Disruption Event}
\acrodef{ztf}[ZTF]{Zwicky Transient Facility}
\acrodef{lsst}[LSST]{Legacy Survey of Space and Time}
\acrodef{bh}[BH]{black hole}
\acrodef{bb}[BB]{black body}
\acrodef{bhmf}[BHMF]{black hole mass function}
\acrodef{gsmf}[GSMF]{galaxy stellar mass function}

\usepackage{natbib}
\bibpunct{(}{)}{;}{a}{}{,}

\begin{document}

\title{Rates of strongly lensed tidal disruption events}
\subtitle{A comprehensive investigation of black hole, luminosity, and temperature dependencies}
\author{E.~Mamuzic \inst{1,2} \and T.~Ryu \inst{2,3,4} \and S.~H.~Suyu \inst{1,2} \and K.~Szekerczes \inst{5,2} \and S.~Huber \inst{2,1} \and L.~Dai \inst{6} \and M.~Oguri \inst{7,8}}
\institute{Technical University of Munich, TUM School of Natural Sciences, Physics Department, James-Franck-Straße 1, 85748 Garching, Germany \\e-mail: elias.mamuzic@tum.de \and Max Planck Institute for Astrophysics, Karl-Schwarzschild Str. 1, 85748 Garching, Germany \\e-mail: emamuzic@MPA-Garching.MPG.DE \and JILA, University of Colorado and National Institute of Standards and Technology, 440 UCB, Boulder, 80308 CO, USA \and Department of Astrophysical and Planetary Sciences, 391 UCB, Boulder, 80309 CO, USA \and Department of Astronomy and Astrophysics, and Institute for Gravitation and the Cosmos, Pennsylvania State University, 525 Davey Lab, 251 Pollock Road, University Park, PA 16802, USA \and Department of Physics, University of Hong Kong, Pokfulam Road, Hong Kong, People's Republic of China \and Center for Frontier Science, Chiba University, 1-33 Yayoicho, Inage, Chiba 263-8522, Japan \and Department of Physics, Graduate School of Science, Chiba University, 1-33 Yayoicho, Inage, Chiba 263-8522, Japan}
\date{Received 04/03/2025; accepted 09/07/2025}

\abstract {In the coming years, surveys such as the Rubin Observatory's \ac{lsst} are expected to increase the number of observed \acp{tde} substantially. We employed Monte Carlo integration to calculate the unlensed and lensed \ac{tde} rate as a function of limiting magnitude in the $u$, $g$, $r$, and $i$ bands. We investigated the impact of multiple luminosity models, \acp{bhmf}, and flare temperatures on the \ac{tde} rate. Notably, this includes a semi-analytical model, which enables the determination of the \ac{tde} temperature in terms of \ac{bh} mass. We predict the highest unlensed \ac{tde} rate to be in the $g$ band. It ranges from $16$ to $5,440\;\mathrm{yr}^{-1}\;(20,000\;\mathrm{deg}^2)^{-1}$ for the \ac{ztf}, and it is more consistent with the observed rate at the low end. For \ac{lsst}, we expect a rate in the $g$ band between $3,580$ and $82,060\;\mathrm{yr}^{-1}\;(20,000\;\mathrm{deg}^2)^{-1}$. A higher theoretical prediction is within reason, as we do not consider observational effects such as completeness. The unlensed and lensed \ac{tde} rates are insensitive to the redshift evolution of the \ac{bhmf}, even for \ac{lsst} limiting magnitudes. The best band for detecting lensed \acp{tde} is also the $g$ band. Its predicted rates range from $0.43$ to $15\;\mathrm{yr}^{-1}\;(20,000\;\mathrm{deg}^2)^{-1}$ for \ac{lsst}. The scatter of predicted rates reduces when we consider the fraction of lensed \acp{tde}; that is, only a few in ten thousand \acp{tde} will be lensed. Despite the large scatter in the rates of lensed \acp{tde}, our comprehensive considerations of multiple models suggest that lensed \acp{tde} will occur in the $10$-year \ac{lsst} lifetime, providing an exciting prospect for detecting such events. We expect the median redshift of a lensed \ac{tde} to be between $1.5$ and $2$. In this paper, we additionally report on lensed \ac{tde} properties, such as the \ac{bh} mass and time delays.}

\maketitle

\section{Introduction}
    When a star ventures too close to a \acf{bh}, tidal forces tear it apart. We can observe such an event as a bright transient, a so-called \aclu{tde} \citep[\ac{tde};][]{clasic_tde,tdes_probe_dormant_galaxies}. These events typically last one to two years for main sequence stars \citep[e.g.,][]{TDE_general_review_article}, but in rare cases, they can extend much longer than that~\citep{tdes_last_one_year}.
    \acp{tde} are multi-messenger transients that produce electromagnetic radiation \citep[e.g.,][]{first_tde_detection,more_tde_detection}, neutrinos \citep[e.g.,][]{tde_neutrino_candidates,multimessenger_tdes}, and gravitational wave signatures \citep[e.g.,][]{gw_background_from_tdes,grav_waves_from_TDEs,multimessenger_tdes}.
    Even today, by analyzing the light of the \ac{tde}, we can infer critical properties such as the mass of the \acp{bh} and the disrupted stars \citep[e.g.,][]{luminosity_temperature_mass_dependence_Ryu_2020} and gain insights into \ac{bh} demographics and accretion physics \citep[e.g.,][]{tdes_probe_dormant_galaxies,tde_as_probes_of_demo_and_phys}.
    Understanding TDE rates and their observational characteristics is vital for detecting them in current and future surveys such as the \aclu{ztf}~\citep[\ac{ztf};][]{ztf_survey} and the Rubin Observatory's \aclu{lsst}~\citep[\ac{lsst};][]{lsst_survey}. Estimating the rate of \acp{tde} across different survey depths allows for better planning of observational campaigns and the design of strategies to maximize scientific returns \citep[e.g.,][]{lsst_rates,wlsst_rates}.
    
    A key advancement in this area of study is the inclusion of lensed \acp{tde} in our analysis. Gravitational lensing provides a powerful mechanism to magnify and detect \acp{tde} from deep within the cosmos. This topic has already been studied with electromagnetic radiation in \citet{TDE_rates_Szekerczes_2024} and \citet{detecability_strongly_lensed_tdes}, upon which we expand here. In the future, we could also expect gravitational waves to be lensed. This has, for example, been studied by \citet{lensed_grav_waves}. Beyond enhancing detectability, lensing offers the potential to probe cosmological parameters such as the Hubble constant \citep{hubble_constant_from_lensing} and the size of the emission region via microlensing effects, as demonstrated for active galactic nuclei \citep[e.g.,][]{agn_microlensing_2,agn_microlensing_1,agn_size_measurement}. Despite these promising applications, the abundance and characteristics of \acp{tde}, lensed or unlensed, remain uncertain \citep[e.g.,][]{identify_tdes}. Addressing this gap, we calculated the \ac{tde} rate for various models and assessed whether these events are numerous enough to serve as a tool for future research.
    
    In this work, we employed Monte Carlo simulations to calculate \ac{tde} rates as a function of limiting magnitude, following the methodology of \citet{TDE_rates_Szekerczes_2024}. Likewise, we generated mock catalogs to analyze the population of unlensed and lensed \acp{tde}. Estimating the \ac{tde} rate requires two primary components: a \ac{tde} luminosity model and the \acf{bhmf}. Building on the calculations of \citet{TDE_rates_Szekerczes_2024}, in addition to the phenomenological fit model~\citep{TDE_L1}, we incorporated the full semi-analytical stream-stream collision model for \acp{tde} \citep{theoretical_L2_model,luminosity_temperature_mass_dependence_Ryu_2020}. This model enables the determination of the \ac{tde} temperature as a function of \ac{bh} mass, eliminating the need to assume a temperature. In addition to the local \ac{bhmf} adopted in \citet{TDE_rates_Szekerczes_2024}, we included two more \acp{bhmf} that account for redshift dependence, enhancing the accuracy of our predictions. For lensed \acp{tde}, we also investigated the fraction of \acp{tde} that are lensed for the different models, leading to a more robust forecast.
    
    The structure of the paper is as follows: In Sect.~\ref{sec:theoryandmethodology}, we provide an overview of the relevant theoretical background, focusing on the \ac{tde} models used to calculate the \ac{tde} magnitude. We also detail our methodology for determining the unlensed \ac{tde} rate and extend it to create a mock \ac{tde} catalog. Sect.~\ref{sec:unlensed_rate} presents the results for the unlensed case, and we analyze the effects of various model parameters and compare them with current observations. In Sect.~\ref{sec:lensed_rate}, we shift our focus to the lensed rate, describing the creation of mock catalogs of lensed \acp{tde}. We then explore the lensed \ac{tde} rate, the fraction of lensed events, and the parameter distributions for the source \ac{tde}, the lens, and the lens system. Finally, we summarize our findings and conclusions in Sect.~\ref{sec:conclusion}. All mock data is published on Zenodo (https://doi.org/10.5281/zenodo.17727178).

\begin{figure*}[htbp]
    \centering
    \includegraphics[width=0.95\textwidth]{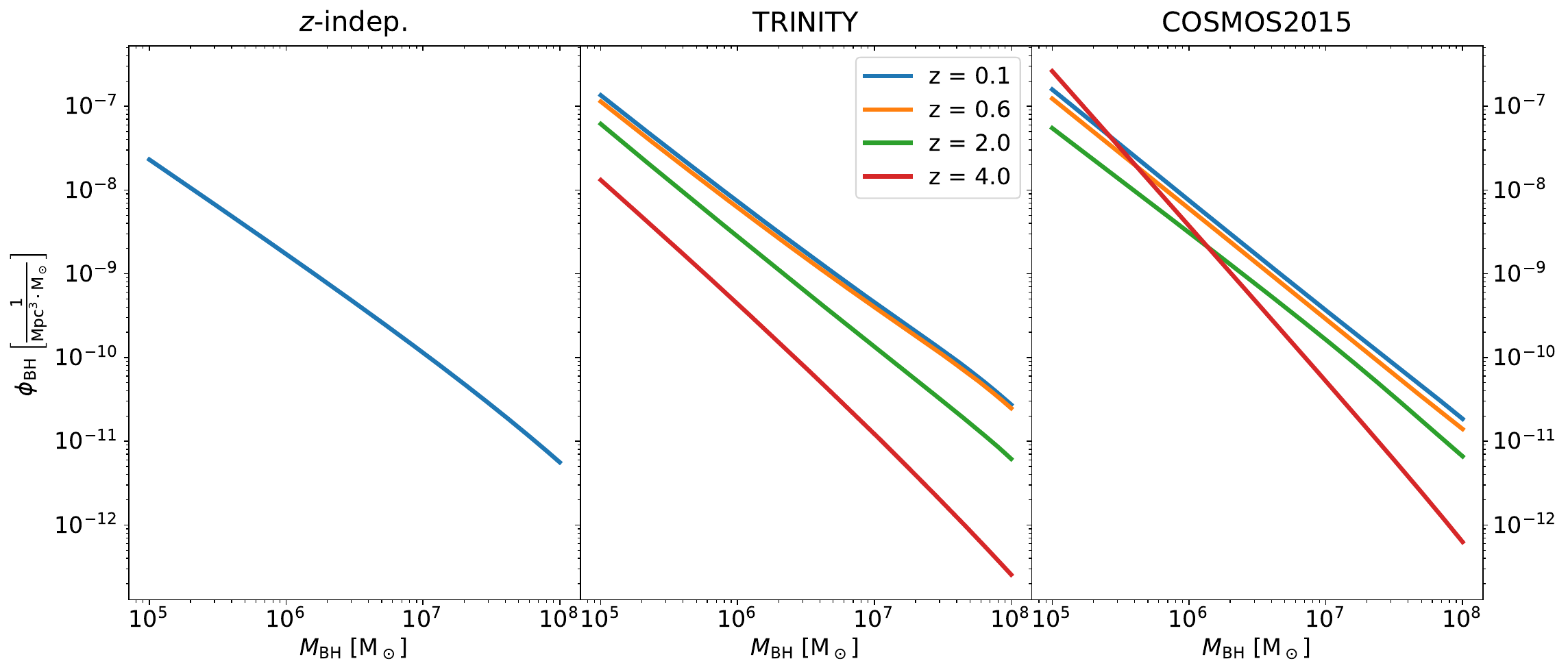}
    \caption{Three different \acp{bhmf} we consider showing significantly different redshift evolutions. Left panel: Redshift independent \ac{bhmf} from~\citet{derivation_of_the_BHMF_Gallo_2019}. Middle panel: \ac{bhmf} derived by the TRINITY model~\citep{z_dep_BHMF_TRINITY_1}. Right panel: \ac{bhmf} calculated by combining a \ac{gsmf}~\citep{GSMF} and a \ac{bh}-to-bulge mass relation~\citep{bh_bulge_mass_relation}. The $z$-independent \ac{bhmf} has a different normalization than the other two. Up to redshift of 2, TRINITY and COSMOS2015 agree closely.}
    \label{fig:all_bhmfs}
\end{figure*}
\section{TDE models and rate derivation}\label{sec:theoryandmethodology}
    \subsection{Rate integral}
    We followed \citet{TDE_rates_Szekerczes_2024} in the derivation of the unlensed TDE rate integral
    \begin{equation}
        N_\mathrm{TDE} = \iint \phi_\mathrm{BH}(M_\mathrm{BH}) \cdot \Gamma(M_\mathrm{BH}) \; \mathrm{d}M_\mathrm{BH} \; \frac{\mathrm{d}V}{\mathrm{d}z} \mathrm{d}z.
        \label{eq:total_tde_rate_eq}
    \end{equation}
    Here $N_\mathrm{TDE}$ is the number of \acp{tde} per year, $M_{\mathrm{BH}}$ is the \ac{bh} mass, $\Gamma$ is the \ac{tde} occurrence rate per galaxy, $\phi_{\mathrm{BH}}$ is the \ac{bhmf}, $V$ is the volume, and $z$ is the redshift. We have already changed the variable from volume to redshift using \citep[e.g.,][]{lensing_code_Oguri_2010}
    \begin{equation}
        \frac{\mathrm{d}V}{\mathrm{d}z} = \Omega \cdot \frac{D_{\mathrm{L}}^{2}}{1+z} \cdot c \cdot \frac{\mathrm{d}t}{\mathrm{d}z},
        \label{eq:volume_redshift_factor}
    \end{equation}
    where $\Omega$ is the solid angle corresponding to the survey area, $D_{\mathrm{L}}$ is the luminosity distance, $c$ is the speed of light, and $\mathrm{d}t/\mathrm{d}z = -1 /(H \cdot (1+z))$, where $H$ is the Hubble parameter.
    We adopted the \ac{tde} occurrence rate 
    \begin{equation}
        \frac{\Gamma(M_\mathrm{BH})}{\mathrm{yr}^{-1}} = 10^{-4.5} \cdot \left( \frac{M_\mathrm{BH}}{10^{6} \, \mathrm{M}_{\odot}} \right)^{-0.14}.
        \label{eq:tde_rate_per_year_per_galaxy}
    \end{equation}
    by \citet{tde_rate_per_year_Pfister_2020}.
    
    In addition to the redshift-independent \ac{bhmf} \citep{derivation_of_the_BHMF_Gallo_2019,black_hole_mass_function_Wong_2022}
    \begin{equation}
    \begin{aligned}
        \log_{10}\left( \frac{\phi_\mathrm{BH}}{\mathrm{Mpc}^{-3} \; \mathrm{M}_{\odot}^{-1}} \right) = &-9.82 - 1.10 \cdot \log_{10}\left( \frac{M_\mathrm{BH}}{10^{7} \; \mathrm{M}_{\odot}} \right) \\
        &- \left( \frac{M_\mathrm{BH}}{128 \cdot 10^{7} \; \mathrm{M}_{\odot}} \right)^{\frac{1}{\log(10)}},
    \end{aligned}
    \label{eq:black_hole_mass_function_z_indep}
    \end{equation}
    which is illustrated in the left panel of Fig.~\ref{fig:all_bhmfs}, we investigated two more. 
    First, we included the \ac{bhmf} derived by the TRINITY model~\citep{z_dep_BHMF_TRINITY_1}. They used a Markov chain Monte Carlo algorithm to fit a star-formation rate to the maximum circular halo velocity relation, a \ac{bh}-to-bulge mass relation, and an active \ac{bh} luminosity distribution to observed data. The derived \ac{bhmf} is shown in the middle panel of Fig.~\ref{fig:all_bhmfs}. This \ac{bhmf} displays an almost uniform redshift evolution across the \ac{bh} mass range. But the evolution only affects the \ac{bhmf} strongly for $z \gtrsim 2$.
    Second, we considered a \ac{bhmf} that is calculated by combining the \acf{gsmf} from~\citet{GSMF} and the \ac{bh}-to-bulge mass relation from~\citet{bh_bulge_mass_relation}
    \begin{equation}
        \frac{M_{\mathrm{BH}}}{10^{9} \; \mathrm{M}_{\odot}} = 0.49 \cdot \left( \frac{M_{\mathrm{bulge}}}{10^{11} \; \mathrm{M}_{\odot}} \right)^{1.16},
        \label{eq:bh_bulge_mass_relation}
    \end{equation}
    with $M_{\mathrm{bulge}}$ as the galaxy bulge mass. It has recently been suggested that this relation is redshift-dependent \citep[e.g.,][]{bh_mass_bulge_mass_redshift_dep}. However, we still assume no redshift dependence because the \ac{gsmf}, which is fitted in bins to the COSMOS2015 catalog \citep{cosmos2015_catalog}, already includes a strong, non-trivial redshift dependence. 
    As illustrated in the right panel of Fig.~\ref{fig:all_bhmfs}, this \ac{bhmf} agrees closely with the TRINITY \ac{bhmf} at $z = 0$, but not for larger redshifts. From $z = 2$ onward, the high-mass end decreases, while the low-mass end turns and increases. This is due to the extrapolation needed to reach such low \ac{bh} masses. The COSMOS2015 catalog only includes data for galaxies of just below $10^{10} \; \mathrm{M}_{\odot}$. This corresponds to a \ac{bh} mass of $10^{7.5} \; \mathrm{M}_{\odot}$, the upper end of our considered mass range. Assuming the fitting function is still valid, we extrapolated the relation down to a \ac{bh} mass of $10^{5} \; \mathrm{M}_{\odot}$.

\subsection{Magnitude bounds}
    The limiting magnitude of the survey gives the bounds to the integral in Eq.~\ref{eq:total_tde_rate_eq}. Hence, calculating the magnitude of a given \ac{tde} is essential. Following \citet{flux_to_magnitude}, as done in \citet{magnitude_and_flux_Huber_2021}, we calculated the magnitude in a band $X$:
    \begin{equation}
    \begin{aligned}
        m_{\mathrm{AB,}X} = &-2.5 \cdot \log_{10}\left( \frac{\int \mathrm{d}\lambda_\mathrm{obs} \; \lambda \cdot S_{X}(\lambda_\mathrm{obs}) \cdot F_{\lambda_\mathrm{obs}}}{\int \mathrm{d}\lambda_\mathrm{obs} \; S_{X}(\lambda_\mathrm{obs}) \cdot \frac{c}{\lambda_\mathrm{obs}}} \cdot \frac{\mathrm{cm}^{2}}{\mathrm{erg}} \right) \\ &- 48.6,
    \end{aligned}
    \label{eq:magnitudes}
    \end{equation}
    where $\lambda_{\mathrm{obs}}$ is the observed wavelength and $S_{X}(\lambda)$ is the band transmission function. We used the same rectangular approximations of the \ac{lsst} transmission functions as \citet{TDE_rates_Szekerczes_2024}. Finally, \citep[e.g.,][]{magnitude_and_flux_Huber_2021}
    \begin{equation}
        F_{\lambda_{\mathrm{obs}}} = \frac{A \cdot I_{\lambda_\mathrm{obs}}}{D_\mathrm{L}^{2} (1 + z)}
        \label{eq:observed_flux}
    \end{equation} 
    is the observed flux for an emitting region of area, $A$, with a spatially constant emitted specific intensity, $I_{\lambda_\mathrm{obs}}$. We also assumed the \ac{tde} to emit as a black body described by the Planck law~\citep{planck_law}
    \begin{equation}
        I_{\lambda_{\mathrm{em}}} = \frac{2 h c^{2}}{\lambda_{\mathrm{em}}^{5}} \cdot \frac{1}{e^{\frac{h c}{\lambda_{\mathrm{em}} k T}} -1},
        \label{eq:planck_law}
    \end{equation}
    where $\lambda_{\mathrm{em}}$ is the emitted wavelength, $T$ is the temperature, $h$ is the Planck constant, and $k$ is the Boltzmann constant. In addition, the Stefan-Boltzmann law \citep[e.g.,][]{stefan_boltzmann_law} applies, and it allows us to rewrite the emission area, 
    \begin{equation}
        A = \frac{L}{4 \sigma T^{4}}
        \label{eq:area_L1}
    \end{equation}
    in terms of luminosity, $L$, and the Stefan-Boltzmann constant, $\sigma$. The full solution to the integral in the numerator of the magnitude is given in Appendix~\ref{sec:mag_int}.

In the following two subsections, we present two different luminosity models for \acp{tde}. They are similar to the ones considered by \citet{TDE_rates_Szekerczes_2024}.

\subsubsection{Phenomenological luminosity model}
\begin{figure}[htbp]
        \centering
        \resizebox{0.9\hsize}{!}{\includegraphics{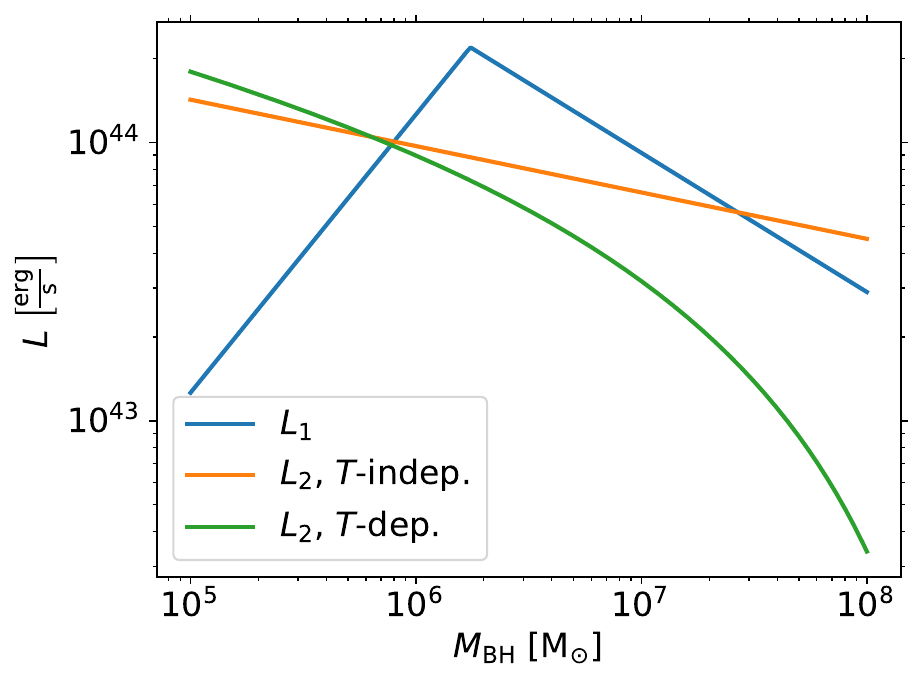}}
        \caption{Three different \ac{tde} luminosity functions dependent on \ac{bh} mass. For low \ac{bh} masses, $L_{1}$ is Eddington limited. The two $L_{2}$ models do not yet differ through their treatment of temperature, as temperature is not involved yet. However, they differ because of a correction factor that accounts for the spread of specific energy in the resulting debris. The $T$-independent model assumes this factor to be constant.}
        \label{fig:tde_luminosity_models}
    \end{figure}

   The first model is an empirical fit between the observed TDE luminosity and the debris fallback rate found by~\citet{TDE_L1}:
    \begin{equation}
        L = 0.01 \cdot \dot M_{\mathrm{fb}} \cdot c^{2}
        \label{eq:L1_of_fallback}
    \end{equation}
    which assumes that the disrupted star has a stellar mass of $0.1 \; \mathrm{M}_{\odot}$. Here
    \begin{equation}
    \begin{aligned}
        \dot M_{\mathrm{fb}} = \; &1.49 \; \mathrm{M}_{\mathbf{\odot}} \; \mathrm{yr}^{-1} \cdot \left( \frac{f}{0.5} \right) \cdot \Xi^{\frac{3}{2}} \\
        &\cdot \left( \frac{M_{\star}}{\mathrm{M}_{\odot}} \right)^{2} \cdot \left( \frac{R_{\star}}{\mathrm{R}_{\odot}} \right)^{-\frac{3}{2}} \cdot \left( \frac{M_{\mathrm{BH}}}{10^{6} \; \mathrm{M}_{\odot}} \right)^{-\frac{1}{2}}
    \end{aligned}
    \label{eq:mass_fallback_rate}
    \end{equation}
    is the mass fallback rate from \citet{mass_fallback_rate}. It depends on the stellar mass, $M_{\star}$; the stellar radius, $R_{\star}$; an energy correction factor, $\Xi$, which we explain in more detail in the next section (Sect.~\ref{sec:theo_lum_model}); and the correction factor, $f$, that accounts for different possible shapes of the energy distribution near the tails of the debris. In this first model, the debris energy spread correction factor is omitted, that is, $\Xi = 1$.
    The correction factor, $f$, is equal to $1$, for $M_{\star} \leq 0.5 \; \mathrm{M}_{\odot}$ and $f = 0.5$ for $M_{\star} > 0.5 \; \mathrm{M}_{\odot}$. Hence, we set $f$ to $1$. Additionally, we assumed a stellar radius of $0.15 \; \mathrm{R}_{\odot}$.
    We also limited the luminosity to the Eddington limit~\citep[e.g.,][]{eddington_limit}:
    \begin{equation}
        L_{\mathrm{Edd}} = 1.26 \cdot 10^{38} \; \mathrm{erg} \; \mathrm{s}^{-1} \cdot \left( \frac{M_{\mathrm{BH}}}{\mathrm{M}_{\odot}} \right).
        \label{eq:eddington_luminosity}
    \end{equation}
    The first \ac{tde} luminosity model, denoted with $L_{1}$, is then
    \begin{equation}
        L_{1} = \min(L, L_{\mathrm{Edd}}).
        \label{eq:L1}
    \end{equation}
    We show this luminosity in Fig.~\ref{fig:tde_luminosity_models}. In this model, we assumed the disk's temperature to be constant. We refer to this assumption as the temperature-independent case. We performed our calculations for five temperatures, $1, 2, 3, 4,$ and $5 \cdot 10^{4} \; \mathrm{K}$, covering the observed range of temperatures of optical \acp{tde}~\citep{observational_zft_tdes}.

\subsubsection{Theoretical luminosity model}\label{sec:theo_lum_model}
    The second model we considered is by~\citet{theoretical_L2_model}, and it was later improved by~\citet{luminosity_temperature_mass_dependence_Ryu_2020}. They propose that shocks within an eccentric accretion flow drive the light emission from a \ac{tde}. The shock is formed when earlier arriving debris, after one orbit around the \ac{bh}, collide with later arriving debris. The peak luminosity,
    \begin{equation}
        L_{\mathrm{max}} = \frac{G M_\mathrm{BH} \dot M_{\mathrm{fb}}}{a_{0}},
        \label{eq:max_luminosity_of_tde}
    \end{equation}
    depends on $a_{0} = (G M_\mathrm{BH})/\Delta E$, which gives the size of the mass flow. Here, $G$ is the gravitational constant, and $\Delta E$ is the spread in specific energy. Specifically, it is the apocenter distance of the most-bound debris.  
    Combining Eqs.~\ref{eq:mass_fallback_rate}~and~\ref{eq:max_luminosity_of_tde}, we obtained the full expression for the second luminosity model:
    \begin{equation}
        L_{2} \equiv L_{\mathrm{max}} = \frac{\sqrt{2} G^{\frac{3}{2}}}{3 \pi} \cdot M_{\star}^{\frac{8}{3}} R_{\star}^{-\frac{5}{2}} \cdot M_\mathrm{BH}^{-\frac{1}{6}} \cdot \Xi^{\frac{5}{2}}.
        \label{eq:l_max_full_expression}
    \end{equation}
    
    We then split this model in two. First, following~\citet{TDE_rates_Szekerczes_2024}, we assumed the temperature to be constant at the same values as for $L_{1}$. Additionally, we assumed all disrupted stars to be solar-like, that is, $M_\star = 1 \; \mathrm{M}_\odot$ and $R_\star = 1 \; \mathrm{R}_\odot$, from which follows $\Xi = 1.49$ and $f = 0.5$. We call this set of assumptions our temperature-independent $L_{2}$ model.
    
    The second option is a case where we include the full temperature dependence in this luminosity model. More specifically, we calculated the temperature in terms of \ac{bh} mass. As this model includes an expression for the \ac{bb} radius, it is possible to calculate an effective emission area, 
    \begin{equation}
        A = 2\pi \cdot a_{0}^{2} = 2\pi \cdot \Xi_{\star}^{-2} M_{\star}^{-\frac{4}{3}} R_{\star}^{2} \cdot \Xi_\mathrm{BH}^{-2} M_\mathrm{BH}^{\frac{4}{3}} \cdot \Xi^{-2}
        \label{eq:tde_area_analytic}
    \end{equation}
    of the \ac{tde}.\footnote{\citet{luminosity_temperature_mass_dependence_Ryu_2020} include a correction factor to account for the uncertainty in the relation between the elliptical emission area and the black body radius. However, for simplicity, we followed \citet{luminosity_temperature_mass_dependence_Ryu_2020} in assuming they are comparable.} 
    Here, we included the full energy correction factor $\Xi = \Delta \epsilon / \Delta E$ that relates the spread in specific energy $\Delta E$ to the order of magnitude estimate $\Delta \epsilon = G M_{\mathrm{BH}} R_{\star}/r_\mathrm{t}^{2}$, where $r_\mathrm{t} = (M_{\mathrm{BH}} / M_{\star})^{1/3} R_{\star}$ is the tidal radius. The energy correction factor $\Xi = \Xi_{\mathrm{BH}} \cdot \Xi_{\star}$ is split into a correction from the star's internal structure, 
    \begin{equation}
        \Xi_{\star}(M_{\star}) = \frac{0.62  + \exp[(M_{\star} /\mathrm{M}_{\odot} - 0.67) /0.21]} {1 + 0.55 \cdot \exp[(M_{\star} /\mathrm{M}_{\odot} - 0.67) /0.21]}
    \end{equation}
    and a correction from relativistic effects close to the \ac{bh}, $\Xi_\mathrm{BH}(M_\mathrm{BH}) = 1.27 - 0.3 (M_\mathrm{BH}/(10^{6} \; \mathrm{M}_{\odot} ))^{0.242}$.
    
\begin{figure}[htbp]
        \centering
        \resizebox{0.9\hsize}{!}{\includegraphics[width=\textwidth]{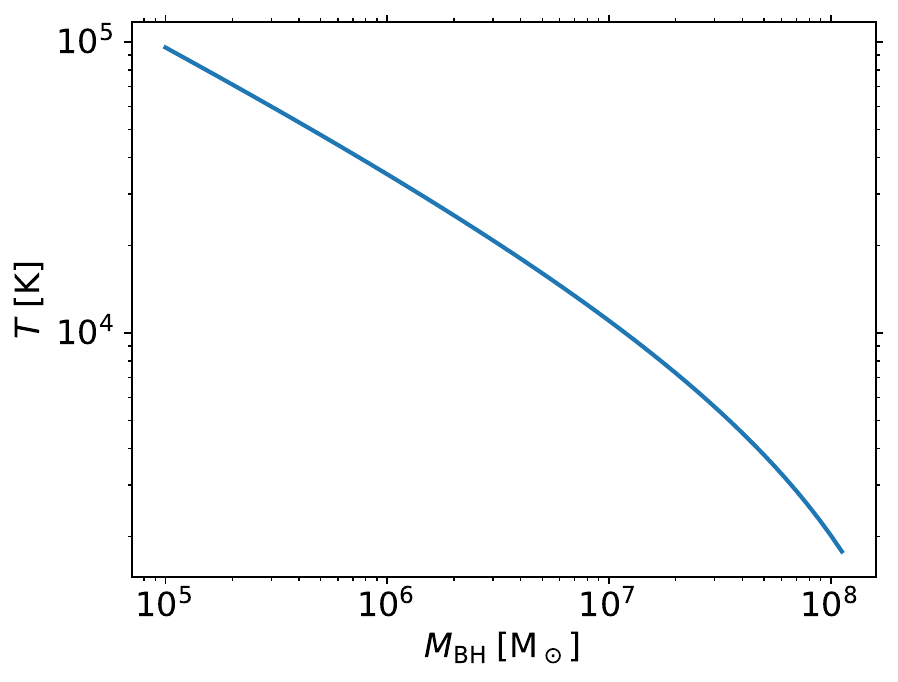}}
        \caption{\ac{tde} temperature dependent on \ac{bh} mass using the $L_{2}$ model. We assumed a solar-like star, that is, $M_\star = 1 \; \mathrm{M}_\odot$ and $R_\star = 1 \; \mathrm{R}_\odot$.}
        \label{fig:temperature_dep_on_bh_mass}
    \end{figure}

    By again using the Stefan-Boltzmann law and the flux-luminosity relation, as in Eq.~\ref{eq:area_L1}, we solved for temperature instead of area: 
    \begin{equation}
        T = \left( \frac{L_{2}}{\sigma \cdot A} \right)^{\frac{1}{4}} = \left( \frac{\sqrt{2} G^{\frac{3}{2}}}{6 \pi^2 \sigma} \right)^{\frac{1}{4}} \cdot \Xi_{\star}^{\frac{9}{8}} M_{\star} R_{\star}^{-\frac{9}{8}} \cdot \Xi_\mathrm{BH}^{\frac{9}{8}} M_\mathrm{BH}^{-\frac{3}{8}}.
        \label{eq:tde_temperature}
    \end{equation}
    We show the dependence of the temperature of a \ac{tde} on the \ac{bh} mass in Fig.~\ref{fig:temperature_dep_on_bh_mass} for a solar-like star being disrupted. 
    The drop-off of the temperature toward the high \ac{bh} mass end is due to the inclusion of the correction factor $\Xi$. The same influence of $\Xi$ is also seen on the \ac{tde} luminosity in Fig.~\ref{fig:tde_luminosity_models}. 

\begin{figure}[htbp]
        \centering
        \begin{subfigure}[b]{0.49\textwidth}
            \centering
            \resizebox{0.95\hsize}{!}{\includegraphics{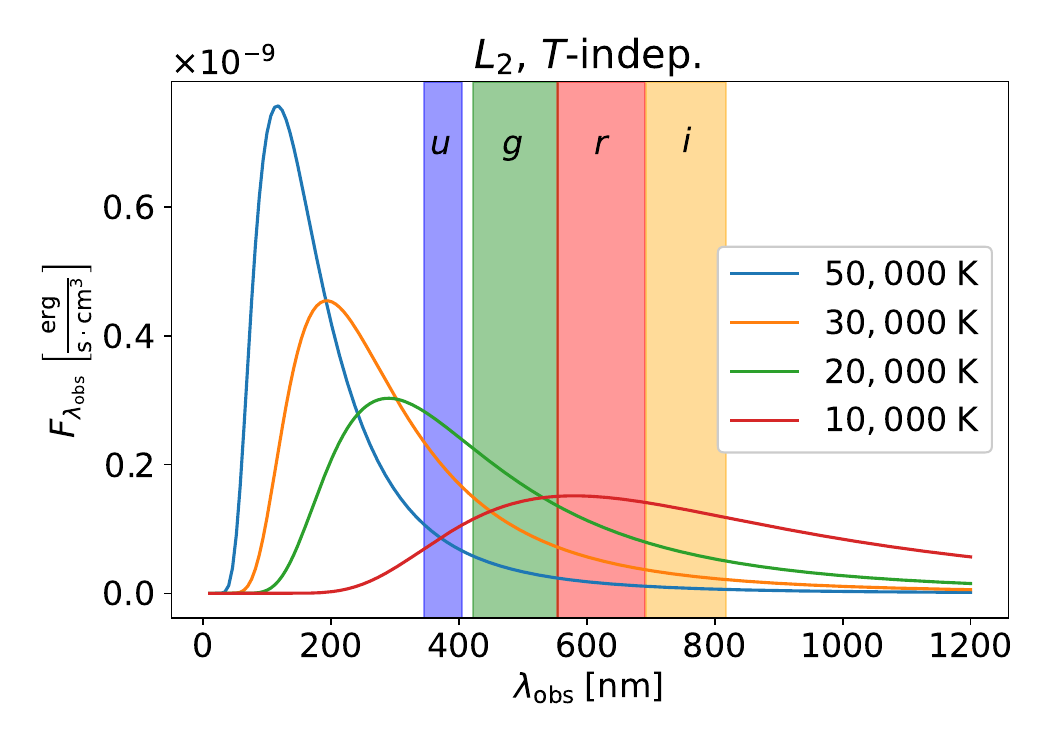}}
            \label{subfig:flux_different_T}
        \end{subfigure}
        \begin{subfigure}[b]{0.49\textwidth}
            \centering
            \resizebox{0.95\hsize}{!}{\includegraphics{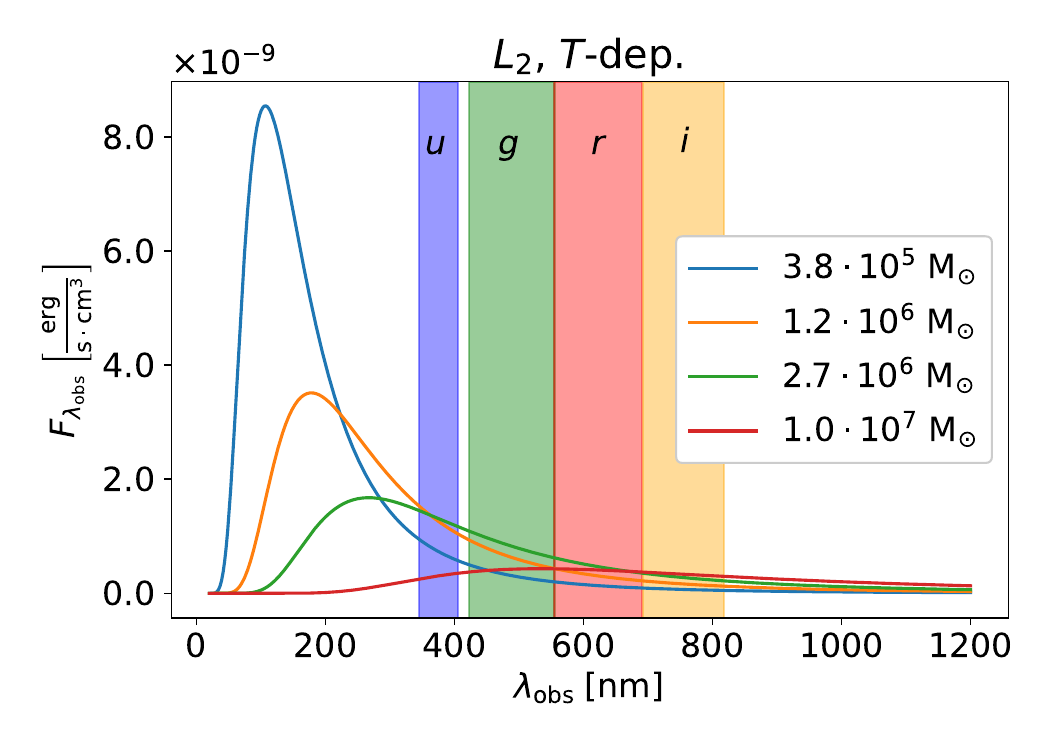}}
            \label{subfig:flux_bh_mass_dep}
        \end{subfigure}
        \caption{Black body flux for different temperatures (upper panel) and \ac{bh} masses (lower panel). For $L_{2}$, $T$-indep. in the upper panel, $F_{\lambda_{\mathrm{obs}}} \propto A_{\mathrm{em}} \propto L_{2} \propto M_{\mathrm{BH}}^{-1/6}$, which means the \ac{bh} mass only changes the normalization of the flux because $T$ is fixed. For $L_{2}$, $T$-dep. in the lower panel, $F_{\lambda_{\mathrm{obs}}}$ has a non-trivial dependence on $M_{\mathrm{BH}}$ and $T \propto M_{\mathrm{BH}}^{-3/8}$, which means the normalization and peak of the flux are coupled through the \ac{bh} mass. For the top panel, we assumed a \ac{bh} mass of $10^{6} \; \mathrm{M}_{\odot}$. In the bottom panel, we chose the \ac{bh} masses such that each line color matches the temperature of the fixed-temperature models. For both panels, we assumed a redshift of $z = 1$.}
        \label{fig:flux_comparison}
        \vspace{-10pt}
    \end{figure}

    As the emitting area is known for the temperature-dependent case, the flux can be directly computed from Eq.~\ref{eq:observed_flux}. The resulting flux is compared to the fixed-temperature $L_{2}$ model in Fig.~\ref{fig:flux_comparison}. For the temperature-independent models, the assumed temperature fixes the location of the \ac{bb} peak, and the \ac{bh} mass only acts as a multiplicative factor. In the temperature-dependent case, the \ac{bh} mass enters into the flux through the emission area as a factor and additionally through the temperature in the emitted specific intensity. This means the location of the \ac{bb} peak is determined by the \ac{bh} mass. One consequence is that the bands are sensitive to different \ac{bh} mass ranges because \acp{tde} can be observed best when the \ac{bb} peak is closely aligned with the observing band.

\subsection{Solving the rate integral}
    The rate integral in Eq.~\ref{eq:total_tde_rate_eq} only depends on two parameters at this point, the redshift and the \ac{bh} mass, since we fixed the stellar mass and radius. These two parameters describe all possible \acp{tde} we have considered. Similar to~\citet{TDE_rates_Szekerczes_2024}, we numerically integrated on a grid in this two-dimensional space but used a logarithmic grid spacing for the BH mass ranging from $10^{5} \; \mathrm{M}_{\odot}$ to $10^{8} \; \mathrm{M}_{\odot}$. We chose this \ac{bh} mass region because on the lower end, the existence of intermediate-mass \acp{bh} is not yet confirmed \citep{ident_of_imbh,search_for_imbh}. The upper end is chosen because for $M_{\mathrm{BH}} \gtrsim 10^{8} \; \mathrm{M}_\odot$, stars would be swallowed whole \citep{stars_swallowed_whole} as the tidal disruption radius becomes smaller than the event horizon.
    To obtain the correct integral region, we calculated the observed magnitude at each grid point and only kept those that produce a magnitude brighter than the observational limit.
    We used the magnitude limits for \ac{ztf} and \ac{lsst} as a reference. For \ac{ztf}, the limits for the ($g$, $r$, $i$) band are ($20.8$, $20.6$, $19.9$), respectively~\citep[e.g.,][]{ztf_limiting_magnitude}. For \ac{lsst}, the limits for ($u$, $g$, $r$, $i$) are ($23.3$, $24.7$, $24.3$, $23.7$), respectively~(e.g.,~\citet{magnitude_and_flux_Huber_2021}; \citet{lsst_limiting_magnitude}). The assumption that a \ac{tde} is observed at peak luminosity was introduced with the mass fallback rate in Eq.~\ref{eq:mass_fallback_rate}. We relaxed this assumption by subtracting $0.7$ from the limiting magnitude of the survey in accordance with \citet{TDE_rates_Szekerczes_2024} and \citet{lensing_code_Oguri_2010}. It accounts for observations not catching the \ac{tde} at its peak brightness but only close to it. The value of $0.7$ gives a time window of tens of days to observe the \ac{tde} \citep[e.g.,][]{tde_40_day_decline,one_tde_lightcurve}. The limiting magnitudes of all surveys have this subtraction incorporated from this point onward.

\begin{figure*}
    \centering
    \includegraphics[width=\textwidth]{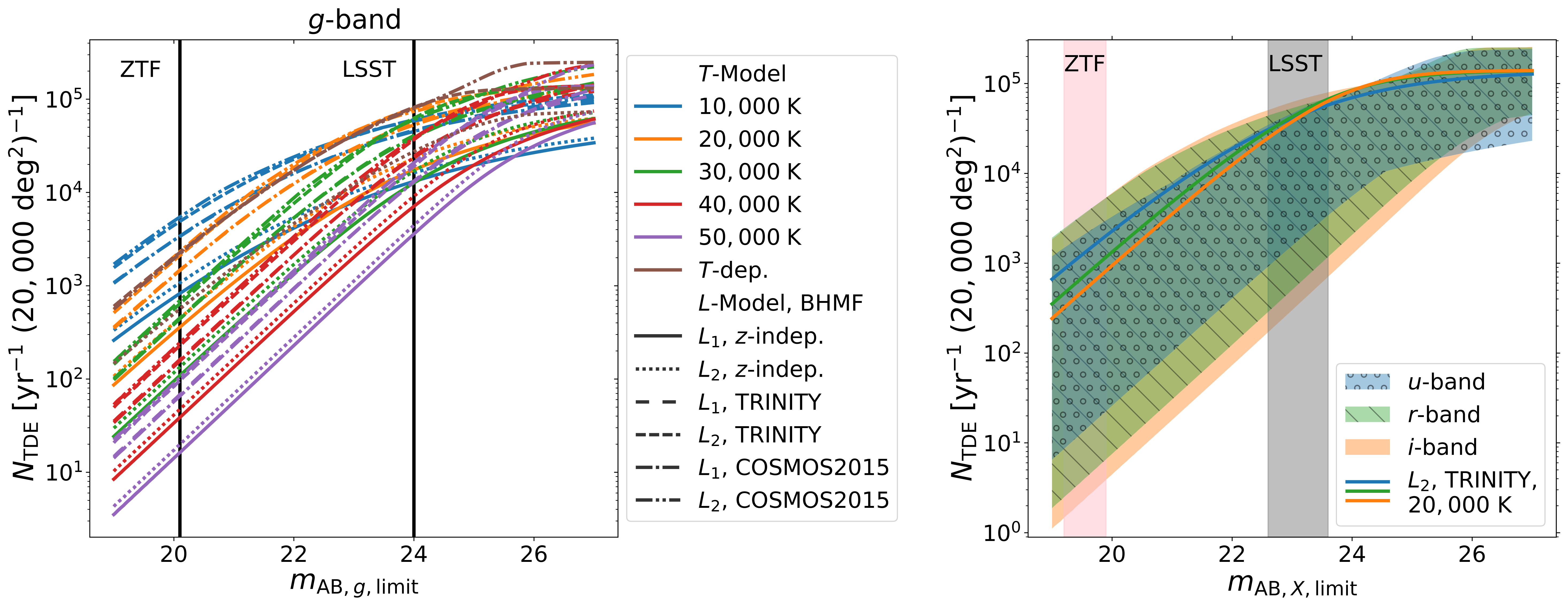}
    \caption{Rate of \acp{tde} as a function of survey limiting magnitude. The left panel shows the detection rate in the $g$ band for the $33$ models individually. We separated the color and line style information to keep the legend short. The right panel shows the detection rate in the $u$, $r$, and $i$ bands, where each shaded region encompasses the spread in the models. We chose $L_{2}$, TRINITY, $20,000$ K as a fiducial model to guide the eye. The $33$ models we considered for the unlensed \ac{tde} rate show a scattering factor of $\mathcal{O}(10^{2})$ that can reach above $\mathcal{O}(10^{3})$ for the $u$ and $i$ bands. The vertical black lines (left panel) and the colorful strips (right panel) show the observational magnitude limit for \ac{ztf} ($g$, $r$, $i$ bands) and \ac{lsst} ($u$, $g$, $r$, $i$ bands).}
    \label{fig:unlensed_rates}
\end{figure*}
\section{Unlensed TDE rate}\label{sec:unlensed_rate}
    We show the results of our unlensed \ac{tde} rate calculations in Fig.~\ref{fig:unlensed_rates}. The vertical lines or bands represent the observational magnitude limit for \ac{ztf} and \ac{lsst}. It is immediately apparent that the predictions of our considered models vary greatly. The spread easily exceeds a factor of $10^{2}$ between the lowest and highest rates. This factor can climb above $10^{3}$ in the $i$ and $u$ bands for bright limiting magnitudes $\sim 19$. For faint limiting magnitudes $\gtrsim 24$, there seems to be a crossing region. Some models flatten out quicker, predicting a lower total \ac{tde} rate. This is mainly due to a smaller \ac{bhmf} at a higher redshift; fewer \acp{bh} produce fewer \acp{tde}.

We tabulated the range of predictions for \ac{ztf} and \ac{lsst} in Table~\ref{tab:ztf_lsst_unlensed_rate_prediction}. Independent of the \ac{tde} model, the $g$ band observes the highest \ac{tde} rate for \ac{ztf} and \ac{lsst}. In large part, this is due to both observatories having the faintest magnitude limit in the $g$ band. 

\begin{table}[h]
    \caption{Predicted range of unlensed \ac{tde} rates.}
    \centering
    \begin{NiceTabular}{c|c|c}
        \CodeBefore
        \rowlistcolors{2}{white,gray!25}[restart]
        \Body
        Band & $N_{\mathrm{TDE}}$ for \ac{ztf} & $N_{\mathrm{TDE}}$ for \ac{lsst} \\
          & $\left[ \mathrm{yr}^{-1} \; (20,000 \; \mathrm{deg}^{2})^{-1} \right]$ & $\left[ \mathrm{yr}^{-1} \; (20,000 \; \mathrm{deg}^{2})^{-1} \right]$ \\
        \hline
        $u$ & - & $920 - 34,630$ \\
        $g$ & $16 - 5,440$ & $3,580 - 82,060$ \\
        $r$ & $6.5 - 5,340$ & $1,180 - 68,320$ \\
        $i$ & $1.6 - 2,350$ & $300 - 62,380$ \\
        \hline
    \end{NiceTabular}
    \tablefoot{For \ac{ztf}, we find a scatter of $\mathcal{O}(10^{3})$. For \ac{lsst}, the scatter is $\mathcal{O}(10^{2})$.}
    \label{tab:ztf_lsst_unlensed_rate_prediction}
    \vspace{-10pt}
\end{table}

In the following, we investigate the impact of the \ac{bhmf}, the luminosity model, and the temperature individually. We find that the assumed temperature is responsible for most of the observed scatter between the different models.

\subsection{Impact of black hole mass function}
\begin{figure}
        \centering
        \resizebox{\hsize}{!}{\includegraphics{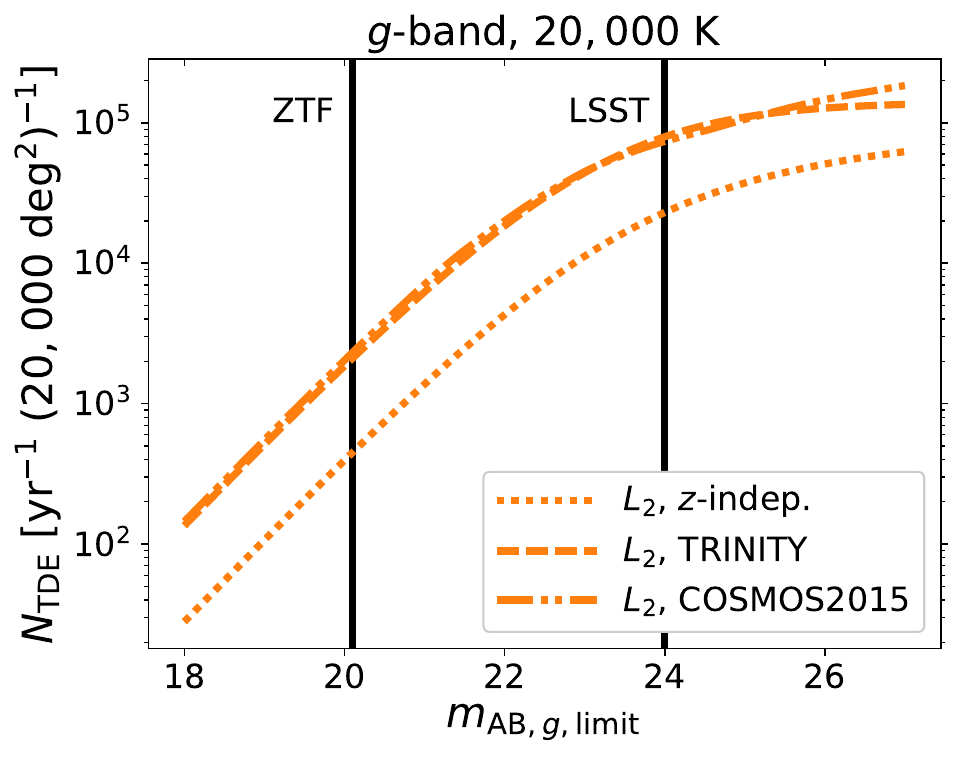}}
        \caption{Rate of \acp{tde} as a function of survey limiting magnitude. Only the local \ac{bhmf} impacts the unlensed \ac{tde} rates observed by current and near-future surveys such as \ac{ztf} and \ac{lsst}. The redshift evolution of the \ac{bhmf} influences the rate from a limiting magnitude of $25$ onward. We have taken a representative subset from the left panel of Fig.~\ref{fig:unlensed_rates} to clearly show the influence of the \ac{bhmf}.}
        \label{fig:unlensed_impact_bhmf}
    \end{figure}

    First, we chose the $L_{2}$ model and fixed the temperature to $20,000 \; \mathrm{K}$. We only varied the \acp{bhmf} (see Fig.~\ref{fig:unlensed_impact_bhmf}). The unlensed \ac{tde} rate we calculated is very similar for the TRINITY and the COSMOS2015 \acp{bhmf} up to a magnitude limit of $25$. This is clear when comparing the two \acp{bhmf} in Fig.~\ref{fig:all_bhmfs}. The two are almost identical for $z \lesssim 2$, where most detectable TDEs are distributed (see Fig.~\ref{fig:unlensed_tde_distributions} for \ac{lsst} limiting magnitudes).
    In contrast, the redshift independent model predicts lower rates by a factor of roughly four, which is due to the \ac{bhmf} having a lower value at low redshift by roughly this factor of four.
    
    The \acp{bhmf} strongly differ at large redshifts, which we can also observe in the predicted rates. They diverge for faint magnitude limits above around $25$ (see Fig.~\ref{fig:unlensed_impact_bhmf}). As the TRINITY \ac{bhmf} decreases at high redshifts, the predicted \ac{tde} rates flatten out, while for the other two, only a slower increase is observed. This slowdown is attributed to the dimming of \acp{tde} due to distance. At some point, even the brightest \acp{tde} are too far away to reach the magnitude limit. Overall, this implies that the redshift evolution of the \ac{bhmf} only becomes important for faint magnitude limits above $25$. These limits will not be reached by \ac{lsst}, even when omitting the $0.7$ reduction of the magnitude limit. However, redder bands detect the redshift evolution at a brighter magnitude than bluer bands. Red bands are naturally better suited to observe objects at high redshift. Hence, for a given magnitude limit, the highest redshift \ac{tde} always appears in the $i$ band, allowing it to probe the redshift evolution of the \ac{bhmf} at a brighter limiting magnitude than, for example, the $u$ band.
    However, the temperature-dependent model can adjust for the alignment of the \ac{bb} peak and the observing band. This adjustment changes the height of the \ac{bb} peak by a factor of $\mathcal{O}(1)$. Hence, through the relation $m_{\mathrm{AB}} \propto \log_{10}(F_{\lambda})$ (see Eq.~\ref{eq:magnitudes}), the magnitude only changes by an additive constant $\simeq 0$. Thus, all bands detect the redshift evolution of the \ac{bhmf} at the same limiting magnitude for the temperature-dependent models.
    
    These results hint at a similar conclusion for the \ac{tde} rate per galaxy, as the \ac{bhmf} and the \ac{tde} occurrence rate per galaxy influence the rate integral in the same way (see Eq.~\ref{eq:total_tde_rate_eq}). We find that the \ac{tde} rate per galaxy only plays a minor role. However, the different occurrence rate can impact the \ac{bh} mass distribution, and through that, the cumulative \ac{tde} rate. But this change is small compared to other factors such as the assumed \ac{tde} temperature (see Sect.~\ref{sec:impact_temperature}). We present the results of choosing the \ac{tde} occurrence rate from \citet{chang_2025_tde_occurrence_rate} in Appendix~\ref{sec:Chang_TDE_occurence}.

\subsection{Impact of luminosity model}
\begin{figure}
        \centering
        \resizebox{0.9\hsize}{!}{\includegraphics{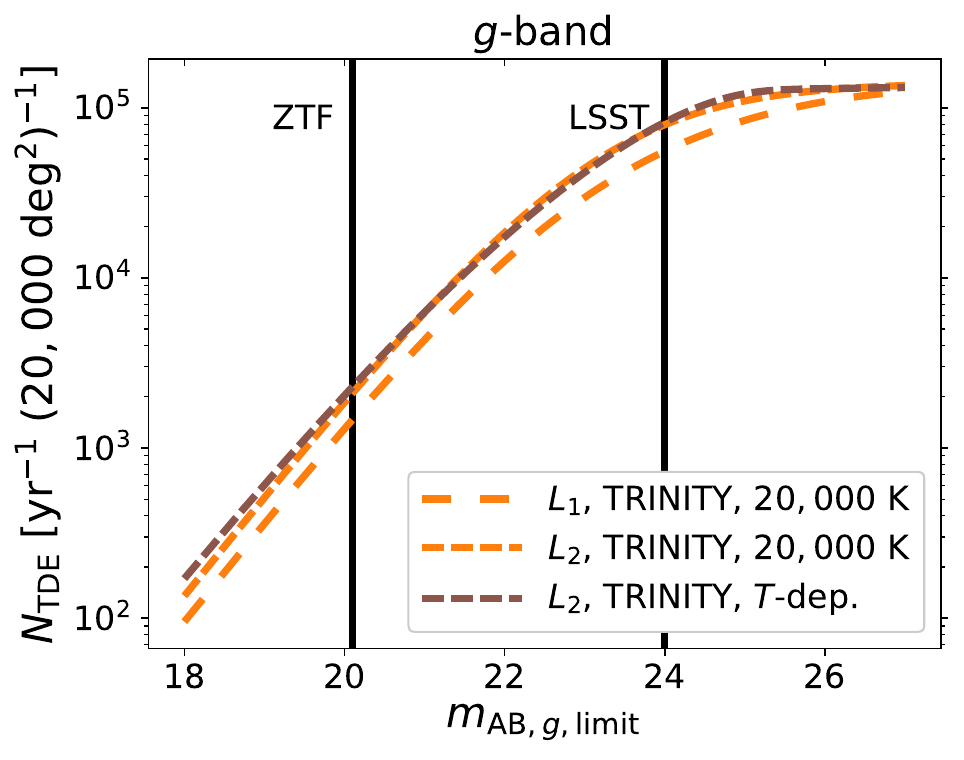}}
        \caption{Rate of \acp{tde} as a function of survey limiting magnitude for different luminosity models. The luminosity model has little impact on the unlensed \ac{tde} rate. It can introduce a dependence of the rate on the observing band for the fixed-temperature models. The temperature-dependent model can compensate for the chosen observing band by selecting the \ac{tde} temperature. Thus, it does not show a strong band dependence. We have taken a representative subset from the left panel of Fig.~\ref{fig:unlensed_rates} to clearly show the influence of the luminosity model.}
        \label{fig:unlensed_impact_lmodel}
    \end{figure}

    To investigate the impact of the luminosity model, we fixed the TRINITY \ac{bhmf} and let the luminosity function vary (see Fig.~\ref{fig:unlensed_impact_lmodel}). The two fixed-temperature luminosity models do not differ strongly in their unlensed \ac{tde} rate predictions. The $L_{2}$ produces higher rates across almost all limiting magnitudes in all bands than $L_{1}$. No matter the \ac{bhmf}, the integrand of the rate integral (see Eq.~\ref{eq:total_tde_rate_eq}) has a small but negative dependence on the \ac{bh} mass. Therefore, \acp{tde} with a lower \ac{bh} mass are preferred, which are brighter for $L_{2}$ compared to $L_{1}$. 
    This trend holds until around a magnitude limit of $27$, where both rates meet. The $L_{2}$ flattens out more quickly than $L_{1}$. This is due to the higher maximal luminosity that $L_{1}$ allows for (see Fig.~\ref{fig:tde_luminosity_models}). At a given magnitude limit, this fact implies that \acp{tde} can be observed at larger redshift for $L_{1}$, meaning the volume in which observable \acp{tde} can take place is larger.
    
    For the temperature-dependent model, the peak luminosity is higher than that for the fixed temperature $L_{2}$ model. However, \acp{tde} with high \ac{bh} masses are suppressed with a lower luminosity. These two effects cancel each other such that almost the same rate of \acp{tde} is observed for the temperature-dependent and the temperature-independent case.

\subsection{Impact of temperature}\label{sec:impact_temperature}
\begin{figure}
        \centering
        \resizebox{0.9\hsize}{!}{\includegraphics{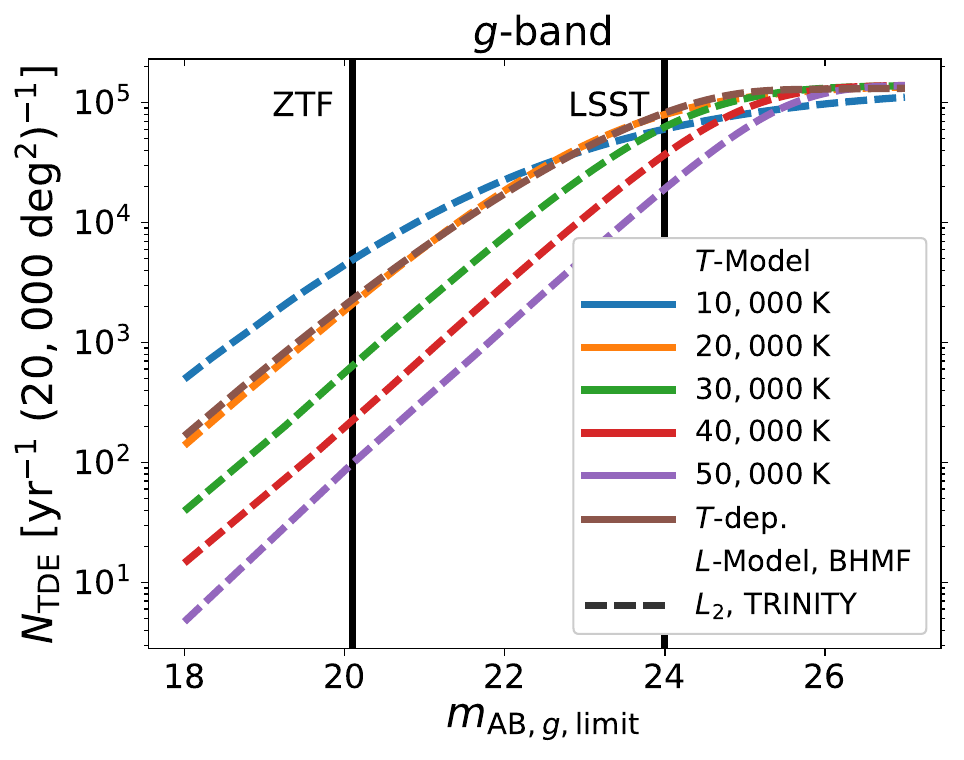}}
        \caption{Rate of \acp{tde} as a function of survey limiting magnitude. The chosen temperature has the largest impact on the unlensed \ac{tde} rate. This parameter accounts for most of the scatter observed between the different models. This implies that it is crucial that the \ac{bb} peak and the observing band are well aligned. We have taken a representative subset from the left panel of Fig.~\ref{fig:unlensed_rates} to clearly show the influence of the temperature.}
        \label{fig:unlensed_impact_temperature}
    \end{figure}

    The assumed temperature contributes the most to the spread of the models (see Fig.~\ref{fig:unlensed_impact_temperature}). The temperature strongly impacts the redshift distribution of the observed \acp{tde}. As explained above, it is crucial that the peak of the \ac{bb} spectrum aligns well with the observing band. At low redshift, the closest alignment is seen for $10,000 \; \mathrm{K}$; hence, models with this temperature show the highest rates at bright limiting magnitudes $\lesssim 22$. 
    The \ac{bb} peak is always shifted into a given observing band at some redshift. Therefore, when a faint limiting magnitude $\gtrsim 26$ is assumed, all bands should roughly observe the same number of \acp{tde}, as every \ac{tde} can be observed in every band. This is exactly what can be observed in Fig.~\ref{fig:unlensed_impact_temperature}. 
    The odd one out is the $10,000 \; \mathrm{K}$ model, as it does not quite reach the same level as the other temperatures. The same deviation is more prominent in the $u$ band but not seen in the $r$ or $i$ bands. For this temperature, the peak of the \ac{bb} spectrum lies at $290 \; \mathrm{nm}$ at $z = 0$, which is just at a smaller wavelength compared to the $u$ band. But the integrand in the numerator of the magnitude is proportional to $\lambda \cdot F_{\lambda_{\mathrm{obs}}}$ (see Eq.~\ref{eq:magnitudes}), meaning longer wavelengths are weighted more. With increasing redshift, the steep rise of the \ac{bb} spectrum with increasing wavelength quickly dims the \acp{tde}, as the peak of the BB shifts out of the $u$ band and even the $g$ band to longer wavelengths. For higher temperatures, the peak of the \ac{bb} spectrum is first shifted into the observing band, counteracting the dimming due to distance. Therefore, \acp{tde} are only observed at fainter magnitudes for $10,000 \; \mathrm{K}$, contributing to this model's shallower slope. Hence, it takes longer to catch up to the models. 
    
\subsection{Comparison to the observed rate}
    \citet{observational_zft_tdes} report that \ac{ztf} observed 30 \acp{tde} in $2.6$ years in multi-band observations using $g$ and $r$ bands. \ac{ztf} observed the entire visible northern sky, roughly speaking around $20,000 \; \mathrm{deg}^{2}$. This means they found an unlensed \ac{tde} rate $N_{\mathrm{TDE}}$ of around $10$ to $15 \; \mathrm{yr}^{-1} \; (20,000 \; \mathrm{deg}^{2})^{-1}$. In the $g$ band, this rate is below all of our models. The lowest prediction is $16.5 \; \mathrm{yr}^{-1} \; (20,000 \; \mathrm{deg}^{2})^{-1}$, and it is from the $L_{1}$ $z$-independent $50,000 \; \mathrm{K}$ model. For the $r$ band, the rate lies between the $L_{1}$ $40,000 \; \mathrm{K}$ and $L_{2}$ $50,000 \; \mathrm{K}$ models, both for the redshift-independent \ac{bhmf}. 

    \citet{observational_zft_tdes} also inferred the temperature and found a range of around $15,000 \; \mathrm{K}$ to $35,000 \; \mathrm{K}$ with an average of $20,783 \; \mathrm{K}$. For this temperature range, our models would predict $108$ ($L_{1}$, $z$-indep., $30,000 \; \mathrm{K}$) to $2,313 \; \mathrm{yr}^{-1} \; (20,000 \; \mathrm{deg}^{2})^{-1}$ ($L_{2}$, COSMOS2015, $20,000 \; \mathrm{K}$) in the $g$ band and $49.5$ ($L_{1}$, $z$-indep., $30,000 \; \mathrm{K}$) to $1,538 \; \mathrm{yr}^{-1} \; (20,000 \; \mathrm{deg}^{2})^{-1}$ ($L_{2}$, TRINITY, $T$-dep.) in the $r$ band. 
    In our calculation, we have not accounted for the effective year length of \ac{ztf}, which might lead to more incomplete observations than our models predict. We have also not accounted for other observational constraints, such as dust extinction or the host galaxy's light. The implicit assumption is that a \ac{tde} is always observable. In practice, some \acp{tde} go undetected because of the bright host galaxy. In addition, we have also not accounted for completeness. The identification of \acp{tde} is an active field of research \citep[e.g.,][]{ambigous_TDEs,identify_TDEs_with_ML,tdescore_identification_TDEs}. On top of that, removing the assumptions that every star is solar-like and every \ac{tde} is a full disruption could notably change the results. In general, lighter stars produce fainter \acp{tde}, while heavier stars produce brighter \acp{tde}. Hence, it is probable that the rate of detectable \acp{tde} in magnitude-limited transient surveys may be lower when a realistic stellar mass distribution favors lower-mass stars. Whether these effects can account for a factor of $10^{2}$ or can only explain part of the overestimation needs further analysis.

    The inferred \ac{bh} mass also yields little insight. \citet{observational_zft_tdes} inferred a range between $10^{6} \; \mathrm{M}_{\odot}$ and $10^{7} \; \mathrm{M}_{\odot}$ using TDEmass \citep{luminosity_temperature_mass_dependence_Ryu_2020} and a range from $2 \cdot 10^{6} \; \mathrm{M}_{\odot}$ to $4 \cdot 10^{7} \; \mathrm{M}_{\odot}$ using MOSFiT \citep{mosfit_1,mosfit_2}. These \ac{bh} mass distributions peak around the same mass as the $L_{1}$ and the temperature-dependent $L_{2}$ models, respectively. But as the \ac{bh} mass is observationally not well constrained, we cannot establish a \ac{tde} luminosity model here. Nevertheless, this may indicate the possibility of inferring the \ac{tde} luminosity model using the \ac{bh} mass distribution. This is especially feasible when better statistics become available with the launch of \ac{lsst}. However, the \ac{bh} mass must be measured robustly for such an investigation.

\section{Lensed TDE rate}\label{sec:lensed_rate}
\begin{figure}[htbp]
        \centering
        \resizebox{0.7\hsize}{!}{\includegraphics{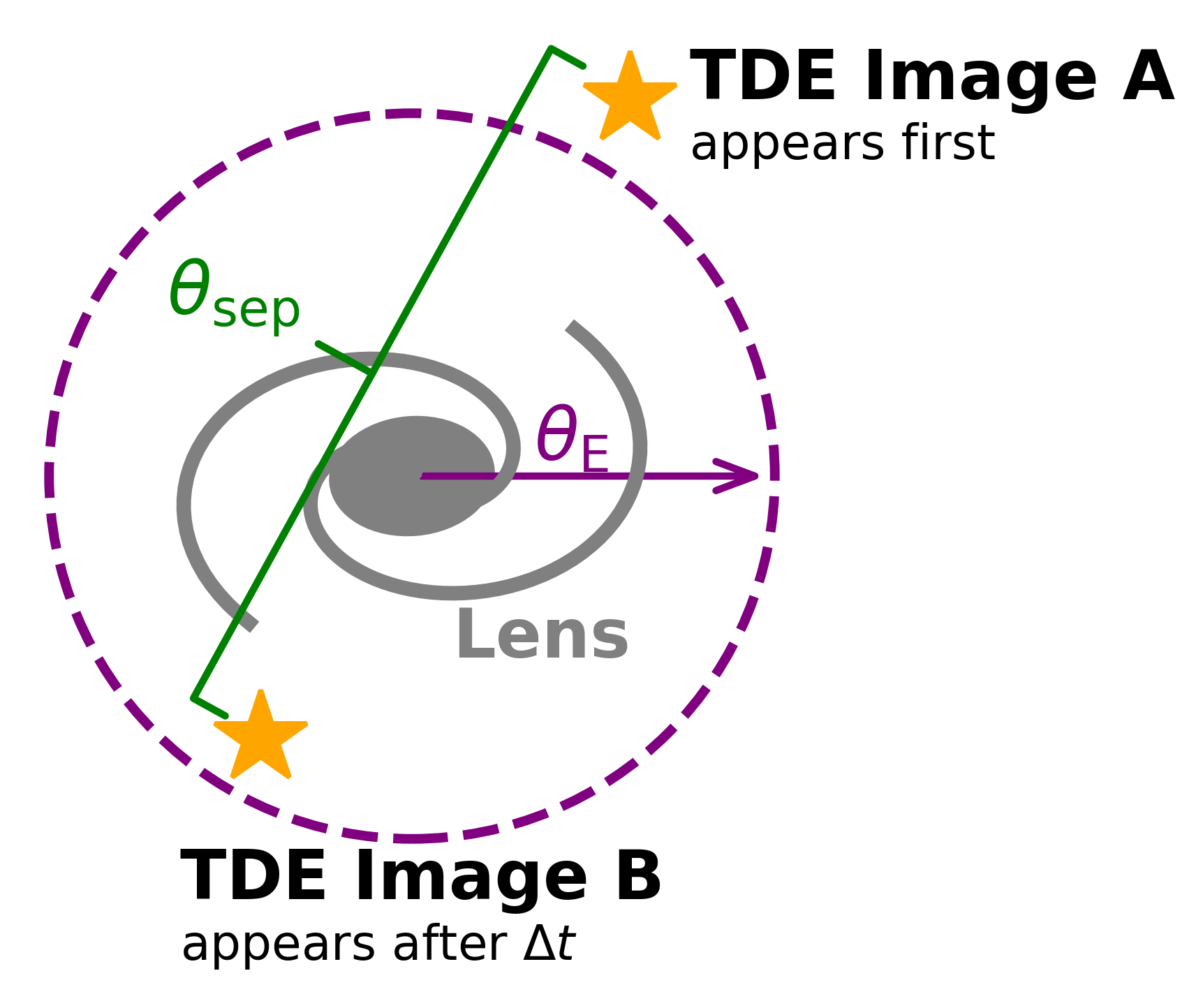}}
        \caption{Illustration of a lensed \ac{tde} with two images, A and B, around a galaxy acting as a lens. The lens has an Einstein radius, $\theta_{\mathrm{E}}$, that gives an estimate of the image separation, $\theta_{\mathrm{sep}}$. Due to the different geometric path lengths and the gravitational time delay \citep{shapiro_delay}, we expect image B to appear later than image A with a time delay of $\Delta t$.}
        \label{fig:double_lensed_tde}
    \end{figure}

    In this section, we outline our methods to calculate the lensed \ac{tde} rate and show our results. In Fig.~\ref{fig:double_lensed_tde} we show an illustration of a lensed \ac{tde} with two images of the \ac{tde}. We refer to such a lensing event as a doubly lensed \ac{tde} or simply as a double.
    For more details on gravitational lensing, we refer to Appendix~\ref{sec:grav_lensing}.

    \subsection{Calculating the lensed rates}
    The lensed TDE rate is calculated with the code developed by~\citet{lensing_code_Oguri_2010}. Some optimizations have been introduced to the code to speed up calculation times significantly. For details, we refer to Appendix~\ref{Implementation}.
    For the calculations, the code from \citet{lensing_code_Oguri_2010} assumes that all lens galaxies are singular isothermal ellipsoids with convergence 
    \begin{equation}
        \kappa = \frac{\theta_{\mathrm{E}}}{2} \cdot \frac{\lambda(e)}{\sqrt{(1 - e)^{-1} \cdot x^{2} + (1-e) \cdot y^{2}}}, 
        \label{eq:convergence_sie}
    \end{equation}
    where $\lambda(e)$ is the dynamical normalization, $e$ is the ellipticity, and 
    \begin{equation}
        \theta_{\mathrm{E}} = 4\pi \cdot \left( \frac{v}{c} \right)^{2} \cdot \frac{D_{\mathrm{ds}}}{D_{\mathrm{s}}}
        \label{eq:einstein_radius_sie}
    \end{equation}
    is the Einstein radius, which is dependent on the velocity dispersion, $v$, of the lens galaxy and the angular diameter distance from the observer to the source, $D_{\mathrm{s}}$, and from the lens to the source, $D_{\mathrm{ds}}$. This is an elliptical profile, which implies that the Einstein radius gives an estimate of the separation of the different images. The lens model is singular and isothermal, meaning the mass distribution drops as $1/r$ and diverges at the origin. These equations and the corresponding deflection of light rays for the singular isothermal ellipsoid were derived by~\citet{SIE_derivation}. 
    
    The code calculates a lensed \ac{tde} mock catalog by taking in an unlensed mock catalog and simulating all possible lens galaxies as elliptical galaxies, which make up $\sim 80\%$ of the lensing probability \citep{elip_lens_1,elip_lens_2,elip_lens_3,elip_lens_4,elip_lens_5,elip_lens_6}. Hence, every lens can be described by its redshift and velocity dispersion in a two-dimensional parameter space. The same technique for generating a mock catalog of unlensed \ac{tde}s can be used to create a catalog of possible lens galaxies.
    By estimating an area on the sky larger than any possible lens system, the lensing code by \citet{lensing_code_Oguri_2010} can calculate a rate of finding a \ac{tde} in the vicinity of a lens galaxy, assuming both lenses and \acp{tde} are spread uniformly on the survey area. This assumption is equivalent to the isotropy of the Universe \citep[e.g.,][]{isotropy_3,isotropy_2,isotropy_4,isotropy_1}. By sampling a Poisson distribution with the close separation rate estimate, they can find the number of candidate systems. The lens equation is then solved for these systems to determine whether it is an actual lens or just a close separation with no multiple images. All systems that produce multiple images are then saved to a mock catalog.
    A \ac{tde} lensed into a double is considered observable when both images are brighter than the limiting magnitude. A schematic diagram of a doubly lensed TDE is shown in Fig.~\ref{fig:double_lensed_tde}. The third brightest image must be brighter than the magnitude limit for a quad lensed \ac{tde}. Calculating rates from such a mock catalog is as simple as counting the entries brighter than the magnitude limit. However, to calculate rates smaller than $1 \; \mathrm{yr}^{-1} \; (20,000 \; \mathrm{deg}^{2})^{-1}$, we need to oversample. We achieved this by calculating $N$ catalogs and, in the end, dividing the summed number of systems by $N$. Thus, every entry in a mock catalog only counts as $1/N$ toward the rate. This mock catalog counting suffers from Poisson noise, so the error scales with $1/\sqrt{N}$.

\subsection{Lensed rate results}
    Following \citet{TDE_rates_Szekerczes_2024}, we only selected lensed events with an image separation in the range $[0.5'', 4'']$ to ensure that the multiple images can be resolved by ground-based surveys such as \ac{lsst}. For a quad, we measured the maximal distance between two images. We included all images of a quad. In a cusp configuration, the fourth image typically sets the largest image separation, but it is also demagnified. However, deeper imaging can reveal it, and we considered the system resolvable. We only considered the lensed rate for \ac{lsst} limiting magnitudes. We did not examine \ac{ztf} limits because many models predict a smaller rate than our numerical limit of $10^{-3} \; \mathrm{yr}^{-1} \; (20,000 \; \mathrm{deg}^{2})^{-1}$, with an oversampling factor of $1000$.

\begin{figure}[htbp]
        \centering
        \resizebox{0.9\hsize}{!}{\includegraphics{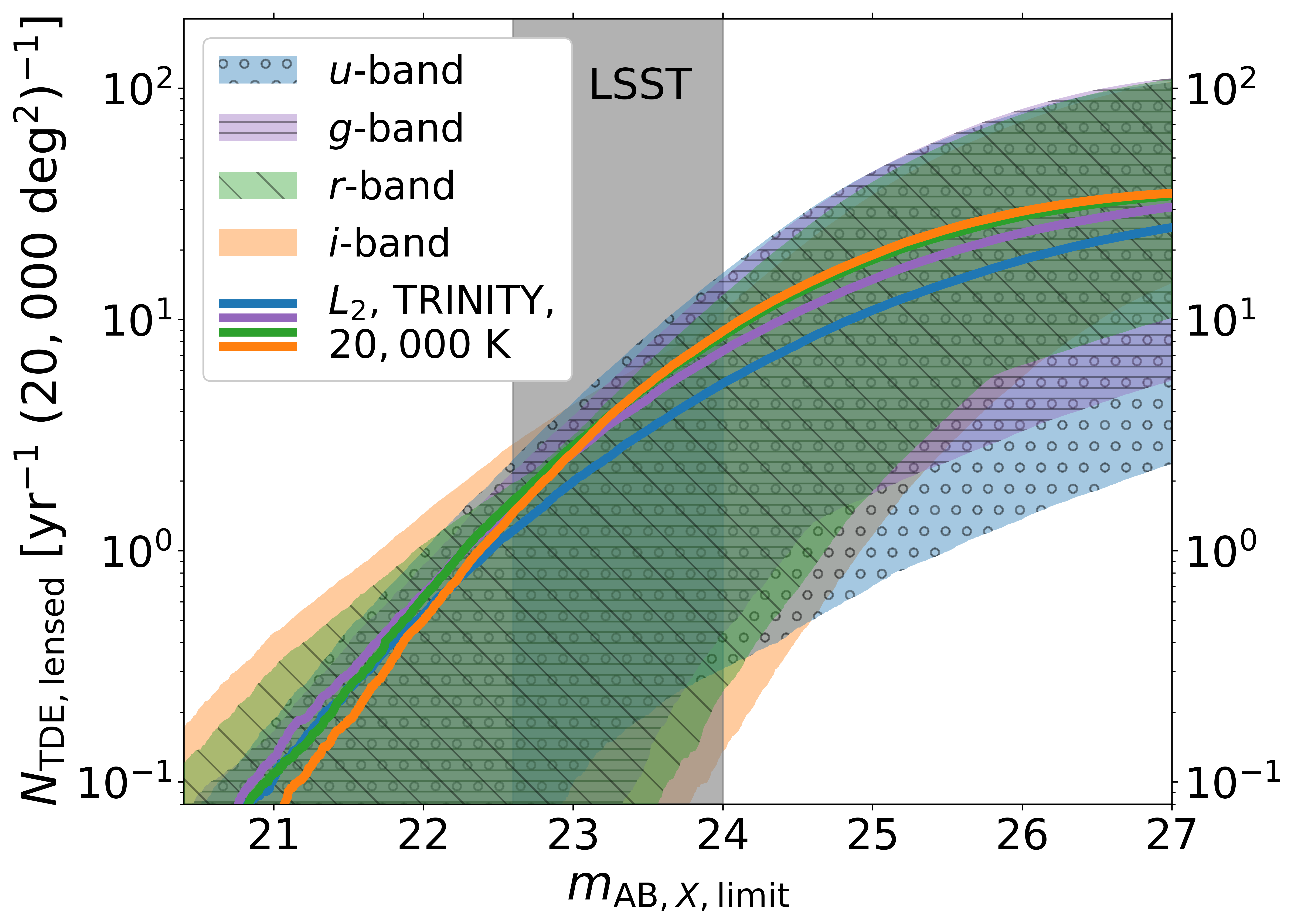}}
        \caption{Rates of lensed \acp{tde} as a function of survey limiting magnitude. The lensed \ac{tde} rates also show a scattering factor of $\mathcal{O}(10^{2})$ that can almost reach $\mathcal{O}(10^{3})$. The gray strip shows the observational magnitude limit for \ac{lsst} ($u$, $g$, $r$, $i$ bands). We display $L_{2}$, TRINITY, $20,000$ K as a fiducial model in solid lines to guide the eye.}
        \label{fig:lensed_rate_range}
    \end{figure}

    The lensed rates show, just like the unlensed ones, a large scatter of $\mathcal{O}(10^{2})$ in the four bands (see Fig.~\ref{fig:lensed_rate_range}). The overall trend is similar to the unlensed rates. Initially, the rate increases steeply and flattens toward fainter limiting magnitudes around $25$ or $26$. For \ac{lsst}, we tabulated the range of predictions in Table~\ref{tab:lsst_lensed_rate_prediction}. Just as for the unlensed rates, only the local \ac{bhmf} is important for the rate prediction. The redshift evolution of the \ac{bhmf} only impacts the rate for magnitude limits fainter than those reached by \ac{lsst}. The lensed rate shows the influence of the redshift evolution at a brighter limiting magnitude of $24$ than the unlensed rate because the magnification due to lensing allows \acp{tde} to be observed at a greater distance from us. The main contribution to the scattering of the different models still stems from the assumed temperature. Just as in the unlensed case, the luminosity model has little impact on the lensed \ac{tde} rate.

    \begin{table}[htbp]
        \caption{Predicted range of lensed \ac{tde} rates.}
        \centering
        \begin{NiceTabular}{c|c}
            \CodeBefore
            \rowlistcolors{2}{white,gray!25}[restart]
            \Body
            Band & $N_{\mathrm{TDE, lensed}}$ for \ac{lsst} \\
              & $\left[ \mathrm{yr}^{-1} \; (20,000 \; \mathrm{deg}^{2})^{-1} \right]$ \\
            \hline
            $u$ & $0.04 - 2.49$ \\
            $g$ & $0.43 - 15$ \\
            $r$ & $0.09 - 7.4$ \\
            $i$ & $0.008 - 4.3$ \\
            \hline
        \end{NiceTabular}
        \tablefoot{For \ac{lsst}, we observe a scatter of $O(10^{2})$ in each band.}
        \label{tab:lsst_lensed_rate_prediction}
    \end{table}

\subsection{Lensed fraction}
\begin{figure*}[htbp]
        \centering
        \includegraphics[width=17cm]{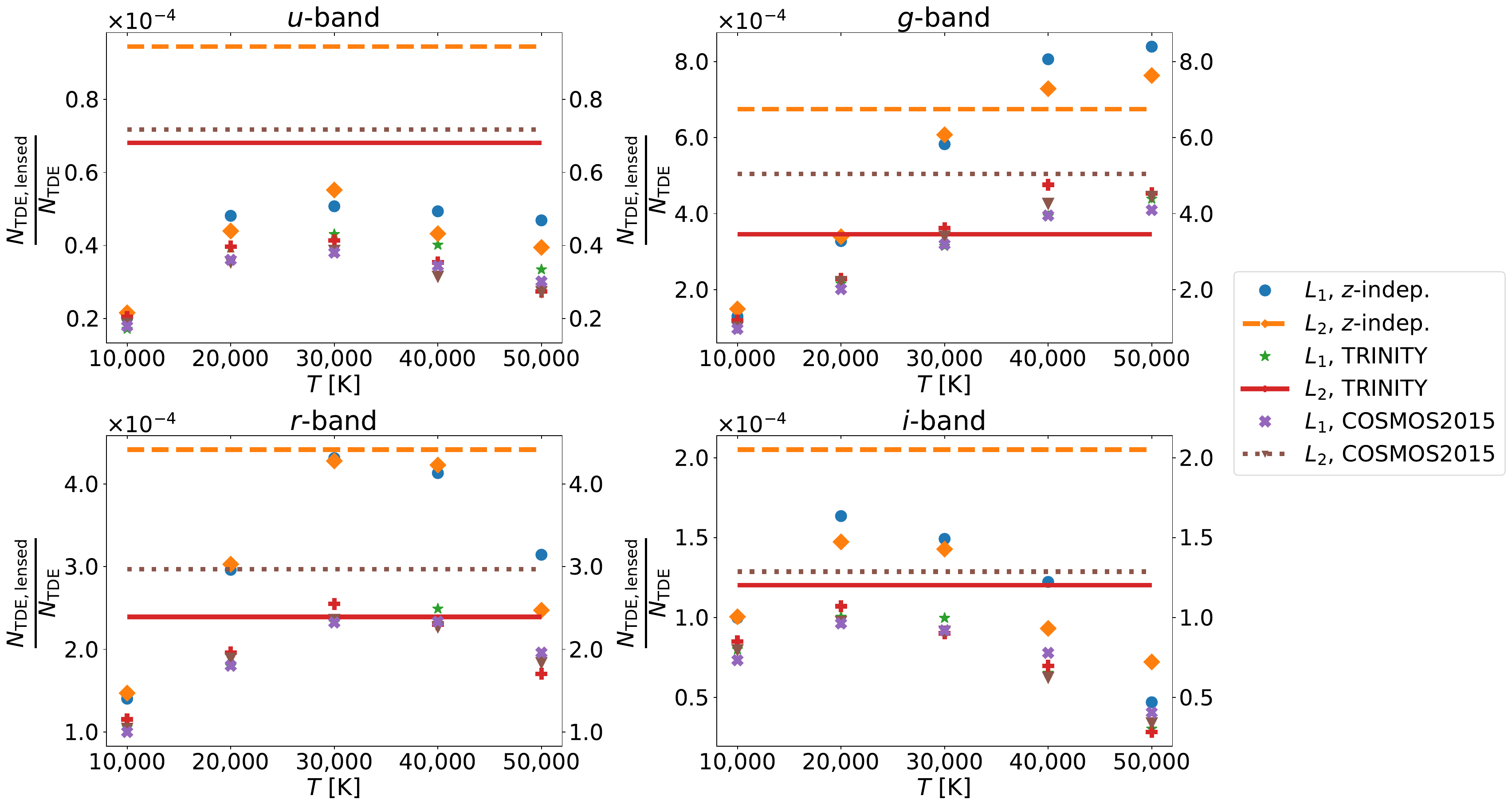}
        \caption{Lensed fraction for the different \ac{tde} models. Each point represents the fraction of \acp{tde} observed as part of a lensing system. The lines are used for the temperature-dependent $L_{2}$ models. The fractions of lensed \acp{tde} show a greatly reduced scatter than the scatter of the unlensed and lensed rates. The fractions lie within a factor of $50$. We calculated these lensed fractions for the \ac{lsst} magnitude limits.}
        \label{fig:lensedFraction}
    \end{figure*}

    The unlensed and lensed \ac{tde} rate predictions both exhibit a broad spread. Therefore, we further investigated the lensed fraction to estimate the fraction of observed \acp{tde} that are part of a lensing system. We present in Fig.~\ref{fig:lensedFraction} the lensed fractions for \ac{lsst} magnitude limits. In the plot, each point or line indicates the ratio of observable lensed \acp{tde} to observable unlensed \acp{tde} given  a \ac{lsst} magnitude limit. The horizontal lines are used for the temperature-dependent $L_{2}$ models, which cover a wide range of possible \ac{tde} temperatures. The lensed fraction is more robust against the different model parameters and displays a smaller scattering factor of $\sim 50$ between the lowest and highest fractions. We predict the fraction of lensed to unlensed \acp{tde} to be between $1.7 \cdot 10^{-5}$ ($L_{1}$, TRINITY, $10,000 \; \mathrm{K}$, $u$ band) and $8.7 \cdot 10^{-4}$ ($L_{1}$, $z$-indep., $50,000 \; \mathrm{K}$, $g$ band). As an order of magnitude, we can say that in order to observe one lensed \ac{tde}, \ac{lsst} has to observe thousands to tens of thousands of unlensed \acp{tde}.
    
    The lensed \ac{tde} population includes many \acp{tde} at a higher redshift than found in the unlensed population. Thus, a larger lensed fraction implies that the fading of \acp{tde} is more strongly counteracted, either due to an increasing \ac{bhmf} with redshift or due to a better \ac{bb} peak alignment. Overall, we can interpret the lensed fraction as a value that compares the number of observable \acp{tde} at low and high redshifts. This becomes clearer when comparing the first row of Fig.~\ref{fig:lensed_tde_hist} to the first row of Fig.~\ref{fig:unlensed_tde_distributions}. The lensed \acp{tde} have a median redshift between $2$ and $3$, while the unlensed \acp{tde} have a median redshift between $0.2$ and $0.7$.

    To better understand Fig.~\ref{fig:lensedFraction}, we considered the observing bands individually. In the $u$ band, we observed the smallest lensed fractions. Here, we know that a temperature of $10,000 \; \mathrm{K}$ aligns the \ac{bb} peak well with the $u$ band at low redshifts, but this alignment is broken at larger redshifts, and only a few \acp{tde} are observed. Thus, we find a small lensed fraction.
    The fraction increases with temperature because the low redshift alignment of the \ac{bb} peak gets worse, while the alignment gets better for a larger redshift. We expect most lensed \acp{tde} to be around $z = 1$ (see the first row of Fig.~\ref{fig:lensed_tde_hist}). At this redshift, a temperature between $20,000 \; \mathrm{K}$ and $30,000 \; \mathrm{K}$ is best aligned with the $u$ band. Hence, we do not expect the increasing trend of the lensed fraction to continue much beyond this temperature.
    The temperature-dependent models can compensate for the \ac{bb} peak alignment and reach higher temperatures and redshifts. Therefore, they predict many \acp{tde} at low redshift. But with the possibility of very high temperatures, they produce many observable \acp{tde} in the $u$ band at a high redshift. Thus, temperature-dependent models yield the highest fraction in the $u$ band.

    We observed the same trend of the lensed fraction in the $g$ band (see the upper right panel of Fig.~\ref{fig:lensedFraction}). But the rise of the fraction continues up to $50,000 \; \mathrm{K}$. This is due to the possibility of observing lensed \acp{tde} up to $z \lesssim 4$ in the $g$ band, while unlensed \acp{tde} fade from view at $z \gtrsim 0.7$.
    
    In the $r$ and $i$ bands, the peak of the lensed fraction moves to lower temperatures because these temperatures perform best in the respective bands at high redshifts. It is worth noting that in the $i$ band at $50,000 \; \mathrm{K}$, we find the lowest lensed fraction despite the fact that the $i$ band observes very few unlensed \acp{tde} at this temperature. However, the $i$ band lies too far away from the \ac{bb} peak. Hence, the dimming due to distance fades any lensed \ac{tde} from view long before the \ac{bb} peak can be aligned with the band. Therefore, the $i$ band does not observe many lensed \acp{tde} either, indicating a small lensed fraction.

    The different \acp{bhmf} seem to amplify the influence of the temperature. Comparing the two redshift-dependent \acp{bhmf}, we observed that the COSMOS2015 \ac{bhmf} increases the fraction compared to the TRINITY \ac{bhmf} only for the temperature-dependent model. This is because the temperature-dependent model reaches the highest redshift (see the first row of Fig.~\ref{fig:lensed_tde_hist}). Therefore, the redshift evolution can no longer be neglected, and the increase of the COSMOS2015 \ac{bhmf} for low-mass \acp{bh} does matter. The redshift-independent \ac{bhmf} amplifies the influence of the temperature the most. This makes sense, as a decreasing \ac{bhmf} with redshift always leads to a decrease in \acp{tde} observed at high redshift. The redshift-independent \ac{bhmf} does not show any evolution and can therefore more easily produce high \ac{tde} rates at high redshift.

\subsection{Lensing parameters}
\begin{figure*}[htbp]
        \centering
        \includegraphics[width=0.98\textwidth]{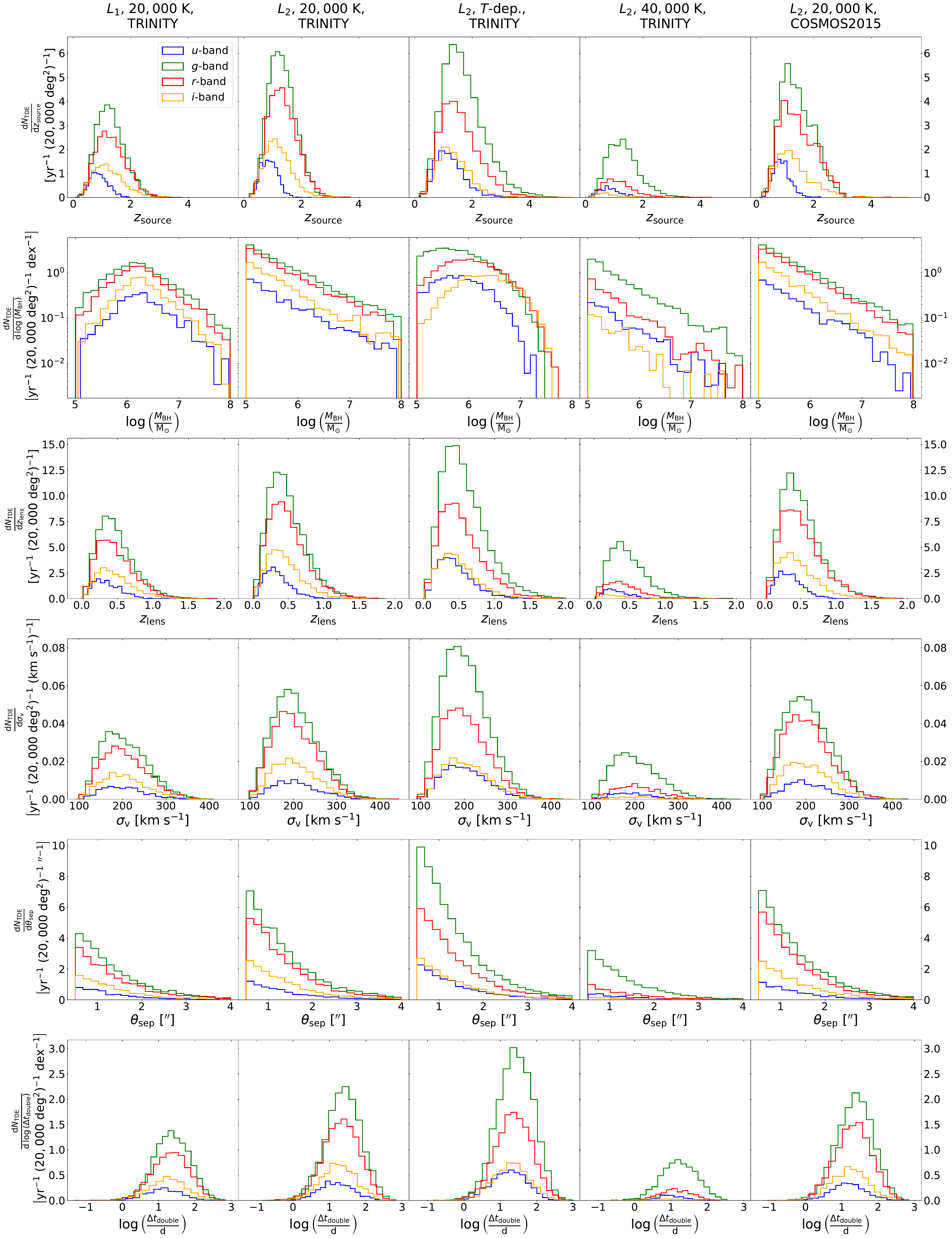}
        \caption{Differential rate of lensed \acp{tde} for five representative models. The rows from top to bottom are the \ac{tde} redshifts, \ac{bh} masses, lens redshifts, lens velocity dispersions, lensed image separations, and time delays of doubles. The three time delays for quads are shown in Appendix~\ref{sec:quads_delays}. These figures were calculated for \ac{lsst} magnitude limits.}
        \label{fig:lensed_tde_hist}
    \end{figure*}

    Lastly, we investigated the different parameters of the lensed \ac{tde} population. We selected a subset of five models to allow each parameter or combination of parameters to be varied. The distributions shown are for \ac{lsst} limiting magnitudes in Fig.~\ref{fig:lensed_tde_hist}.
    
    First, we considered the \ac{tde} parameters: source redshift and \ac{bh} mass (see the first and second rows of Fig.~\ref{fig:lensed_tde_hist}), respectively. We expected most lensed \acp{tde} to be observed between redshift $1$ and $2$. Overall, the $g$ band is always best suited to observe \acp{tde} because it has the faintest limiting magnitude. As expected, the \ac{bh} mass distribution follows the shape of the magnitude functions. The brighter \acp{tde} at a given \ac{bh} mass, the more of these \acp{tde} are observed.

    Second, we assessed the lens parameters, which are the lens redshift and lens velocity dispersion (see the third and fourth rows of Fig.~\ref{fig:lensed_tde_hist}, respectively). As expected, the \ac{tde} model has little to no effect on the lenses. For the temperature-dependent and the $40,000 \; \mathrm{K}$ models, we observed lenses at a higher redshift, but this is solely due to the fact that in these models, \acp{tde} can also be observed at a larger redshift. Most lenses sit at $z \lesssim 1$ and have a velocity dispersion between $100$ and $300 \; \mathrm{km} \; \mathrm{s}^{-1}$.

    Third, we examined the parameters of the resulting lens system. The image separation is shown in the fifth row and the time delay of doubles is shown in the sixth row of Fig.~\ref{fig:lensed_tde_hist}. Additionally, we tabulated the ratio of double to quad lenses in Table~\ref{tab:doubles_quads_number_images}. For quads, the maximal image separation was taken. These parameters are also largely independent of the chosen \ac{tde} model. Only the fact that some models predict \acp{tde} at higher redshifts influences the distributions. Most images have an image separation below $3''$, but the smaller the image separation, the more probable such a system is. As mentioned earlier, the image separation was cut at $0.5''$. Similarly independent is the time delay of doubles. Most lens systems display a time delay ranging from a few days to around $100$ days. Models that predict \acp{tde} at higher redshifts also observe longer time delays. Quads show a very similar time-delay distribution for the longest time delay. The second and third time-delay distributions are shifted to shorter time delays. Their peak is around $4$ days instead of around $11$ days. Their distributions are shown in Appendix~\ref{sec:quads_delays}. 
    
    Our analysis suggests around three or four doubles for one quad system. In Table~\ref{tab:doubles_quads_number_images}, the ratio of doubles to quads falls into a range from $0.96$ to $5.41$. Here, the maximal redshift of the different \ac{tde} models plays a large role in the ratio of doubles to quads. Doubles typically prefer lower redshifts, while quads prefer higher redshifts. The exemplary $L_{2}$, COSMOS2015, $20,000 \; \mathrm{K}$ model clearly shows that the quad lenses peak around redshift $2$, while the double lenses peak approximately at $z = 1$ in the $g$, $r$ and $i$ bands. For the $u$ band, quads also peak at a higher redshift than doubles. This difference in the preferred redshift is due to the fact that the tangential caustic typically provides a higher magnification than the radial caustic~\citep{dq_ratio_z_dep1,dq_ratio_z_dep2}. Therefore, quads are more likely to receive a large boost in brightness and are more likely to be observed at great distances. If a given \ac{tde} model predicts more lensed \acp{tde} at low redshifts compared to another model, we expect the former one to display a larger ratio. In Table~\ref{tab:doubles_quads_number_images}, we also calculate the fraction of quad lenses relative to the total number of lensed \acp{tde}. We find that between $16$ to $39\%$ of lensed \acp{tde} will be a quad system. \citet{lensing_code_Oguri_2010} predict in their catalogs for \ac{lsst} that $13\%$ of lensed quasars and $30$ to $32\%$ of lensed supernovae will be observed as quad lenses. Hence, our results are comparable to their findings.
    
    \begin{table}[htbp]
        \caption{Rate of double and quad lensed \acp{tde}, the ratio of double to quad lenses, and the fraction of quad to total lenses.}
        \centering
        \begin{NiceTabular}{c|c|c|c|c|c}
            \CodeBefore
            \rowlistcolors{2}{white,gray!25}[restart,cols={2-6}]
            \Body
            Model & Band & Doubles & Quads & $\frac{N_{\mathrm{doubles}}}{N_{\mathrm{quads}}}$ & $\frac{N_{\mathrm{quads}}}{N_{\mathrm{TDE, lensed}}}$ \\ 
             &  & Rate & Rate &  & $[\%]$ \\
            \hline
            $L_{1}$,  & $u$ & 0.63 & 0.2 & 3.18 & 23.91 \\
            TRINITY, & $g$ & 3.86 & 0.77 & 5.03 & 16.58 \\
            $20 \, 000 \; \mathrm{K}$ & $r$ & 2.82 & 0.69 & 4.12 & 19.54 \\
             & $i$ & 1.26 & 0.41 & 3.07 & 24.58 \\
            \hline
            $L_{2}$,  & $u$ & 0.97 & 0.26 & 3.78 & 20.92 \\
            TRINITY, & $g$ & 6.15 & 1.14 & 5.41 & 15.6 \\
            $20 \, 000 \; \mathrm{K}$ & $r$ & 4.61 & 1.15 & 4.02 & 19.92 \\
             & $i$ & 1.98 & 0.67 & 2.94 & 25.4 \\
            \hline
            $L_{2}$,  & $u$ & 1.7 & 0.6 & 2.81 & 26.22 \\
            TRINITY & $g$ & 8.63 & 1.8 & 4.81 & 17.22 \\
            $T$-dep. & $r$ & 4.98 & 1.28 & 3.89 & 20.43 \\
             & $i$ & 2.06 & 0.67 & 3.09 & 24.46 \\
            \hline
            $L_{2}$,  & $u$ & 0.22 & 0.14 & 1.56 & 39.13 \\
            TRINITY, & $g$ & 2.26 & 0.85 & 2.65 & 27.41 \\
            $40 \, 000 \; \mathrm{K}$ & $r$ & 0.56 & 0.35 & 1.58 & 38.71 \\
             & $i$ & 0.08 & 0.09 & 0.96 & 50.9 \\
            \hline
            $L_{2}$,  & $u$ & 0.9 & 0.24 & 3.75 & 21.04 \\
            COSMOS2015, & $g$ & 5.92 & 1.38 & 4.29 & 18.91 \\
            $20 \, 000 \; \mathrm{K}$ & $r$ & 4.42 & 1.37 & 3.24 & 23.59 \\
             & $i$ & 1.78 & 0.78 & 2.28 & 30.5 \\
            \hline
        \end{NiceTabular}
        \tablefoot{The rates are given in per year and $20,000 \; \mathrm{deg}^{2}$. We expect to find a few doubles for every quad. This table is calculated for \ac{lsst} magnitude limits.}
        \label{tab:doubles_quads_number_images}
    \end{table}

\section{Conclusion}\label{sec:conclusion}
    In this paper, we have numerically calculated the predicted unlensed and lensed \ac{tde} rate as a function of limiting magnitude in four bands. We covered the ongoing \ac{ztf} and the future \ac{lsst} surveys and different model parameters such as the \ac{bhmf}, \ac{tde} luminosity model, and assumed flare temperature. 
    For the unlensed rates, our models predict between $16$ and $5,440  \; \mathrm{yr}^{-1} \; (20,000 \; \mathrm{deg}^{2})^{-1}$ for \ac{ztf} in the $g$ band. For \ac{lsst}, the rate is in the range from $3,580$ to $82,060 \; \mathrm{yr}^{-1} \; (20,000 \; \mathrm{deg}^{2})^{-1}$. In both surveys, the $g$ band stands out as the best option to observe \acp{tde} because it has the faintest limiting magnitude.
    The sizable scattering factor of $\mathcal{O}(10^{2})$ up to $\sim 10^{3}$ that the models display can be attributed to the assumed \ac{tde} temperature, making it the most critical assumption for \ac{tde} rate estimates.
    Despite the uncertain predictions, we have demonstrated that only the local \ac{bhmf} is relevant for current and near-future surveys. Our redshift-dependent \ac{bhmf} only showed a significant deviation from the redshift-independent model at a limiting magnitude of $25$ or fainter. As \ac{lsst} will only reach about magnitude $24$, the redshift evolution of the \ac{bhmf} is not important to \ac{lsst}. 
    Compared to the observational unlensed \ac{tde} rate around $10$ to $15 \; \mathrm{yr}^{-1} \; (20,000 \; \mathrm{deg}^{2})^{-1}$ obtained from \ac{ztf}, most of our models overestimate the rate. It should be expected that we overestimate the observed rate, as we have not accounted for completeness or any observational effects in our calculations. For a more accurate estimate in the future, we could include the host galaxy light or dust extinction to better match observations. Furthermore, removing the assumption that every star is solar-like would yield even more accurate results. Last, certain physical processes in TDEs can also make the detected TDE rates significantly deviate from the theoretical rates \citep{black_hole_mass_function_Wong_2022}.

    From our investigation into the lensed fraction, we can approximate that for every $10,000$ unlensed \acp{tde}, we expect \ac{lsst} to find a few lensed events. The results show a reduced scatter to around a factor of $50$. As our unlensed rates only poorly match observational ones, it seems improbable that a large sample of hundreds of lensed \acp{tde} can be obtained. Yet, \ac{lsst} will observe a few lensed \acp{tde} over its $10$-year survey time.
    Lensed \acp{tde} are also best observed in the $g$ band due to its faint limiting magnitude. For \ac{lsst}, our predictions range from one every two years to more than ten yearly, assuming $20,000 \; \mathrm{deg}^{2}$ in the $g$ band, still showing much of the scatter also observed for the unlensed rate. However, analogously to the unlensed rate, we were able to show that the \ac{tde} temperature is the most important assumption and only the \ac{bhmf} is relevant even beyond \ac{lsst} limiting magnitudes.

\begin{acknowledgements}
    We thank Volker Springel for the helpful discussion on \acp{bhmf}. EM and SHS thank the Max Planck Society for support through the Max Planck Fellowship for SHS. KS acknowledges support through a Fulbright grant of the German-American Fulbright Commission. LD acknowledges the support from the Hong Kong Research Grants Council (17304821, 17314822, 17305124). This work was supported by JSPS KAKENHI Grant Numbers JP23K22531, JP22K21349, JP19KK0076.
\end{acknowledgements}

\bibliographystyle{aa}
\bibliography{bib}

@article{black_hole_mass_function_Wong_2022,
       author = {{Wong}, Thomas Hong Tsun and {Pfister}, Hugo and {Dai}, L.},
        title = "{Revisiting the Rates and Demographics of Tidal Disruption Events: Effects of the Disk Formation Efficiency}",
      journal = {\apjl},
         year = 2022,
        month = mar,
       volume = {927},
       number = {1},
          eid = {L19},
        pages = {L19},
          doi = {10.3847/2041-8213/ac5823},
archivePrefix = {arXiv},
       eprint = {2111.09173},
 primaryClass = {astro-ph.HE},
       adsurl = {https://ui.adsabs.harvard.edu/abs/2022ApJ...927L..19W}
}

@article{tde_rate_per_year_Pfister_2020,
       author = {{Pfister}, Hugo and {Volonteri}, Marta and {Dai}, Lixin and {Colpi}, Monica}, 
        title = "{Enhancement of the tidal disruption event rate in galaxies with a nuclear star cluster: from dwarfs to ellipticals}",
      journal = {\mnras},
         year = 2020,
        month = sep,
       volume = {497},
       number = {2},
        pages = {2276-2285},
          doi = {10.1093/mnras/staa1962},
archivePrefix = {arXiv},
       eprint = {2003.08133},
 primaryClass = {astro-ph.GA},
       adsurl = {https://ui.adsabs.harvard.edu/abs/2020MNRAS.497.2276P}
}

@article{magnitude_and_flux_Huber_2021,
       author = {{Huber}, S. and {Suyu}, S. and {Noebauer}, U.~M. and {Chan}, J.~H.~H. and {Kromer}, M. and {Sim}, S.~A. and {Sluse}, D. and {Taubenberger}, S.},
        title = "{HOLISMOKES. III. Achromatic phase of strongly lensed Type Ia supernovae}",
      journal = {\aap},
         year = 2021,
        month = feb,
       volume = {646},
          eid = {A110},
        pages = {A110},
          doi = {10.1051/0004-6361/202039218},
archivePrefix = {arXiv},
       eprint = {2008.10393},
 primaryClass = {astro-ph.HE},
       adsurl = {https://ui.adsabs.harvard.edu/abs/2021A&A...646A.110H}
}

@article{luminosity_temperature_mass_dependence_Ryu_2020,
       author = {{Ryu}, Taeho and {Krolik}, Julian and {Piran}, Tsvi},
        title = "{Measuring Stellar and Black Hole Masses of Tidal Disruption Events}",
      journal = {\apj},
         year = 2020,
        month = nov,
       volume = {904},
       number = {1},
          eid = {73},
        pages = {73},
          doi = {10.3847/1538-4357/abbf4d},
archivePrefix = {arXiv},
       eprint = {2007.13765},
 primaryClass = {astro-ph.HE},
       adsurl = {https://ui.adsabs.harvard.edu/abs/2020ApJ...904...73R}
}

@article{lensing_code_Oguri_2010,
       author = {{Oguri}, Masamune and {Marshall}, Philip J.},
        title = "{Gravitationally lensed quasars and supernovae in future wide-field optical imaging surveys}",
      journal = {\mnras},
         year = 2010,
        month = jul,
       volume = {405},
       number = {4},
        pages = {2579-2593},
          doi = {10.1111/j.1365-2966.2010.16639.x},
archivePrefix = {arXiv},
       eprint = {1001.2037},
 primaryClass = {astro-ph.CO},
       adsurl = {https://ui.adsabs.harvard.edu/abs/2010MNRAS.405.2579O}
}

@article{derivation_of_the_BHMF_Gallo_2019,
       author = {{Gallo}, Elena and {Sesana}, Alberto},
        title = "{Exploring the Local Black Hole Mass Function below {}10$^{6}$ Solar Masses}",
      journal = {\apjl},
         year = 2019,
        month = sep,
       volume = {883},
       number = {1},
          eid = {L18},
        pages = {L18},
          doi = {10.3847/2041-8213/ab40c6},
archivePrefix = {arXiv},
       eprint = {1909.02585},
 primaryClass = {astro-ph.HE},
       adsurl = {https://ui.adsabs.harvard.edu/abs/2019ApJ...883L..18G}
}

@article{TDE_rates_Szekerczes_2024,
       author = {{Szekerczes}, K. and {Ryu}, T. and {Suyu}, S.~H. and {Huber}, S. and {Oguri}, M. and {Dai}, L.},
        title = "{Strong lensing of tidal disruption events: Detection rates in imaging surveys}",
      journal = {\aap},
         year = 2024,
        month = oct,
       volume = {690},
          eid = {A384},
        pages = {A384},
          doi = {10.1051/0004-6361/202449481},
archivePrefix = {arXiv},
       eprint = {2402.03443},
 primaryClass = {astro-ph.HE},
       adsurl = {https://ui.adsabs.harvard.edu/abs/2024A&A...690A.384S}
}

@article{z_dep_BHMF_TRINITY_1,
       author = {{Zhang}, Haowen and {Behroozi}, Peter and {Volonteri}, Marta and {Silk}, Joseph and {Fan}, Xiaohui and {Hopkins}, Philip F. and {Yang}, Jinyi and {Aird}, James},
        title = "{TRINITY I: self-consistently modelling the dark matter halo-galaxy-supermassive black hole connection from z = 0-10}",
      journal = {\mnras},
         year = 2023,
        month = jan,
       volume = {518},
       number = {2},
        pages = {2123-2163},
          doi = {10.1093/mnras/stac2633},
archivePrefix = {arXiv},
       eprint = {2105.10474},
 primaryClass = {astro-ph.GA},
       adsurl = {https://ui.adsabs.harvard.edu/abs/2023MNRAS.518.2123Z}
}

@ARTICLE{tdes_last_one_year,
       author = {{Lin}, Dacheng and {Guillochon}, James and {Komossa}, S. and {Ramirez-Ruiz}, Enrico and {Irwin}, Jimmy A. and {Maksym}, W. Peter and {Grupe}, Dirk and {Godet}, Olivier and {Webb}, Natalie A. and {Barret}, Didier and {Zauderer}, B. Ashley and {Duc}, Pierre-Alain and {Carrasco}, Eleazar R. and {Gwyn}, Stephen D.~J.},
        title = "{A likely decade-long sustained tidal disruption event}",
      journal = {\na},
         year = 2017,
        month = feb,
       volume = {1},
          eid = {0033},
        pages = {0033},
          doi = {10.1038/s41550-016-0033},
archivePrefix = {arXiv},
       eprint = {1702.00792},
 primaryClass = {astro-ph.HE},
       adsurl = {https://ui.adsabs.harvard.edu/abs/2017NatAs...1E..33L}
}

@ARTICLE{tdes_probe_dormant_galaxies,
       author = {{Rees}, Martin J.},
        title = "{Tidal disruption of stars by black holes of {}10$^{6}$-{}10$^{8}$ solar masses in nearby galaxies}",
      journal = {\nat},
         year = 1988,
        month = jun,
       volume = {333},
       number = {6173},
        pages = {523-528},
          doi = {10.1038/333523a0},
       adsurl = {https://ui.adsabs.harvard.edu/abs/1988Natur.333..523R}
}

@INPROCEEDINGS{tde_as_probes_of_demo_and_phys,
       author = {{Gezari}, Suvi},
        title = "{Tidal Disruption Events: New Transient Probes of Accretion, Jet Physics, and Black Hole Demographics}",
    booktitle = {American Astronomical Society Meeting Abstracts \#221},
         year = 2013,
       series = {American Astronomical Society Meeting Abstracts},
       volume = {221},
        month = jan,
          eid = {131.02},
        pages = {131.02},
       adsurl = {https://ui.adsabs.harvard.edu/abs/2013AAS...22113102G}
}

@ARTICLE{hubble_constant_from_lensing,
       author = {{Refsdal}, S.},
        title = "{On the possibility of determining Hubble's parameter and the masses of galaxies from the gravitational lens effect}",
      journal = {\mnras},
         year = 1964,
        month = jan,
       volume = {128},
        pages = {307},
          doi = {10.1093/mnras/128.4.307},
       adsurl = {https://ui.adsabs.harvard.edu/abs/1964MNRAS.128..307R}
}

@ARTICLE{agn_size_measurement,
       author = {{Kochanek}, C.~S.},
        title = "{Quantitative Interpretation of Quasar Microlensing Light Curves}",
      journal = {\apj},
         year = 2004,
        month = apr,
       volume = {605},
       number = {1},
        pages = {58-77},
          doi = {10.1086/382180},
archivePrefix = {arXiv},
       eprint = {astro-ph/0307422},
 primaryClass = {astro-ph},
       adsurl = {https://ui.adsabs.harvard.edu/abs/2004ApJ...605...58K}
}

@ARTICLE{GSMF,
       author = {{Davidzon}, I. and {Ilbert}, O. and {Laigle}, C. and {Coupon}, J. and {McCracken}, H.~J. and {Delvecchio}, I. and {Masters}, D. and {Capak}, P. and {Hsieh}, B.~C. and {Le F{\`e}vre}, O. and {Tresse}, L. and {Bethermin}, M. and {Chang}, Y. -Y. and {Faisst}, A.~L. and {Le Floc'h}, E. and {Steinhardt}, C. and {Toft}, S. and {Aussel}, H. and {Dubois}, C. and {Hasinger}, G. and {Salvato}, M. and {Sanders}, D.~B. and {Scoville}, N. and {Silverman}, J.~D.},
        title = "{The COSMOS2015 galaxy stellar mass function . Thirteen billion years of stellar mass assembly in ten snapshots}",
      journal = {\aap},
         year = 2017,
        month = sep,
       volume = {605},
          eid = {A70},
        pages = {A70},
          doi = {10.1051/0004-6361/201730419},
archivePrefix = {arXiv},
       eprint = {1701.02734},
 primaryClass = {astro-ph.GA},
       adsurl = {https://ui.adsabs.harvard.edu/abs/2017A&A...605A..70D}
}

@ARTICLE{bh_bulge_mass_relation,
       author = {{Kormendy}, John and {Ho}, Luis C.},
        title = "{Coevolution (Or Not) of Supermassive Black Holes and Host Galaxies}",
      journal = {\araa},
         year = 2013,
        month = aug,
       volume = {51},
       number = {1},
        pages = {511-653},
          doi = {10.1146/annurev-astro-082708-101811},
archivePrefix = {arXiv},
       eprint = {1304.7762},
 primaryClass = {astro-ph.CO},
       adsurl = {https://ui.adsabs.harvard.edu/abs/2013ARA&A..51..511K}
}

@ARTICLE{bh_mass_bulge_mass_redshift_dep,
       author = {{Shimizu}, Tatsuki and {Oogi}, Taira and {Okamoto}, Takashi and {Nagashima}, Masahiro and {Enoki}, Motohiro},
        title = "{The evolution of supermassive black hole mass-bulge mass relation by a semi-analytical model, {\ensuremath{\nu}}$^{2}$GC}",
      journal = {\mnras},
         year = 2024,
        month = jun,
       volume = {531},
       number = {1},
        pages = {851-858},
          doi = {10.1093/mnras/stae1226},
archivePrefix = {arXiv},
       eprint = {2405.07461},
 primaryClass = {astro-ph.GA},
       adsurl = {https://ui.adsabs.harvard.edu/abs/2024MNRAS.531..851S}
}

@ARTICLE{flux_to_magnitude,
       author = {{Bessell}, Michael and {Murphy}, Simon},
        title = "{Spectrophotometric Libraries, Revised Photonic Passbands, and Zero Points for UBVRI, Hipparcos, and Tycho Photometry}",
      journal = {\pasp},
         year = 2012,
        month = feb,
       volume = {124},
       number = {912},
        pages = {140},
          doi = {10.1086/664083},
archivePrefix = {arXiv},
       eprint = {1112.2698},
 primaryClass = {astro-ph.SR},
       adsurl = {https://ui.adsabs.harvard.edu/abs/2012PASP..124..140B}
}

@ARTICLE{planck_law,
       author = {{Planck}, Max},
        title = "{Ueber das Gesetz der Energieverteilung im Normalspectrum}",
      journal = {Annalen der Physik},
         year = 1901,
        month = jan,
       volume = {309},
       number = {3},
        pages = {553-563},
          doi = {10.1002/andp.19013090310},
       adsurl = {https://ui.adsabs.harvard.edu/abs/1901AnP...309..553P}
}

@BOOK{stefan_boltzmann_law,
  title = "{Thermodynamik}",
  subtitle = {Grundlagen und technische Anwendungen},
  author = {{Baehr}, H.~D. and {Kabelac}, S.},
  isbn={978-3-662-49567-4 },
  year = {2016},
  PUBLISHER = {Springer},
  doi={10.1007/978-3-662-49568-1}
}

@ARTICLE{TDE_L1,
       author = {{Thomsen}, Lars L. and {Kwan}, Tom M. and {Dai}, Lixin and {Wu}, Samantha C. and {Roth}, Nathaniel and {Ramirez-Ruiz}, Enrico},
        title = "{Dynamical Unification of Tidal Disruption Events}",
      journal = {\apjl},
         year = 2022,
        month = oct,
       volume = {937},
       number = {2},
          eid = {L28},
        pages = {L28},
          doi = {10.3847/2041-8213/ac911f},
archivePrefix = {arXiv},
       eprint = {2206.02804},
 primaryClass = {astro-ph.HE},
       adsurl = {https://ui.adsabs.harvard.edu/abs/2022ApJ...937L..28T}
}

@ARTICLE{mass_fallback_rate,
       author = {{Ryu}, Taeho and {Krolik}, Julian and {Piran}, Tsvi and {Noble}, Scott C.},
        title = "{Tidal Disruptions of Main-sequence Stars. I. Observable Quantities and Their Dependence on Stellar and Black Hole Mass}",
      journal = {\apj},
         year = 2020,
        month = dec,
       volume = {904},
       number = {2},
          eid = {98},
        pages = {98},
          doi = {10.3847/1538-4357/abb3cf},
archivePrefix = {arXiv},
       eprint = {2001.03501},
 primaryClass = {astro-ph.HE},
       adsurl = {https://ui.adsabs.harvard.edu/abs/2020ApJ...904...98R}
}

@BOOK{eddington_limit,
  title = "{Radiative processes in astrophysics}",
  author = {{Rybicki}, George B. and {Lightman}, Alan P.},
  isbn={978-0-471-82759-7},
  year = 1979,
  PUBLISHER = {John Wiley \& Sons}
}

@ARTICLE{theoretical_L2_model,
       author = {{Piran}, Tsvi and {Svirski}, Gilad and {Krolik}, Julian and {Cheng}, Roseanne M. and {Shiokawa}, Hotaka},
        title = "{Disk Formation Versus Disk Accretion - What Powers Tidal Disruption Events?}", 
      journal = {\apj},
         year = 2015,
        month = jun,
       volume = {806},
       number = {2},
          eid = {164},
        pages = {164},
          doi = {10.1088/0004-637X/806/2/164},
archivePrefix = {arXiv},
       eprint = {1502.05792},
 primaryClass = {astro-ph.HE},
       adsurl = {https://ui.adsabs.harvard.edu/abs/2015ApJ...806..164P}
}

@ARTICLE{ident_of_imbh,
       author = {{Davis}, Benjamin L. and {Graham}, Alister W. and {Soria}, Roberto and {Jin}, Zehao and {Karachentsev}, Igor D. and {Karachentseva}, Valentina E. and {D'Onghia}, Elena},
        title = "{Identification of Intermediate-mass Black Hole Candidates among a Sample of Sd Galaxies}",
      journal = {\apj},
         year = 2024,
        month = aug,
       volume = {971},
       number = {2},
          eid = {123},
        pages = {123},
          doi = {10.3847/1538-4357/ad55eb},
archivePrefix = {arXiv},
       eprint = {2406.05778},
 primaryClass = {astro-ph.GA},
       adsurl = {https://ui.adsabs.harvard.edu/abs/2024ApJ...971..123D}
}

@ARTICLE{search_for_imbh,
       author = {{Pomeroy}, Richard T. and {Norris}, Mark A.},
        title = "{A search for intermediate-mass black holes in compact stellar systems through optical emissions from tidal disruption events}",
      journal = {\mnras},
         year = 2024,
        month = may,
       volume = {530},
       number = {3},
        pages = {3043-3050},
          doi = {10.1093/mnras/stae960},
archivePrefix = {arXiv},
       eprint = {2404.09144},
 primaryClass = {astro-ph.GA},
       adsurl = {https://ui.adsabs.harvard.edu/abs/2024MNRAS.530.3043P}
}

@ARTICLE{stars_swallowed_whole,
       author = {{Hills}, J.~G.},
        title = "{Possible power source of Seyfert galaxies and QSOs}",
      journal = {\nat},
         year = 1975,
        month = mar,
       volume = {254},
       number = {5498},
        pages = {295-298},
          doi = {10.1038/254295a0},
       adsurl = {https://ui.adsabs.harvard.edu/abs/1975Natur.254..295H}
}

@ARTICLE{observational_zft_tdes,
       author = {{Hammerstein}, Erica and {van Velzen}, Sjoert and {Gezari}, Suvi and {Cenko}, S. Bradley and {Yao}, Yuhan and {Ward}, Charlotte and {Frederick}, Sara and {Villanueva}, Natalia and {Somalwar}, Jean J. and {Graham}, Matthew J. and {Kulkarni}, Shrinivas R. and {Stern}, Daniel and {Andreoni}, Igor and {Bellm}, Eric C. and {Dekany}, Richard and {Dhawan}, Suhail and {Drake}, Andrew J. and {Fremling}, Christoffer and {Gatkine}, Pradip and {Groom}, Steven L. and {Ho}, Anna Y.~Q. and {Kasliwal}, Mansi M. and {Karambelkar}, Viraj and {Kool}, Erik C. and {Masci}, Frank J. and {Medford}, Michael S. and {Perley}, Daniel A. and {Purdum}, Josiah and {van Roestel}, Jan and {Sharma}, Yashvi and {Sollerman}, Jesper and {Taggart}, Kirsty and {Yan}, Lin},
        title = "{The Final Season Reimagined: 30 Tidal Disruption Events from the ZTF-I Survey}",
      journal = {\apj},
         year = 2023,
        month = jan,
       volume = {942},
       number = {1},
          eid = {9},
        pages = {9},
          doi = {10.3847/1538-4357/aca283},
archivePrefix = {arXiv},
       eprint = {2203.01461},
 primaryClass = {astro-ph.HE},
       adsurl = {https://ui.adsabs.harvard.edu/abs/2023ApJ...942....9H}
}

@ARTICLE{ambigous_TDEs,
       author = {{Zabludoff}, Ann and {Arcavi}, Iair and {LaMassa}, Stephanie and {Perets}, Hagai B. and {Trakhtenbrot}, Benny and {Zauderer}, B. Ashley and {Auchettl}, Katie and {Dai}, Jane L. and {French}, K. Decker and {Hung}, Tiara and {Kara}, Erin and {Lodato}, Giuseppe and {Maksym}, W. Peter and {Qin}, Yujing and {Ramirez-Ruiz}, Enrico and {Roth}, Nathaniel and {Runnoe}, Jessie C. and {Wevers}, Thomas},
        title = "{Distinguishing Tidal Disruption Events from Impostors}",
      journal = {\ssr},
         year = 2021,
        month = jun,
       volume = {217},
       number = {4},
          eid = {54},
        pages = {54},
          doi = {10.1007/s11214-021-00829-4},
archivePrefix = {arXiv},
       eprint = {2103.12150},
 primaryClass = {astro-ph.HE},
       adsurl = {https://ui.adsabs.harvard.edu/abs/2021SSRv..217...54Z}
}

@ARTICLE{identify_TDEs_with_ML,
       author = {{Gomez}, Sebastian and {Villar}, V. Ashley and {Berger}, Edo and {Gezari}, Suvi and {van Velzen}, Sjoert and {Nicholl}, Matt and {Blanchard}, Peter K. and {Alexander}, Kate. D.},
        title = "{Identifying Tidal Disruption Events with an Expansion of the FLEET Machine-learning Algorithm}",
      journal = {\apj},
         year = 2023,
        month = jun,
       volume = {949},
       number = {2},
          eid = {113},
        pages = {113},
          doi = {10.3847/1538-4357/acc535},
archivePrefix = {arXiv},
       eprint = {2210.10810},
 primaryClass = {astro-ph.HE},
       adsurl = {https://ui.adsabs.harvard.edu/abs/2023ApJ...949..113G}
}

@ARTICLE{tdescore_identification_TDEs,
       author = {{Stein}, Robert and {Mahabal}, Ashish and {Reusch}, Simeon and {Graham}, Matthew and {Kasliwal}, Mansi M. and {Kowalski}, Marek and {Gezari}, Suvi and {Hammerstein}, Erica and {Nakoneczny}, Szymon J. and {Nicholl}, Matt and {Sollerman}, Jesper and {van Velzen}, Sjoert and {Yao}, Yuhan and {Laher}, Russ R. and {Rusholme}, Ben},
        title = "{tdescore: An Accurate Photometric Classifier for Tidal Disruption Events}",
      journal = {\apjl},
         year = 2024,
        month = apr,
       volume = {965},
       number = {2},
          eid = {L14},
        pages = {L14},
          doi = {10.3847/2041-8213/ad3337},
archivePrefix = {arXiv},
       eprint = {2312.00139},
 primaryClass = {astro-ph.IM},
       adsurl = {https://ui.adsabs.harvard.edu/abs/2024ApJ...965L..14S}
}

@ARTICLE{SIE_derivation,
       author = {{Kormann}, R. and {Schneider}, P. and {Bartelmann}, M.},
        title = "{Isothermal elliptical gravitational lens models.}",
      journal = {\aap},
         year = 1994,
        month = apr,
       volume = {284},
        pages = {285-299},
       adsurl = {https://ui.adsabs.harvard.edu/abs/1994A&A...284..285K}
}

@ARTICLE{tde_40_day_decline,
       author = {{Hinkle}, Jason T. and {Holoien}, Thomas W. -S. and {Shappee}, Benjamin. J. and {Auchettl}, Katie and {Kochanek}, Christopher S. and {Stanek}, K.~Z. and {Payne}, Anna V. and {Thompson}, Todd A.},
        title = "{Examining a Peak-luminosity/Decline-rate Relationship for Tidal Disruption Events}",
      journal = {\apjl},
         year = 2020,
        month = may,
       volume = {894},
       number = {1},
          eid = {L10},
        pages = {L10},
          doi = {10.3847/2041-8213/ab89a2},
archivePrefix = {arXiv},
       eprint = {2001.08215},
 primaryClass = {astro-ph.HE},
       adsurl = {https://ui.adsabs.harvard.edu/abs/2020ApJ...894L..10H}
}

@ARTICLE{one_tde_lightcurve,
       author = {{Holoien}, Thomas W. -S. and {Auchettl}, Katie and {Tucker}, Michael A. and {Shappee}, Benjamin J. and {Patel}, Shannon G. and {Miller-Jones}, James C.~A. and {Mockler}, Brenna and {Groenewald}, Dani{\`e}l N. and {Hinkle}, Jason T. and {Brown}, Jonathan S. and {Kochanek}, Christopher S. and {Stanek}, K.~Z. and {Chen}, Ping and {Dong}, Subo and {Prieto}, Jose L. and {Thompson}, Todd A. and {Beaton}, Rachael L. and {Connor}, Thomas and {Cowperthwaite}, Philip S. and {Dahmen}, Linnea and {French}, K. Decker and {Morrell}, Nidia and {Buckley}, David A.~H. and {Gromadzki}, Mariusz and {Roy}, Rupak and {Coulter}, David A. and {Dimitriadis}, Georgios and {Foley}, Ryan J. and {Kilpatrick}, Charles D. and {Piro}, Anthony L. and {Rojas-Bravo}, C{\'e}sar and {Siebert}, Matthew R. and {van Velzen}, Sjoert},
        title = "{The Rise and Fall of ASASSN-18pg: Following a TDE from Early to Late Times}",
      journal = {\apj},
         year = 2020,
        month = aug,
       volume = {898},
       number = {2},
          eid = {161},
        pages = {161},
          doi = {10.3847/1538-4357/ab9f3d},
archivePrefix = {arXiv},
       eprint = {2003.13693},
 primaryClass = {astro-ph.HE},
       adsurl = {https://ui.adsabs.harvard.edu/abs/2020ApJ...898..161H}
}

@ARTICLE{lsst_limiting_magnitude,
       author = {{Lochner}, Michelle and {Scolnic}, Dan and {Almoubayyed}, Husni and {Anguita}, Timo and {Awan}, Humna and {Gawiser}, Eric and {A Gontcho}, Satya Gontcho and {Graham}, Melissa L. and {Gris}, Philippe and {Huber}, Simon and {Jha}, Saurabh W. and {Lynne Jones}, R. and {Kim}, Alex G. and {Mandelbaum}, Rachel and {Marshall}, Phil and {Petrushevska}, Tanja and {Regnault}, Nicolas and {Setzer}, Christian N. and {Suyu}, Sherry H. and {Yoachim}, Peter and {Biswas}, Rahul and {Blaineau}, Tristan and {Hook}, Isobel and {Moniez}, Marc and {Neilsen}, Eric and {Peiris}, Hiranya and {Rothchild}, Daniel and {Stubbs}, Christopher and {LSST Dark Energy Science Collaboration}},
        title = "{The Impact of Observing Strategy on Cosmological Constraints with LSST}",
      journal = {\apjs},
         year = 2022,
        month = apr,
       volume = {259},
       number = {2},
          eid = {58},
        pages = {58},
          doi = {10.3847/1538-4365/ac5033},
archivePrefix = {arXiv},
       eprint = {2104.05676},
 primaryClass = {astro-ph.CO},
       adsurl = {https://ui.adsabs.harvard.edu/abs/2022ApJS..259...58L}
}

@ARTICLE{ztf_limiting_magnitude,
       author = {{Bellm}, Eric C. and {Kulkarni}, Shrinivas R. and {Graham}, Matthew J. and {Dekany}, Richard and {Smith}, Roger M. and {Riddle}, Reed and {Masci}, Frank J. and {Helou}, George and {Prince}, Thomas A. and {Adams}, Scott M. and {Barbarino}, C. and {Barlow}, Tom and {Bauer}, James and {Beck}, Ron and {Belicki}, Justin and {Biswas}, Rahul and {Blagorodnova}, Nadejda and {Bodewits}, Dennis and {Bolin}, Bryce and {Brinnel}, Valery and {Brooke}, Tim and {Bue}, Brian and {Bulla}, Mattia and {Burruss}, Rick and {Cenko}, S. Bradley and {Chang}, Chan-Kao and {Connolly}, Andrew and {Coughlin}, Michael and {Cromer}, John and {Cunningham}, Virginia and {De}, Kishalay and {Delacroix}, Alex and {Desai}, Vandana and {Duev}, Dmitry A. and {Eadie}, Gwendolyn and {Farnham}, Tony L. and {Feeney}, Michael and {Feindt}, Ulrich and {Flynn}, David and {Franckowiak}, Anna and {Frederick}, S. and {Fremling}, C. and {Gal-Yam}, Avishay and {Gezari}, Suvi and {Giomi}, Matteo and {Goldstein}, Daniel A. and {Golkhou}, V. Zach and {Goobar}, Ariel and {Groom}, Steven and {Hacopians}, Eugean and {Hale}, David and {Henning}, John and {Ho}, Anna Y.~Q. and {Hover}, David and {Howell}, Justin and {Hung}, Tiara and {Huppenkothen}, Daniela and {Imel}, David and {Ip}, Wing-Huen and {Ivezi{\'c}}, {\v{Z}}eljko and {Jackson}, Edward and {Jones}, Lynne and {Juric}, Mario and {Kasliwal}, Mansi M. and {Kaspi}, S. and {Kaye}, Stephen and {Kelley}, Michael S.~P. and {Kowalski}, Marek and {Kramer}, Emily and {Kupfer}, Thomas and {Landry}, Walter and {Laher}, Russ R. and {Lee}, Chien-De and {Lin}, Hsing Wen and {Lin}, Zhong-Yi and {Lunnan}, Ragnhild and {Giomi}, Matteo and {Mahabal}, Ashish and {Mao}, Peter and {Miller}, Adam A. and {Monkewitz}, Serge and {Murphy}, Patrick and {Ngeow}, Chow-Choong and {Nordin}, Jakob and {Nugent}, Peter and {Ofek}, Eran and {Patterson}, Maria T. and {Penprase}, Bryan and {Porter}, Michael and {Rauch}, Ludwig and {Rebbapragada}, Umaa and {Reiley}, Dan and {Rigault}, Mickael and {Rodriguez}, Hector and {van Roestel}, Jan and {Rusholme}, Ben and {van Santen}, Jakob and {Schulze}, S. and {Shupe}, David L. and {Singer}, Leo P. and {Soumagnac}, Maayane T. and {Stein}, Robert and {Surace}, Jason and {Sollerman}, Jesper and {Szkody}, Paula and {Taddia}, F. and {Terek}, Scott and {Van Sistine}, Angela and {van Velzen}, Sjoert and {Vestrand}, W. Thomas and {Walters}, Richard and {Ward}, Charlotte and {Ye}, Quan-Zhi and {Yu}, Po-Chieh and {Yan}, Lin and {Zolkower}, Jeffry},
        title = "{The Zwicky Transient Facility: System Overview, Performance, and First Results}",
      journal = {\pasp},
         year = 2019,
        month = jan,
       volume = {131},
       number = {995},
        pages = {018002},
          doi = {10.1088/1538-3873/aaecbe},
archivePrefix = {arXiv},
       eprint = {1902.01932},
 primaryClass = {astro-ph.IM},
       adsurl = {https://ui.adsabs.harvard.edu/abs/2019PASP..131a8002B}
}

@ARTICLE{cosmos2015_catalog,
       author = {{Laigle}, C. and {McCracken}, H.~J. and {Ilbert}, O. and {Hsieh}, B.~C. and {Davidzon}, I. and {Capak}, P. and {Hasinger}, G. and {Silverman}, J.~D. and {Pichon}, C. and {Coupon}, J. and {Aussel}, H. and {Le Borgne}, D. and {Caputi}, K. and {Cassata}, P. and {Chang}, Y. -Y. and {Civano}, F. and {Dunlop}, J. and {Fynbo}, J. and {Kartaltepe}, J.~S. and {Koekemoer}, A. and {Le F{\`e}vre}, O. and {Le Floc'h}, E. and {Leauthaud}, A. and {Lilly}, S. and {Lin}, L. and {Marchesi}, S. and {Milvang-Jensen}, B. and {Salvato}, M. and {Sanders}, D.~B. and {Scoville}, N. and {Smolcic}, V. and {Stockmann}, M. and {Taniguchi}, Y. and {Tasca}, L. and {Toft}, S. and {Vaccari}, Mattia and {Zabl}, J.},
        title = "{The COSMOS2015 Catalog: Exploring the 1 < z < 6 Universe with Half a Million Galaxies}",
      journal = {\apjs},
         year = 2016,
        month = jun,
       volume = {224},
       number = {2},
          eid = {24},
        pages = {24},
          doi = {10.3847/0067-0049/224/2/24},
archivePrefix = {arXiv},
       eprint = {1604.02350},
 primaryClass = {astro-ph.GA},
       adsurl = {https://ui.adsabs.harvard.edu/abs/2016ApJS..224...24L}
}

@ARTICLE{clasic_tde,
       author = {{Hills}, J.~G.},
        title = "{Hyper-velocity and tidal stars from binaries disrupted by a massive Galactic black hole}",
      journal = {\nat},
         year = 1988,
        month = feb,
       volume = {331},
       number = {6158},
        pages = {687-689},
          doi = {10.1038/331687a0},
       adsurl = {https://ui.adsabs.harvard.edu/abs/1988Natur.331..687H}
}

@ARTICLE{TDE_general_review_article,
       author = {{Gezari}, Suvi},
        title = "{Tidal Disruption Events}",
      journal = {\araa},
         year = 2021,
        month = sep,
       volume = {59},
        pages = {21-58},
          doi = {10.1146/annurev-astro-111720-030029},
archivePrefix = {arXiv},
       eprint = {2104.14580},
 primaryClass = {astro-ph.HE},
       adsurl = {https://ui.adsabs.harvard.edu/abs/2021ARA&A..59...21G}
}

@PROCEEDINGS{ztf_survey,
        title = "{The Third Hot-wiring the Transient Universe Workshop (HTU-III)}",
    booktitle = {The Third Hot-wiring the Transient Universe Workshop},
         year = 2014,
       editor = {{Wozniak}, P.~R. and {Graham}, M.~J. and {Mahabal}, A.~A. and {Seaman}, R.},
        month = oct,
       adsurl = {https://ui.adsabs.harvard.edu/abs/2014htu..conf.....W}
}

@ARTICLE{lsst_survey,
       author = {{Ivezi{\'c}}, {\v{Z}}eljko and {Kahn}, Steven M. and {Tyson}, J. Anthony and {Abel}, Bob and {Acosta}, Emily and {Allsman}, Robyn and {Alonso}, David and {AlSayyad}, Yusra and {Anderson}, Scott F. and {Andrew}, John and {Angel}, James Roger P. and {Angeli}, George Z. and {Ansari}, Reza and {Antilogus}, Pierre and {Araujo}, Constanza and {Armstrong}, Robert and {Arndt}, Kirk T. and {Astier}, Pierre and {Aubourg}, {\'E}ric and {Auza}, Nicole and {Axelrod}, Tim S. and {Bard}, Deborah J. and {Barr}, Jeff D. and {Barrau}, Aurelian and {Bartlett}, James G. and {Bauer}, Amanda E. and {Bauman}, Brian J. and {Baumont}, Sylvain and {Bechtol}, Ellen and {Bechtol}, Keith and {Becker}, Andrew C. and {Becla}, Jacek and {Beldica}, Cristina and {Bellavia}, Steve and {Bianco}, Federica B. and {Biswas}, Rahul and {Blanc}, Guillaume and {Blazek}, Jonathan and {Blandford}, Roger D. and {Bloom}, Josh S. and {Bogart}, Joanne and {Bond}, Tim W. and {Booth}, Michael T. and {Borgland}, Anders W. and {Borne}, Kirk and {Bosch}, James F. and {Boutigny}, Dominique and {Brackett}, Craig A. and {Bradshaw}, Andrew and {Brandt}, William Nielsen and {Brown}, Michael E. and {Bullock}, James S. and {Burchat}, Patricia and {Burke}, David L. and {Cagnoli}, Gianpietro and {Calabrese}, Daniel and {Callahan}, Shawn and {Callen}, Alice L. and {Carlin}, Jeffrey L. and {Carlson}, Erin L. and {Chandrasekharan}, Srinivasan and {Charles-Emerson}, Glenaver and {Chesley}, Steve and {Cheu}, Elliott C. and {Chiang}, Hsin-Fang and {Chiang}, James and {Chirino}, Carol and {Chow}, Derek and {Ciardi}, David R. and {Claver}, Charles F. and {Cohen-Tanugi}, Johann and {Cockrum}, Joseph J. and {Coles}, Rebecca and {Connolly}, Andrew J. and {Cook}, Kem H. and {Cooray}, Asantha and {Covey}, Kevin R. and {Cribbs}, Chris and {Cui}, Wei and {Cutri}, Roc and {Daly}, Philip N. and {Daniel}, Scott F. and {Daruich}, Felipe and {Daubard}, Guillaume and {Daues}, Greg and {Dawson}, William and {Delgado}, Francisco and {Dellapenna}, Alfred and {de Peyster}, Robert and {de Val-Borro}, Miguel and {Digel}, Seth W. and {Doherty}, Peter and {Dubois}, Richard and {Dubois-Felsmann}, Gregory P. and {Durech}, Josef and {Economou}, Frossie and {Eifler}, Tim and {Eracleous}, Michael and {Emmons}, Benjamin L. and {Fausti Neto}, Angelo and {Ferguson}, Henry and {Figueroa}, Enrique and {Fisher-Levine}, Merlin and {Focke}, Warren and {Foss}, Michael D. and {Frank}, James and {Freemon}, Michael D. and {Gangler}, Emmanuel and {Gawiser}, Eric and {Geary}, John C. and {Gee}, Perry and {Geha}, Marla and {Gessner}, Charles J.~B. and {Gibson}, Robert R. and {Gilmore}, D. Kirk and {Glanzman}, Thomas and {Glick}, William and {Goldina}, Tatiana and {Goldstein}, Daniel A. and {Goodenow}, Iain and {Graham}, Melissa L. and {Gressler}, William J. and {Gris}, Philippe and {Guy}, Leanne P. and {Guyonnet}, Augustin and {Haller}, Gunther and {Harris}, Ron and {Hascall}, Patrick A. and {Haupt}, Justine and {Hernandez}, Fabio and {Herrmann}, Sven and {Hileman}, Edward and {Hoblitt}, Joshua and {Hodgson}, John A. and {Hogan}, Craig and {Howard}, James D. and {Huang}, Dajun and {Huffer}, Michael E. and {Ingraham}, Patrick and {Innes}, Walter R. and {Jacoby}, Suzanne H. and {Jain}, Bhuvnesh and {Jammes}, Fabrice and {Jee}, M. James and {Jenness}, Tim and {Jernigan}, Garrett and {Jevremovi{\'c}}, Darko and {Johns}, Kenneth and {Johnson}, Anthony S. and {Johnson}, Margaret W.~G. and {Jones}, R. Lynne and {Juramy-Gilles}, Claire and {Juri{\'c}}, Mario and {Kalirai}, Jason S. and {Kallivayalil}, Nitya J. and {Kalmbach}, Bryce and {Kantor}, Jeffrey P. and {Karst}, Pierre and {Kasliwal}, Mansi M. and {Kelly}, Heather and {Kessler}, Richard and {Kinnison}, Veronica and {Kirkby}, David and {Knox}, Lloyd and {Kotov}, Ivan V. and {Krabbendam}, Victor L. and {Krughoff}, K. Simon and {Kub{\'a}nek}, Petr and {Kuczewski}, John and {Kulkarni}, Shri and {Ku}, John and {Kurita}, Nadine R. and {Lage}, Craig S. and {Lambert}, Ron and {Lange}, Travis and {Langton}, J. Brian and {Le Guillou}, Laurent and {Levine}, Deborah and {Liang}, Ming and {Lim}, Kian-Tat and {Lintott}, Chris J. and {Long}, Kevin E. and {Lopez}, Margaux and {Lotz}, Paul J. and {Lupton}, Robert H. and {Lust}, Nate B. and {MacArthur}, Lauren A. and {Mahabal}, Ashish and {Mandelbaum}, Rachel and {Markiewicz}, Thomas W. and {Marsh}, Darren S. and {Marshall}, Philip J. and {Marshall}, Stuart and {May}, Morgan and {McKercher}, Robert and {McQueen}, Michelle and {Meyers}, Joshua and {Migliore}, Myriam and {Miller}, Michelle and {Mills}, David J. and {Miraval}, Connor and {Moeyens}, Joachim and {Moolekamp}, Fred E. and {Monet}, David G. and {Moniez}, Marc and {Monkewitz}, Serge and {Montgomery}, Christopher and {Morrison}, Christopher B. and {Mueller}, Fritz and {Muller}, Gary P. and {Mu{\~n}oz Arancibia}, Freddy and {Neill}, Douglas R. and {Newbry}, Scott P. and {Nief}, Jean-Yves and {Nomerotski}, Andrei and {Nordby}, Martin and {O'Connor}, Paul and {Oliver}, John and {Olivier}, Scot S. and {Olsen}, Knut and {O'Mullane}, William and {Ortiz}, Sandra and {Osier}, Shawn and {Owen}, Russell E. and {Pain}, Reynald and {Palecek}, Paul E. and {Parejko}, John K. and {Parsons}, James B. and {Pease}, Nathan M. and {Peterson}, J. Matt and {Peterson}, John R. and {Petravick}, Donald L. and {Libby Petrick}, M.~E. and {Petry}, Cathy E. and {Pierfederici}, Francesco and {Pietrowicz}, Stephen and {Pike}, Rob and {Pinto}, Philip A. and {Plante}, Raymond and {Plate}, Stephen and {Plutchak}, Joel P. and {Price}, Paul A. and {Prouza}, Michael and {Radeka}, Veljko and {Rajagopal}, Jayadev and {Rasmussen}, Andrew P. and {Regnault}, Nicolas and {Reil}, Kevin A. and {Reiss}, David J. and {Reuter}, Michael A. and {Ridgway}, Stephen T. and {Riot}, Vincent J. and {Ritz}, Steve and {Robinson}, Sean and {Roby}, William and {Roodman}, Aaron and {Rosing}, Wayne and {Roucelle}, Cecille and {Rumore}, Matthew R. and {Russo}, Stefano and {Saha}, Abhijit and {Sassolas}, Benoit and {Schalk}, Terry L. and {Schellart}, Pim and {Schindler}, Rafe H. and {Schmidt}, Samuel and {Schneider}, Donald P. and {Schneider}, Michael D. and {Schoening}, William and {Schumacher}, German and {Schwamb}, Megan E. and {Sebag}, Jacques and {Selvy}, Brian and {Sembroski}, Glenn H. and {Seppala}, Lynn G. and {Serio}, Andrew and {Serrano}, Eduardo and {Shaw}, Richard A. and {Shipsey}, Ian and {Sick}, Jonathan and {Silvestri}, Nicole and {Slater}, Colin T. and {Smith}, J. Allyn and {Smith}, R. Chris and {Sobhani}, Shahram and {Soldahl}, Christine and {Storrie-Lombardi}, Lisa and {Stover}, Edward and {Strauss}, Michael A. and {Street}, Rachel A. and {Stubbs}, Christopher W. and {Sullivan}, Ian S. and {Sweeney}, Donald and {Swinbank}, John D. and {Szalay}, Alexander and {Takacs}, Peter and {Tether}, Stephen A. and {Thaler}, Jon J. and {Thayer}, John Gregg and {Thomas}, Sandrine and {Thornton}, Adam J. and {Thukral}, Vaikunth and {Tice}, Jeffrey and {Trilling}, David E. and {Turri}, Max and {Van Berg}, Richard and {Vanden Berk}, Daniel and {Vetter}, Kurt and {Virieux}, Francoise and {Vucina}, Tomislav and {Wahl}, William and {Walkowicz}, Lucianne and {Walsh}, Brian and {Walter}, Christopher W. and {Wang}, Daniel L. and {Wang}, Shin-Yawn and {Warner}, Michael and {Wiecha}, Oliver and {Willman}, Beth and {Winters}, Scott E. and {Wittman}, David and {Wolff}, Sidney C. and {Wood-Vasey}, W. Michael and {Wu}, Xiuqin and {Xin}, Bo and {Yoachim}, Peter and {Zhan}, Hu},
        title = "{LSST: From Science Drivers to Reference Design and Anticipated Data Products}",
      journal = {\apj},
         year = 2019,
        month = mar,
       volume = {873},
       number = {2},
          eid = {111},
        pages = {111},
          doi = {10.3847/1538-4357/ab042c},
archivePrefix = {arXiv},
       eprint = {0805.2366},
 primaryClass = {astro-ph},
       adsurl = {https://ui.adsabs.harvard.edu/abs/2019ApJ...873..111I}
}

@ARTICLE{lsst_rates,
       author = {{Bricman}, Katja and {Gomboc}, Andreja},
        title = "{The Prospects of Observing Tidal Disruption Events with the Large Synoptic Survey Telescope}",
      journal = {\apj},
         year = 2020,
        month = feb,
       volume = {890},
       number = {1},
          eid = {73},
        pages = {73},
          doi = {10.3847/1538-4357/ab6989},
archivePrefix = {arXiv},
       eprint = {1906.08235},
 primaryClass = {astro-ph.HE},
       adsurl = {https://ui.adsabs.harvard.edu/abs/2020ApJ...890...73B}
}

@ARTICLE{wlsst_rates,
       author = {{Lin}, Zheyu and {Jiang}, Ning and {Kong}, Xu},
        title = "{The prospects of finding tidal disruption events with 2.5-m Wide-Field Survey Telescope based on mock observations}",
      journal = {\mnras},
         year = 2022,
        month = jun,
       volume = {513},
       number = {2},
        pages = {2422-2436},
          doi = {10.1093/mnras/stac946},
archivePrefix = {arXiv},
       eprint = {2204.01615},
 primaryClass = {astro-ph.HE},
       adsurl = {https://ui.adsabs.harvard.edu/abs/2022MNRAS.513.2422L}
}

@ARTICLE{elip_lens_1,
       author = {{Turner}, E.~L. and {Ostriker}, J.~P. and {Gott}, III, J.~R.},
        title = "{The statistics of gravitational lenses : the distributions of image angular separations and lens redshifts.}",
      journal = {\apj},
         year = 1984,
        month = sep,
       volume = {284},
        pages = {1-22},
          doi = {10.1086/162379},
       adsurl = {https://ui.adsabs.harvard.edu/abs/1984ApJ...284....1T}
}

@ARTICLE{elip_lens_2,
       author = {{Fukugita}, M. and {Futamase}, T. and {Kasai}, M. and {Turner}, E.~L.},
        title = "{Statistical Properties of Gravitational Lenses with a Nonzero Cosmological Constant}",
      journal = {\apj},
         year = 1992,
        month = jul,
       volume = {393},
        pages = {3},
          doi = {10.1086/171481},
       adsurl = {https://ui.adsabs.harvard.edu/abs/1992ApJ...393....3F}
}

@ARTICLE{elip_lens_3,
       author = {{Kochanek}, Christopher S.},
        title = "{Is There a Cosmological Constant?}",
      journal = {\apj},
         year = 1996,
        month = aug,
       volume = {466},
        pages = {638},
          doi = {10.1086/177538},
archivePrefix = {arXiv},
       eprint = {astro-ph/9510077},
 primaryClass = {astro-ph},
       adsurl = {https://ui.adsabs.harvard.edu/abs/1996ApJ...466..638K}
}

@ARTICLE{elip_lens_4,
       author = {{Chae}, Kyu-Hyun},
        title = "{The Cosmic Lens All-Sky Survey: statistical strong lensing, cosmological parameters, and global properties of galaxy populations}",
      journal = {\mnras},
         year = 2003,
        month = dec,
       volume = {346},
       number = {3},
        pages = {746-772},
          doi = {10.1111/j.1365-2966.2003.07092.x},
archivePrefix = {arXiv},
       eprint = {astro-ph/0211244},
 primaryClass = {astro-ph},
       adsurl = {https://ui.adsabs.harvard.edu/abs/2003MNRAS.346..746C}
}

@ARTICLE{elip_lens_5,
       author = {{Oguri}, Masamune},
        title = "{The image separation distribution of strong lenses: halo versus subhalo populations}",
      journal = {\mnras},
         year = 2006,
        month = apr,
       volume = {367},
       number = {3},
        pages = {1241-1250},
          doi = {10.1111/j.1365-2966.2006.10043.x},
archivePrefix = {arXiv},
       eprint = {astro-ph/0508528},
 primaryClass = {astro-ph},
       adsurl = {https://ui.adsabs.harvard.edu/abs/2006MNRAS.367.1241O}
}

@ARTICLE{elip_lens_6,
       author = {{M{\"o}ller}, Ole and {Kitzbichler}, Manfred and {Natarajan}, Priyamvada},
        title = "{Strong lensing statistics in large, z <\raisebox{-0.5ex}\textasciitilde 0.2, surveys: bias in the lens galaxy population}",
      journal = {\mnras},
         year = 2007,
        month = aug,
       volume = {379},
       number = {3},
        pages = {1195-1208},
          doi = {10.1111/j.1365-2966.2007.12004.x},
archivePrefix = {arXiv},
       eprint = {astro-ph/0607032},
 primaryClass = {astro-ph},
       adsurl = {https://ui.adsabs.harvard.edu/abs/2007MNRAS.379.1195M}
}

@ARTICLE{isotropy_1,
       author = {{Bengaly}, Carlos A.~P. and {Alcaniz}, Jailson S. and {Pigozzo}, C{\'a}ssio},
        title = "{Testing the isotropy of cosmic acceleration with the Pantheon + and SH0ES datasets: A cosmographic analysis}",
      journal = {\prd},
         year = 2024,
        month = jun,
       volume = {109},
       number = {12},
          eid = {123533},
        pages = {123533},
          doi = {10.1103/PhysRevD.109.123533},
archivePrefix = {arXiv},
       eprint = {2402.17741},
 primaryClass = {astro-ph.CO},
       adsurl = {https://ui.adsabs.harvard.edu/abs/2024PhRvD.109l3533B}
}

@ARTICLE{isotropy_2,
       author = {{Clarkson}, Chris and {Maartens}, Roy},
        title = "{Inhomogeneity and the foundations of concordance cosmology}",
      journal = {Classical and Quantum Gravity},
         year = 2010,
        month = jun,
       volume = {27},
       number = {12},
          eid = {124008},
        pages = {124008},
          doi = {10.1088/0264-9381/27/12/124008},
archivePrefix = {arXiv},
       eprint = {1005.2165},
 primaryClass = {astro-ph.CO},
       adsurl = {https://ui.adsabs.harvard.edu/abs/2010CQGra..27l4008C}
}

@ARTICLE{isotropy_3,
       author = {{Goodman}, Jeremy},
        title = "{Geocentrism reexamined}",
      journal = {\prd},
         year = 1995,
        month = aug,
       volume = {52},
       number = {4},
        pages = {1821-1827},
          doi = {10.1103/PhysRevD.52.1821},
archivePrefix = {arXiv},
       eprint = {astro-ph/9506068},
 primaryClass = {astro-ph},
       adsurl = {https://ui.adsabs.harvard.edu/abs/1995PhRvD..52.1821G}
}

@ARTICLE{isotropy_4,
       author = {{Kumar Aluri}, Pavan and {Cea}, Paolo and {Chingangbam}, Pravabati and {Chu}, Ming-Chung and {Clowes}, Roger G. and {Hutsem{\'e}kers}, Damien and {Kochappan}, Joby P. and {Lopez}, Alexia M. and {Liu}, Lang and {Martens}, Niels C.~M. and {Martins}, C.~J.~A.~P. and {Migkas}, Konstantinos and {{\'O} Colg{\'a}in}, Eoin and {Pranav}, Pratyush and {Shamir}, Lior and {Singal}, Ashok K. and {Sheikh-Jabbari}, M.~M. and {Wagner}, Jenny and {Wang}, Shao-Jiang and {Wiltshire}, David L. and {Yeung}, Shek and {Yin}, Lu and {Zhao}, Wen},
        title = "{Is the observable Universe consistent with the cosmological principle?}",
      journal = {Classical and Quantum Gravity},
         year = 2023,
        month = may,
       volume = {40},
       number = {9},
          eid = {094001},
        pages = {094001},
          doi = {10.1088/1361-6382/acbefc},
       adsurl = {https://ui.adsabs.harvard.edu/abs/2023CQGra..40i4001K}
}

@ARTICLE{mosfit_1,
       author = {{Guillochon}, James and {Nicholl}, Matt and {Villar}, V. Ashley and {Mockler}, Brenna and {Narayan}, Gautham and {Mandel}, Kaisey S. and {Berger}, Edo and {Williams}, Peter K.~G.},
        title = "{MOSFiT: Modular Open Source Fitter for Transients}",
      journal = {\apjs},
         year = 2018,
        month = may,
       volume = {236},
       number = {1},
          eid = {6},
        pages = {6},
          doi = {10.3847/1538-4365/aab761},
archivePrefix = {arXiv},
       eprint = {1710.02145},
 primaryClass = {astro-ph.IM},
       adsurl = {https://ui.adsabs.harvard.edu/abs/2018ApJS..236....6G}
}

@ARTICLE{mosfit_2,
       author = {{Mockler}, Brenna and {Guillochon}, James and {Ramirez-Ruiz}, Enrico},
        title = "{Weighing Black Holes Using Tidal Disruption Events}",
      journal = {\apj},
         year = 2019,
        month = feb,
       volume = {872},
       number = {2},
          eid = {151},
        pages = {151},
          doi = {10.3847/1538-4357/ab010f},
archivePrefix = {arXiv},
       eprint = {1801.08221},
 primaryClass = {astro-ph.HE},
       adsurl = {https://ui.adsabs.harvard.edu/abs/2019ApJ...872..151M}
}

@ARTICLE{dq_ratio_z_dep1,
       author = {{Venumadhav}, Tejaswi and {Dai}, Liang and {Miralda-Escud{\'e}}, Jordi},
        title = "{Microlensing of Extremely Magnified Stars near Caustics of Galaxy Clusters}",
      journal = {\apj},
         year = 2017,
        month = nov,
       volume = {850},
       number = {1},
          eid = {49},
        pages = {49},
          doi = {10.3847/1538-4357/aa9575},
archivePrefix = {arXiv},
       eprint = {1707.00003},
 primaryClass = {astro-ph.CO},
       adsurl = {https://ui.adsabs.harvard.edu/abs/2017ApJ...850...49V}
}

@ARTICLE{dq_ratio_z_dep2,
       author = {{Meena}, Ashish Kumar and {Arad}, Ofir and {Zitrin}, Adi},
        title = "{An efficient method for simulating light curves of cosmological microlensing and caustic crossing events}",
      journal = {\mnras},
         year = 2022,
        month = aug,
       volume = {514},
       number = {2},
        pages = {2545-2560},
          doi = {10.1093/mnras/stac1511},
archivePrefix = {arXiv},
       eprint = {2203.08131},
 primaryClass = {astro-ph.IM},
       adsurl = {https://ui.adsabs.harvard.edu/abs/2022MNRAS.514.2545M}
}

@BOOK{gravitational_lensing_book,
  title = "{Gravitational Lensing: Strong, Weak and Micro}",
  author = {{Schneider}, P. and {Kochanek}, C. and {Wambsganss}, J.},
  isbn={978-3-540-30309-1},
  series={Saas-Fee Advanced Course},
  year = {2006},
  PUBLISHER = {Springer Berlin, Heidelberg},
  doi={10.1007/978-3-540-30310-7}
}

@ARTICLE{shapiro_delay,
       author = {{Shapiro}, Irwin I.},
        title = "{Fourth Test of General Relativity}",
      journal = {\prl},
         year = 1964,
        month = dec,
       volume = {13},
       number = {26},
        pages = {789-791},
          doi = {10.1103/PhysRevLett.13.789},
       adsurl = {https://ui.adsabs.harvard.edu/abs/1964PhRvL..13..789S}
}

@ARTICLE{chang_2025_tde_occurrence_rate,
       author = {{Chang}, Janet N.~Y. and {Dai}, Lixin and {Pfister}, Hugo and {Kar Chowdhury}, Rudrani and {Natarajan}, Priyamvada},
        title = "{Rates of Stellar Tidal Disruption Events around Intermediate-mass Black Holes}",
      journal = {\apjl},
         year = 2025,
        month = feb,
       volume = {980},
       number = {2},
          eid = {L22},
        pages = {L22},
          doi = {10.3847/2041-8213/adace7},
archivePrefix = {arXiv},
       eprint = {2407.09339},
 primaryClass = {astro-ph.HE},
       adsurl = {https://ui.adsabs.harvard.edu/abs/2025ApJ...980L..22C}
}

@ARTICLE{grav_waves_from_TDEs,
       author = {{Pfister}, Hugo and {Toscani}, Martina and {Wong}, Thomas Hong Tsun and {Dai}, Jane Lixin and {Lodato}, Giuseppe and {Rossi}, Elena M.},
        title = "{Observable gravitational waves from tidal disruption events and their electromagnetic counterpart}",
      journal = {\mnras},
         year = 2022,
        month = feb,
       volume = {510},
       number = {2},
        pages = {2025-2040},
          doi = {10.1093/mnras/stab3387},
archivePrefix = {arXiv},
       eprint = {2103.05883},
 primaryClass = {astro-ph.HE},
       adsurl = {https://ui.adsabs.harvard.edu/abs/2022MNRAS.510.2025P}
}

@ARTICLE{lensed_grav_waves,
       author = {{Toscani}, Martina and {Rossi}, Elena M. and {Tamanini}, Nicola and {Cusin}, Giulia},
        title = "{Lensing of gravitational waves from tidal disruption events}",
      journal = {\mnras},
         year = 2023,
        month = aug,
       volume = {523},
       number = {3},
        pages = {3863-3873},
          doi = {10.1093/mnras/stad1633},
archivePrefix = {arXiv},
       eprint = {2301.01804},
 primaryClass = {astro-ph.HE},
       adsurl = {https://ui.adsabs.harvard.edu/abs/2023MNRAS.523.3863T}
}

@BOOK{grav_lenses_book,
       author = {{Schneider}, Peter and {Ehlers}, J{\"u}rgen and {Falco}, Emilio E.},
        title = "{Gravitational Lenses}",
         year = 1992,
          doi = {10.1007/978-3-662-03758-4},
       adsurl = {https://ui.adsabs.harvard.edu/abs/1992grle.book.....S},
    publisher = {Springer Berlin, Heidelberg}
}

@BOOK{gr_grav_lensing_book,
       author = {{Cheng}, Ta-Pei},
        title = "{Relativity, gravitation and cosmology. A basic introduction}",
         year = 2005,
       adsurl = {https://ui.adsabs.harvard.edu/abs/2005rgc..book.....C},
    publisher = {Oxford University Press}
}

@BOOK{grav_lensing_python,
       author = {{Meneghetti}, Massimo},
        title = "{Introduction to Gravitational Lensing: With Python Examples}",
         year = 2021,
       adsurl = {https://ui.adsabs.harvard.edu/abs/2022iglp.book.....M},
    publisher = {Springer Cham}
}

@ARTICLE{multimessenger_tdes,
       author = {{Wevers}, Thomas and {Ryu}, Taeho},
        title = "{Multi-messenger astronomy with black holes: tidal disruption events}",
      journal = {arXiv e-prints},
         year = 2023,
        month = oct,
          eid = {arXiv:2310.16879},
        pages = {arXiv:2310.16879},
          doi = {10.48550/arXiv.2310.16879},
archivePrefix = {arXiv},
       eprint = {2310.16879},
 primaryClass = {astro-ph.HE},
       adsurl = {https://ui.adsabs.harvard.edu/abs/2023arXiv231016879W}
}

@ARTICLE{tde_neutrino_candidates,
       author = {{Reusch}, Simeon and {Stein}, Robert and {Kowalski}, Marek and {van Velzen}, Sjoert and {Franckowiak}, Anna and {Lunardini}, Cecilia and {Murase}, Kohta and {Winter}, Walter and {Miller-Jones}, James C.~A. and {Kasliwal}, Mansi M. and {Gilfanov}, Marat and {Garrappa}, Simone and {Paliya}, Vaidehi S. and {Ahumada}, Tom{\'a}s and {Anand}, Shreya and {Barbarino}, Cristina and {Bellm}, Eric C. and {Brinnel}, Val{\'e}ry and {Buson}, Sara and {Cenko}, S. Bradley and {Coughlin}, Michael W. and {De}, Kishalay and {Dekany}, Richard and {Frederick}, Sara and {Gal-Yam}, Avishay and {Gezari}, Suvi and {Giroletti}, Marcello and {Graham}, Matthew J. and {Karambelkar}, Viraj and {Kimura}, Shigeo S. and {Kong}, Albert K.~H. and {Kool}, Erik C. and {Laher}, Russ R. and {Medvedev}, Pavel and {Necker}, Jannis and {Nordin}, Jakob and {Perley}, Daniel A. and {Rigault}, Mickael and {Rusholme}, Ben and {Schulze}, Steve and {Schweyer}, Tassilo and {Singer}, Leo P. and {Sollerman}, Jesper and {Strotjohann}, Nora Linn and {Sunyaev}, Rashid and {van Santen}, Jakob and {Walters}, Richard and {Zhang}, B. Theodore and {Zimmerman}, Erez},
        title = "{Candidate Tidal Disruption Event AT2019fdr Coincident with a High-Energy Neutrino}",
      journal = {\prl},
         year = 2022,
        month = jun,
       volume = {128},
       number = {22},
          eid = {221101},
        pages = {221101},
          doi = {10.1103/PhysRevLett.128.221101},
archivePrefix = {arXiv},
       eprint = {2111.09390},
 primaryClass = {astro-ph.HE},
       adsurl = {https://ui.adsabs.harvard.edu/abs/2022PhRvL.128v1101R}
}

@ARTICLE{detecability_strongly_lensed_tdes,
       author = {{Chen}, Zhiwei and {Lu}, Youjun and {Chen}, Yunfeng},
        title = "{Detectability of Strongly Gravitationally Lensed Tidal Disruption Events}",
      journal = {\apj},
         year = 2024,
        month = feb,
       volume = {962},
       number = {1},
          eid = {3},
        pages = {3},
          doi = {10.3847/1538-4357/ad19d3},
archivePrefix = {arXiv},
       eprint = {2401.00992},
 primaryClass = {astro-ph.HE},
       adsurl = {https://ui.adsabs.harvard.edu/abs/2024ApJ...962....3C}
}

@ARTICLE{gw_background_from_tdes,
       author = {{Toscani}, Martina and {Rossi}, Elena M. and {Lodato}, Giuseppe},
        title = "{The gravitational wave background signal from tidal disruption events}",
      journal = {\mnras},
         year = 2020,
        month = oct,
       volume = {498},
       number = {1},
        pages = {507-516},
          doi = {10.1093/mnras/staa2290},
archivePrefix = {arXiv},
       eprint = {2007.13225},
 primaryClass = {astro-ph.HE},
       adsurl = {https://ui.adsabs.harvard.edu/abs/2020MNRAS.498..507T}
}

@ARTICLE{agn_microlensing_1,
       author = {{Yonehara}, A. and {Mineshige}, S. and {Fukue}, J. and {Umemura}, M. and {Turner}, E.~L.},
        title = "{Microlens diagnostics of accretion disks in active galactic nuclei}",
      journal = {\aap},
         year = 1999,
        month = mar,
       volume = {343},
        pages = {41-50},
       adsurl = {https://ui.adsabs.harvard.edu/abs/1999A&A...343...41Y}
}

@ARTICLE{agn_microlensing_2,
       author = {{Rauch}, Kevin P. and {Blandford}, Roger D.},
        title = "{Microlensing and the Structure of Active Galactic Nucleus Accretion Disks}",
      journal = {\apjl},
         year = 1991,
        month = nov,
       volume = {381},
        pages = {L39},
          doi = {10.1086/186191},
       adsurl = {https://ui.adsabs.harvard.edu/abs/1991ApJ...381L..39R}
}

@ARTICLE{identify_tdes,
       author = {{Gomez}, Sebastian and {Villar}, V. Ashley and {Berger}, Edo and {Gezari}, Suvi and {van Velzen}, Sjoert and {Nicholl}, Matt and {Blanchard}, Peter K. and {Alexander}, Kate. D.},
        title = "{Identifying Tidal Disruption Events with an Expansion of the FLEET Machine-learning Algorithm}",
      journal = {\apj},
         year = 2023,
        month = jun,
       volume = {949},
       number = {2},
          eid = {113},
        pages = {113},
          doi = {10.3847/1538-4357/acc535},
archivePrefix = {arXiv},
       eprint = {2210.10810},
 primaryClass = {astro-ph.HE},
       adsurl = {https://ui.adsabs.harvard.edu/abs/2023ApJ...949..113G}
}

@ARTICLE{first_tde_detection,
       author = {{Komossa}, Stefanie and {Halpern}, Jules and {Schartel}, Norbert and {Hasinger}, G{\"u}nther and {Santos-Lleo}, Maria and {Predehl}, Peter},
        title = "{A Huge Drop in the X-Ray Luminosity of the Nonactive Galaxy RX J1242.6-1119A, and the First Postflare Spectrum: Testing the Tidal Disruption Scenario}",
      journal = {\apjl},
         year = 2004,
        month = mar,
       volume = {603},
       number = {1},
        pages = {L17-L20},
          doi = {10.1086/382046},
archivePrefix = {arXiv},
       eprint = {astro-ph/0402468},
 primaryClass = {astro-ph},
       adsurl = {https://ui.adsabs.harvard.edu/abs/2004ApJ...603L..17K}
}

@ARTICLE{more_tde_detection,
       author = {{Gezari}, Suvi and {Heckman}, Tim and {Cenko}, S. Bradley and {Eracleous}, Michael and {Forster}, Karl and {Gon{\c{c}}alves}, Thiago S. and {Martin}, D. Chris and {Morrissey}, Patrick and {Neff}, Susan G. and {Seibert}, Mark and {Schiminovich}, David and {Wyder}, Ted K.},
        title = "{Luminous Thermal Flares from Quiescent Supermassive Black Holes}",
      journal = {\apj},
         year = 2009,
        month = jun,
       volume = {698},
       number = {2},
        pages = {1367-1379},
          doi = {10.1088/0004-637X/698/2/1367},
archivePrefix = {arXiv},
       eprint = {0904.1596},
 primaryClass = {astro-ph.CO},
       adsurl = {https://ui.adsabs.harvard.edu/abs/2009ApJ...698.1367G}
}

\clearpage

\begin{appendix}
\section{Magnitude integral}\label{sec:mag_int}
    To calculate the flux integral in the numerator of Eq.~\ref{eq:magnitudes}, we first introduced the variables $a$ and $b$ to simplify calculations. The flux then takes the form 
    \begin{equation*}
        F_{\lambda_{\mathrm{obs}}, \mathrm{o}} = \frac{a}{\lambda_{\mathrm{obs}}^{5}} \cdot \frac{1}{e^{\frac{b}{\lambda_{\mathrm{obs}}}} -1}.
    \end{equation*}
    We obtained this expression by plugging Eq.~\ref{eq:planck_law} into Eq.~\ref{eq:observed_flux}. The integral could then be computed as follows:
    \begin{align*}
        \int_{0}^{\infty} \mathrm{d}\lambda \; \lambda \cdot S_{\mathrm{X}} \cdot F_{\lambda, \mathrm{o}} &= \int_{\lambda_{\mathrm{min}}}^{\lambda_{\mathrm{max}}} \mathrm{d}\lambda \; \lambda \cdot \frac{a}{\lambda^{5}} \cdot \frac{1}{e^{\frac{b}{\lambda}} -1} = \\[5pt]
        &= a \int_{\lambda_{\mathrm{min}}}^{\lambda_{\mathrm{max}}} \mathrm{d}\lambda \; \frac{1}{\lambda^{4}} \cdot \frac{1}{e^{\frac{b}{\lambda} -1}} = \\[5pt]
        \left[ \mathrm{Subs.} \; \frac{b}{\lambda} = x \right] \;\;\; &= a \int_{\frac{b}{\lambda_{\mathrm{min}}}}^{\frac{b}{\lambda_{\mathrm{max}}}} \mathrm{d}x \; \left( -\frac{b}{x^{2}} \right) \cdot \left( \frac{b}{x} \right)^{-4} \cdot \frac{1}{e^{x} -1} = \\[5pt]
        &= -\frac{a}{b^{3}} \int_{\frac{b}{\lambda_{\mathrm{min}}}}^{\frac{b}{\lambda_{\mathrm{max}}}} \mathrm{d}x \; \frac{x^{2}}{e^{x} -1}.
    \end{align*}
    To solve this integral, we first solved
    \begin{equation*}
        \int \mathrm{d}x \; \frac{1}{e^{x} -1}
    \end{equation*}
    by substituting $u = 1- e^{-x}$, from which follows
    \begin{equation*}
        \int \frac{\mathrm{d}u}{1 - u} \cdot \frac{1}{\frac{1}{1 - u} -1} = \int \frac{\mathrm{d}u}{1 - u} \cdot \frac{1 - u}{u} = \int \mathrm{d}u \; \frac{1}{u} = \ln(u).
    \end{equation*}
    By re-substituting, we found
    \begin{equation*}
        \int \mathrm{d}x \; \frac{1}{e^{x} -1} = \ln (1 - e^{-x}).
    \end{equation*}
    We could then solve the flux integral by applying partial integration ($\mathrm{PI}$) twice:
    \begin{align*}
        \int \mathrm{d}x \; \frac{x^{2}}{e^{x} -1} &\stackrel{\mathrm{PI}}= \left[ x^{2} \cdot \ln  (1 - e^{-x}) \right] - \int \mathrm{d}x \; 2x \ln (1 - e^{-x}) = \\[5pt]
        &= -x^{2} \cdot \mathrm{Li}_{1}(e^{-x}) + 2\int \mathrm{d}x \; x \mathrm{Li}_{1}(e^{-x}) = \\[5pt]
        \left[ \mathrm{Subs.} \; u = e^{-x} \right] \;\;\; &= -x^{2} \cdot \mathrm{Li}_{1}(e^{-x}) + 2\int \mathrm{d}u \; \frac{\ln(u)}{u} \mathrm{Li}_{1}(u) = \\[5pt]
        &\stackrel{\mathrm{PI}}= -x^{2} \cdot \mathrm{Li}_{1}(e^{-x}) + 2\left[ \ln(u) \cdot \mathrm{Li}_{2}(u) \right] \\
        & \quad - 2\int \mathrm{d}u \; \frac{1}{u} \mathrm{Li}_{2}(u) = \\[5pt]
        &= -x^{2} \cdot \mathrm{Li}_{1}(e^{-x}) + 2 \ln(u) \cdot \mathrm{Li}_{2}(u) \\[3pt]
        & \quad - 2\left[ \mathrm{Li}_{3}(u) \right] = \\[5pt]
        \left[ \mathrm{Re-subs.} \right] \;\;\; &= -x^{2} \cdot \mathrm{Li}_{1}(e^{-x}) - 2 x \cdot \mathrm{Li}_{2}(e^{-x}) - 2 \mathrm{Li}_{3}(e^{-x}),
    \end{align*}
    where $\mathrm{Li}_{n}(x) = \int_{0}^{x} \mathrm{d}t \; \mathrm{Li}_{n-1}(t) \, /t$ is the polylogarithm, with $\mathrm{Li}_{1}(x) = -\ln(1-x)$. The final result could then be expressed as
    \begin{align*}
        &\int_{0}^{\infty} \mathrm{d}\lambda \; \lambda \cdot S_{\mathrm{X}} \cdot F_{\lambda, \mathrm{o}} = \\[-3pt]
        &= \frac{a}{b^{3}} \left[ \left( \frac{b}{\lambda} \right)^{2} \cdot \mathrm{Li}_{1} \left( e^{-\frac{b}{\lambda}} \right) + 2 \frac{b}{\lambda} \cdot \mathrm{Li}_{2} \left( e^{-\frac{b}{\lambda}} \right) + 2 \mathrm{Li}_{3} \left( e^{-\frac{b}{\lambda}} \right) \right]^{\frac{b}{\lambda_{\mathrm{max}}}}_{\frac{b}{\lambda_{\mathrm{min}}}}.
    \end{align*}

\section{Impact of the TDE occurrence rate}\label{sec:Chang_TDE_occurence}
    \begin{figure}[htbp]
        \centering
        \includegraphics[width=\linewidth]{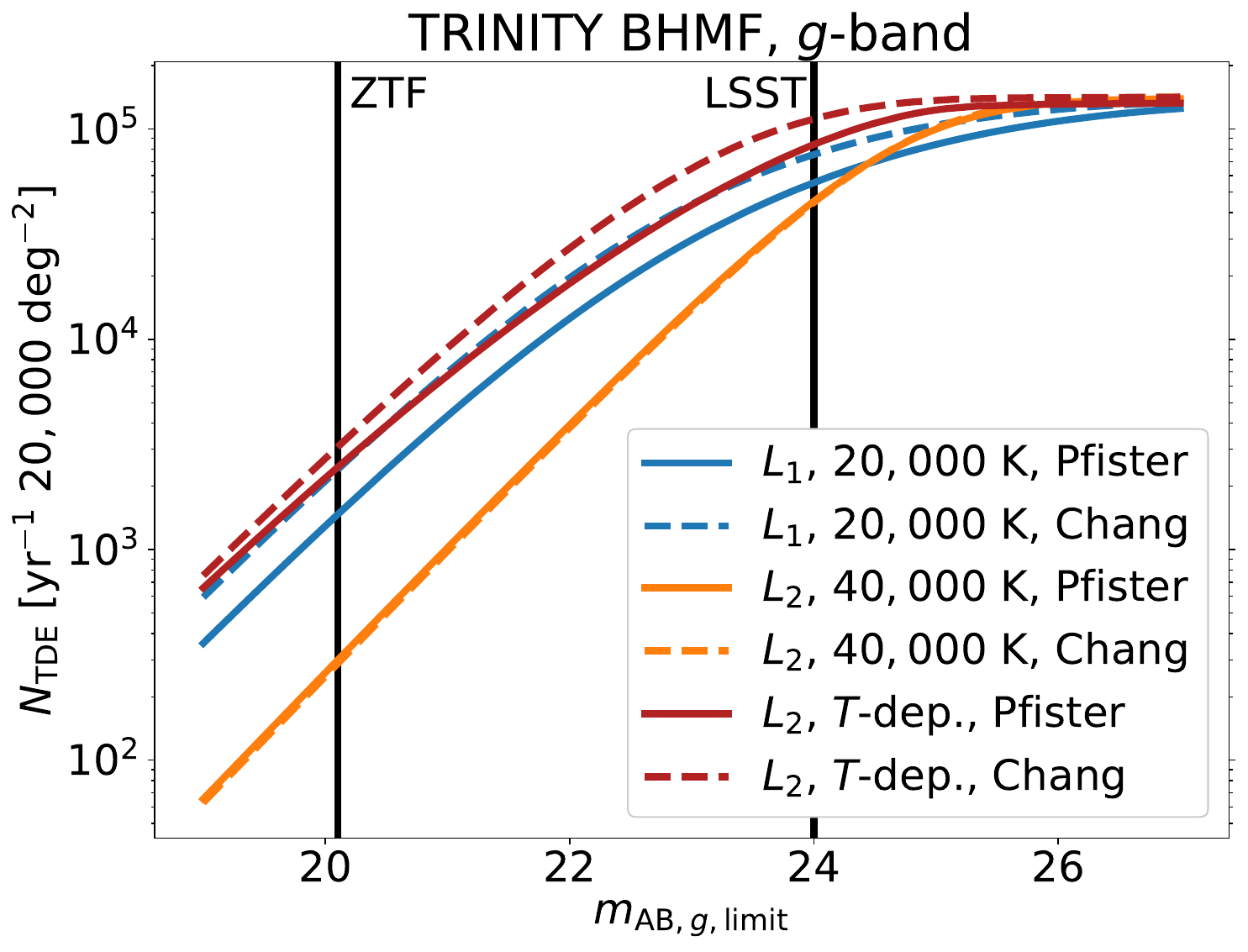}
        \caption{Impact of a different \ac{tde} occurrence rate per galaxy on the cumulative, unlensed \ac{tde} rate. The \ac{tde} occurrence rate mainly acts as a constant offset with an overall weak influence on the \ac{tde} rate.}
        \label{fig:impact_tde_occurence_rate}
    \end{figure}
    \noindent
    In order to examine the impact of the \ac{tde} occurrence rate, we considered a different \ac{tde} rate per galaxy \citep{chang_2025_tde_occurrence_rate}
    \begin{equation}
        \Gamma = \begin{cases}
            		1.2 \cdot 10^{-4} \left( \frac{M_{\mathrm{BH}}}{10^{6} \, \mathrm{M}_{\odot}} \right)^{1.2} \; \mathrm{yr}^{-1} & \text{for } M_{\mathrm{BH}} \leq 10^{6} \; \mathrm{M}_{\odot} \\
                    1.2 \cdot 10^{-4} \left( \frac{M_{\mathrm{BH}}}{10^{6} \, \mathrm{M}_{\odot}} \right)^{-1.2} \; \mathrm{yr}^{-1} & \text{for } M_{\mathrm{BH}} > 10^{6} \; \mathrm{M}_{\odot}
            	   \end{cases}.
        \label{eq:chang_tde_occurrence_rate}
    \end{equation}
    Compared to the uniform power law from \citet{tde_rate_per_year_Pfister_2020}, this function peaks at $M_{\mathrm{BH}}\simeq10^{6} \; \mathrm{M}_{\odot}$ and covers a larger \ac{tde} occurrence rate range, due to the larger exponents of $\pm 1.2$. It peaks above the one from \citet{tde_rate_per_year_Pfister_2020} but falls off more steeply toward smaller or larger \ac{bh} masses. In Fig.~\ref{fig:impact_tde_occurence_rate} we show the impact of the \ac{tde} occurrence rate on the cumulative, unlensed \ac{tde} rate. We have fixed the \ac{bhmf} to the TRINITY model as a representative case. Overall, the impact of the \ac{tde} occurrence rate is small, especially when compared to the fixed temperature case. The impact of the \ac{tde} occurrence rate is mainly driven through the alteration of the \ac{bh} mass distribution.
    \begin{figure}
        \centering
        \includegraphics[width=\linewidth]{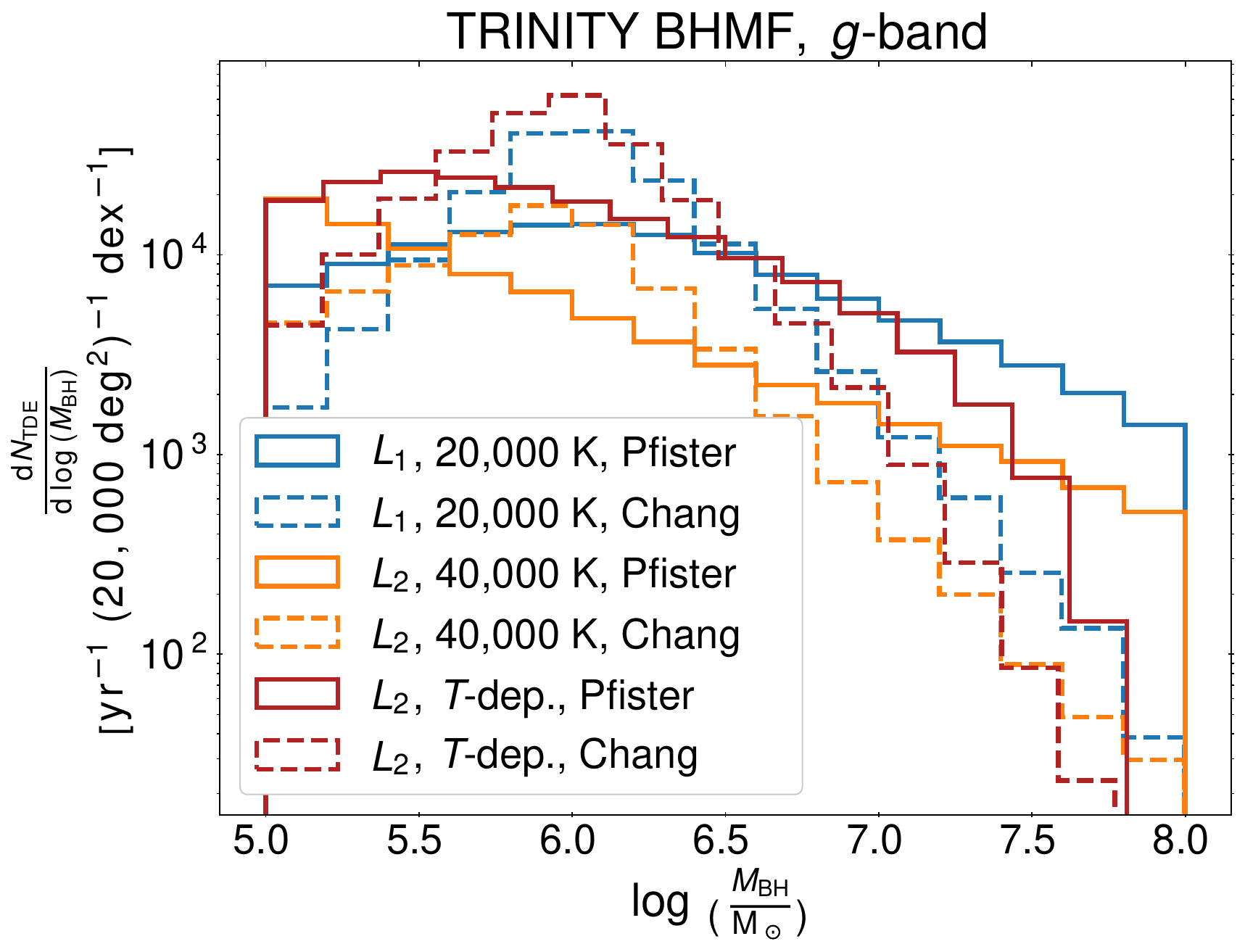}
        \caption{\ac{bh} mass distribution for different \ac{tde} occurrence rates. These distributions are calculated for \ac{lsst} magnitude limits.}
        \label{fig:bh_mass_dist_tde_occurence_rate}
    \end{figure}
    Fig.~\ref{fig:bh_mass_dist_tde_occurence_rate} shows that the \ac{bh} mass distribution reflects the \ac{tde} occurrence rate. The \ac{bh} mass distribution with the occurrence rate by \cite{chang_2025_tde_occurrence_rate} shows a peak around $10^{6} \; \mathrm{M}_{\odot}$ and drops off more steeply toward lower or higher \ac{bh} masses. A change in \ac{bh} mass distribution also means a difference in \ac{tde} luminosity distribution. This new \ac{bh} mass distribution has more \acp{tde} at $\sim 10^{6} \; \mathrm{M}_{\odot}$ where the $L_{1}$ luminosity peaks. Hence, for the $L_{1}$ model, more of these luminous \acp{tde} are detected with a resulting increase in the number of observed \acp{tde} shown in Fig.~\ref{fig:impact_tde_occurence_rate}.
    In contrast, a shift in the \ac{bh} mass only changes the luminosity of the fixed temperature $L_{2}$ model marginally. Hence, we do not observe a different cumulative \ac{tde} rate. In the temperature-dependent case, the luminosity decreases only slightly between $10^{5} \; \mathrm{M}_{\odot}$ and $10^{6} \; \mathrm{M}_{\odot}$, but falls off steeply toward higher \ac{bh} masses. A lowering of the number of \acp{bh} at these higher \ac{bh} masses ($\sim 10^{7}$ to $10^{8} \; \mathrm{M}_{\odot}$) would therefore only mildly decrease the number of \acp{tde}, whereas the higher number of \acp{bh} at $\sim10^{6} \; \mathrm{M}_{\odot}$ would increase the number of \acp{tde} more significantly. Hence, the shift in \ac{bh} mass distribution leads to an overall increase in the cumulative \ac{tde} rate, even if the most luminous \acp{tde} are less abundant.

\section{Gravitational lensing}\label{sec:grav_lensing}
    \setcounter{figure}{0}
    \begin{figure}[htbp]
        \centering
        \resizebox{\hsize}{!}{\includegraphics{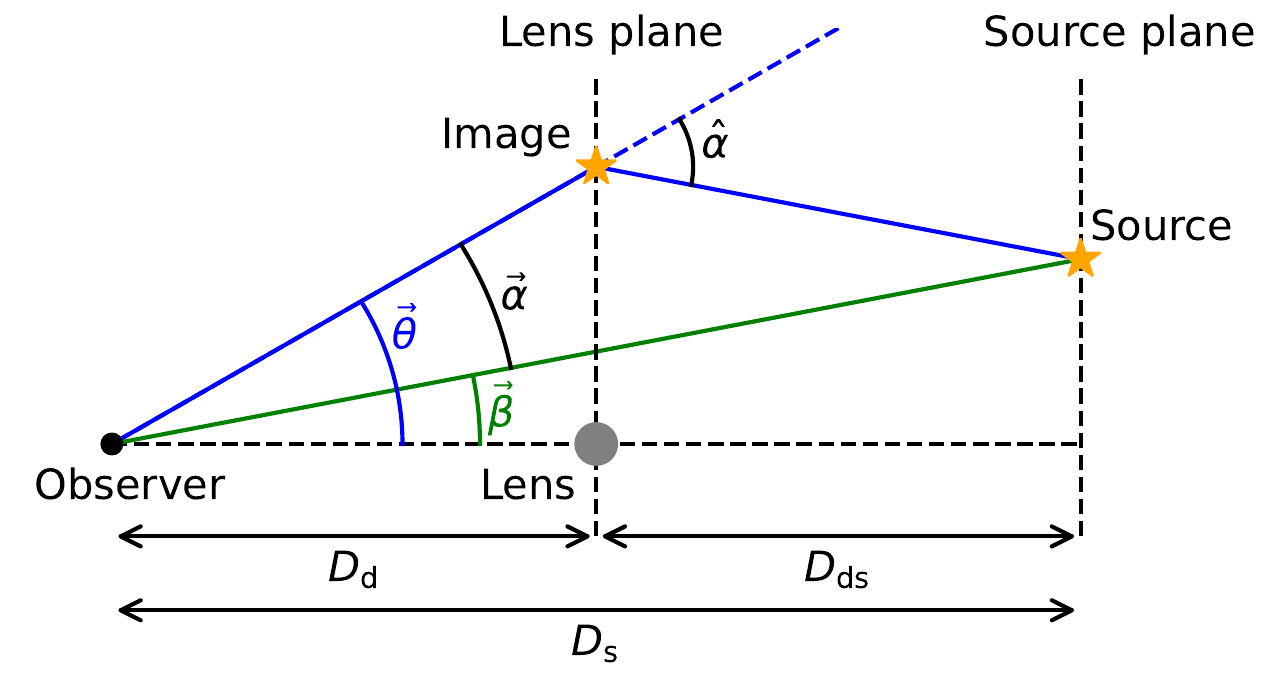}}
        \caption{Schematic view of a gravitational lens. The light travels along the blue path, meaning we observe the image at a different angular position $\vec \theta$ compared to the true angular source position $\vec \beta$ indicated with the green line.}
        \label{fig:grav_lens_setup}
    \end{figure}
    For our calculations involving gravitational lensing, we use the code from \citet{lensing_code_Oguri_2010}. They follow a standard approach to gravitational lensing using the thin-lens approximation \citep[e.g.][]{grav_lenses_book,gravitational_lensing_book,gr_grav_lensing_book,grav_lensing_python}. As the distances between the observer, lens, and source are much greater than the extent of the lens and the source, both the lens and the source are projected onto 2D planes, the lens and source plane, respectively. We show this basic setup in Fig.~\ref{fig:grav_lens_setup}. We can rescale the coordinates in the respective planes to angles measured by the observer using the distances $D_{X}$. We denote the distances between the observer and the lens, the observer and the source, and the lens and the source with $D_{\mathrm{d}}$, $D_{\mathrm{s}}$, and $D_{\mathrm{ds}}$, respectively. As these distances are large, the small-angle approximation holds, meaning the angles can be treated as two-dimensional Euclidean coordinates. The governing equation, also known as the lens equation
    \begin{equation}
        \vec \beta = \vec \theta - \hat \alpha \cdot \frac{D_{\mathrm{ds}}}{D_{\mathrm{s}}} = \vec \theta - \vec \alpha,
        \label{eq:lens_equation}
    \end{equation}
    can then be geometrically derived. Here, $\vec \beta$ is the true source position on the source plane, $\vec \theta$ is the observed position of the lensed source on the lens plane, and $\hat \alpha$ is the deflection angle of the light at the lens plane. By absorbing the distance fraction, the deflection angle is rescaled to the scaled deflection angle 
    \begin{equation}
        \vec \alpha = \hat \alpha \cdot \frac{D_{\mathrm{ds}}}{D_{\mathrm{s}}} = \frac{1}{\pi} \int_{\mathbb{R}^{2}} \mathrm{d}^{2}\theta' \; \kappa(\vec \theta') \frac{\vec \theta - \vec \theta'}{\mid \vec \theta - \vec \theta' \mid^{2}},
        \label{eq:scaled_def_angle}
    \end{equation}
    where $\kappa = \Sigma(D_{\mathrm{d}} \vec \theta) / \Sigma_{\mathrm{crit}}$ denotes the convergence, a dimensionless quantity, related to the projected mass density $\Sigma$. It is scaled to the critical mass density $\Sigma_{\mathrm{crit}} = c^{2} /(4\pi G) \cdot D_{\mathrm{s}} /(D_{\mathrm{d}} D_{\mathrm{ds}})$.

    An important quantity in lensing is the magnification. The flux from a source is increased by this factor due to the light being focused by the lens. The magnification 
    \begin{equation}
        \mu = \frac{1}{\det A_{ij}} = \left[ \det \left( \frac{\partial \beta_{i}}{\partial \theta_{j}} \right) \right]^{-1}
        \label{eq:magnification}
    \end{equation}
    can be calculated using the Jacobian matrix $A_{ij}$, which gives the distortion of the lensed images. The determinant can vanish, leading to an infinite magnification. The curves where this happens are called critical curves on the lens plane and caustics on the source plane. In practice, the magnifications of the lensed sources near critical curves are high yet finite because geometric optics (thin-lens approximation) is no longer applicable in this regime, and wave optics is needed.

\section{Implementation}\label{Implementation}
    In the original implementation from~\citet{lensing_code_Oguri_2010}, all possible lens galaxies are described by two parameters, the redshift and velocity dispersion. Hence, the code calculates through a two dimensional grid. At each grid point, the number of galaxies is calculated, and for every lens galaxy, the number of possible \acp{tde} is drawn from a Poisson distribution. As an example, when considering $20,000 \; \mathrm{deg}^{2}$ on the sky, we calculate above $358$ billion galaxies up to a redshift of $2$. Drawing $1$ billion samples from a Poisson distribution takes on the order of one minute, which means a lot of time is spent on this portion of the code. 

    As all lens galaxies are considered the same at a given grid point, we only need to calculate the total number of possible \acp{tde} for $N$ galaxies. Here we can use the fact that the sum of two independent Poisson distributed variables is also Poisson distributed. When $X \sim \mathrm{Poisson}(\lambda)$ and $Y \sim \mathrm{Poisson}(\mu)$ then $X + Y \sim \mathrm{Poisson}(\lambda + \mu)$. Therefore, we do not need to draw $N$ times from a Poisson distribution with expectation value $\lambda$ but we can draw from a Poisson distribution with $N \cdot \lambda$ as its expectation value a single time, significantly speeding up calculations. 
    
    Other optimizations have been made by removing the need to instantiate large data arrays not needed for this application of the code, yielding significant performance improvement without changing the code's fundamental logic. These speed increases are needed as the rate of lensed \acp{tde} is close to or even below unity. This creates the need to over-sample the detection area by up to a factor of $1,000$ to find a few samples at faint magnitudes and bring the Monte Carlo error down.

\onecolumn
\section{Unlensed TDE distributions}\label{sec:unlensed_tde_hist}
\begin{figure}[htbp]
        \centering
        \includegraphics[width=\textwidth]{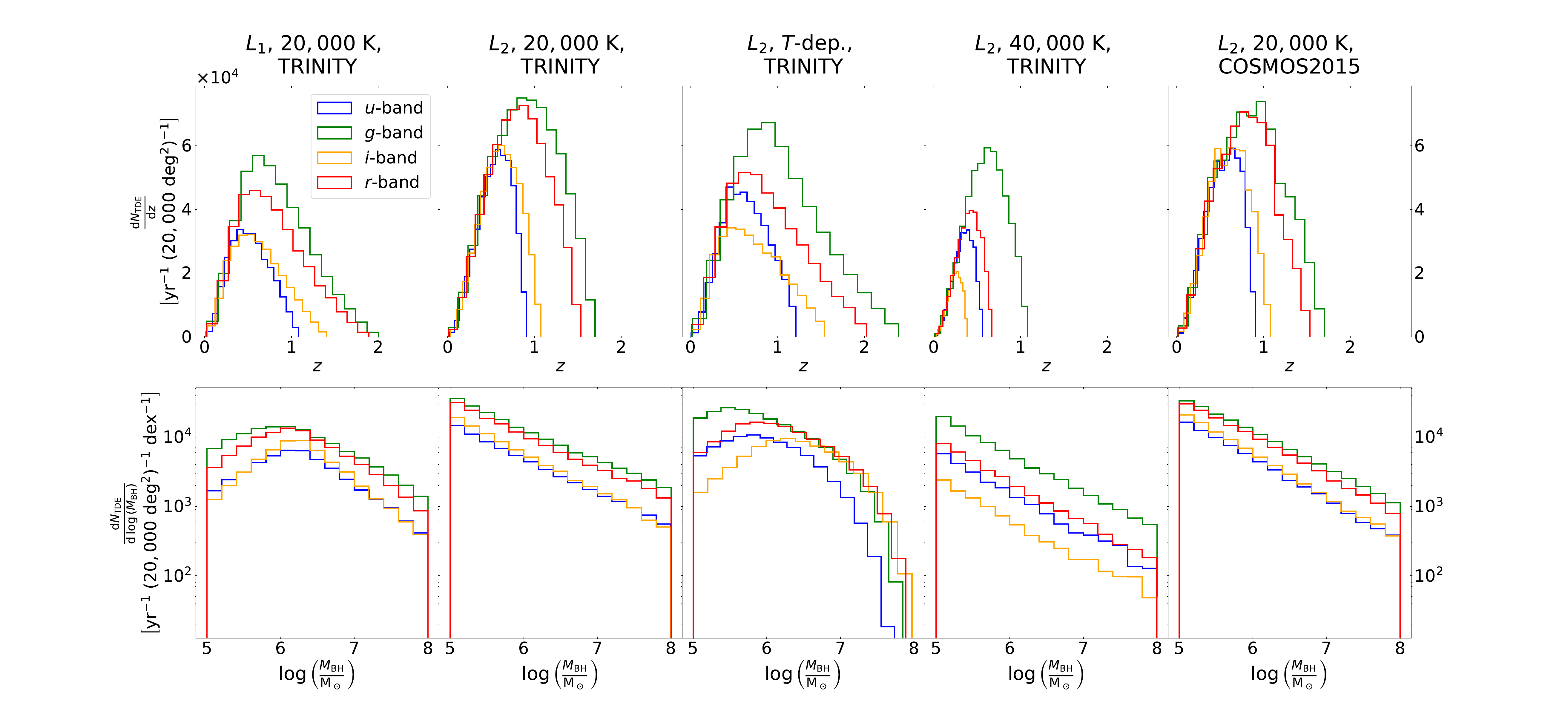}
        \caption{Differential rate of unlensed \acp{tde} for five representative models. The first row shows the \ac{tde} redshift, and the second row shows the \ac{bh} mass. These figures are calculated for \ac{lsst} magnitude limits.}
        \label{fig:unlensed_tde_distributions}
    \end{figure}

    In Fig.~\ref{fig:unlensed_tde_distributions}, we show the expected redshift and \ac{bh} mass distributions for five representative models calculated for \ac{lsst} limiting magnitudes. We expect \ac{lsst} to find most unlensed \acp{tde} at $z \lesssim 1.5$. $r$ and $g$-band are best suited to observe high redshift \acp{tde}, while $u$ and $i$-band only see \acp{tde} for $z \lesssim 1$. The differential \ac{tde} rate over the \ac{bh} mass follows the magnitude function. This shows clearly that the temperature-independent $L_{2}$ model prefers lower \ac{bh} masses, while the $L_{1}$ and the temperature-dependent models peak around $10^{6} \; \mathrm{M}_{\odot}$.

\section{Quads time delay distributions}\label{sec:quads_delays}
\begin{figure}[htbp]
        \centering
        \includegraphics[width=\textwidth]{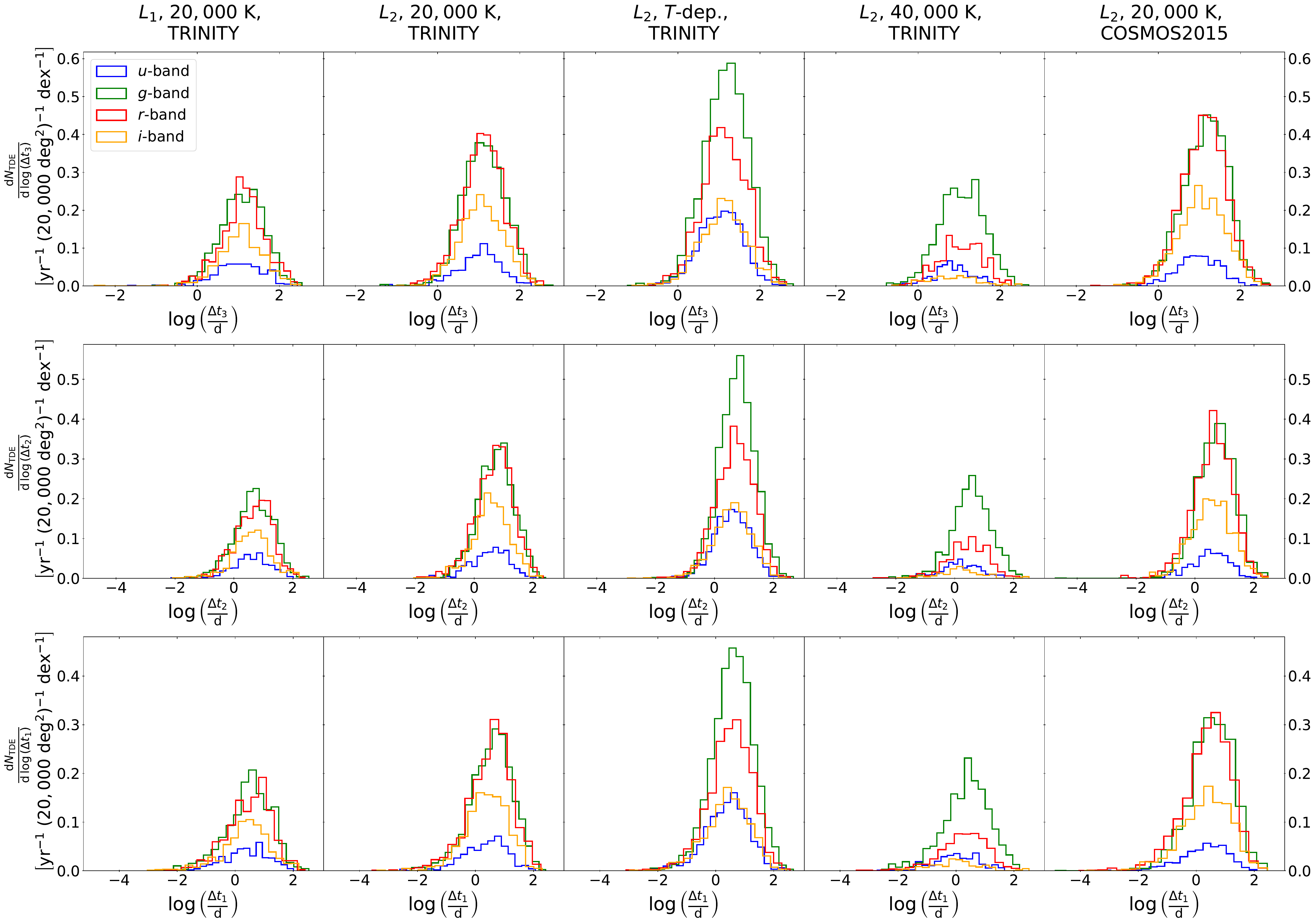}
        \caption{Histograms show the differential rate of \acp{tde} for five representative models. $\Delta t_{1}$ is the time delay between the first and second arriving image. $\Delta t_{2}$ is between the first and third image. $\Delta t_{3}$ is the first and last image. These figures are calculated for \ac{lsst} magnitude limits.}
        \label{fig:lensed_tde_hist_quads_delay}
    \end{figure}

    In Fig.~\ref{fig:lensed_tde_hist_quads_delay}, we show the time delay distributions for quad lenses calculated for \ac{lsst} limiting magnitudes. The earliest image is taken as the reference image. The time delay distributions for double lenses are shown in the last row of Fig.~\ref{fig:lensed_tde_hist}. The delay between the first and second image is $\Delta t_{1}$. The distributions show that we expect this time delay to be approximately $11$ days. The delay between the first and third image is $\Delta t_{2}$. This time delay is approximately $5$ days. The delay between the first and fourth image is $\Delta t_{3}$. This time delay is approximately $3$ days. We find a very similar time delay distribution to that of the double lenses. The quads exhibit signs of a low time delay tail that is not observed for the doubles. However, such systems seem rare and are only observed at numerical uncertainty.

\end{appendix}

\end{document}